\documentclass[a4paper,11pt]{article}

\usepackage[margin=.8in]{geometry}

\usepackage{etex}
\usepackage{array,tabularx,booktabs,geometry,graphicx,longtable}
\usepackage{mathrsfs}
\usepackage{amsmath}
\usepackage{amssymb}
\usepackage{textcomp}
\usepackage{array}
\usepackage[usenames, dvipsnames]{color}
\usepackage[font=small]{caption}
\usepackage{cite}
\usepackage[utf8]{inputenc}
\usepackage{amsthm}
\usepackage{multicol}
\usepackage{bbold}
\usepackage{cancel}
\usepackage[hyperfootnotes=false, linktocpage=true, colorlinks, citecolor=blue, linkcolor=blue, urlcolor=Maroon]{hyperref}
\usepackage{cleveref}
\usepackage{arydshln}

\numberwithin{equation}{section}

\newtheorem*{UnLemma}{Lemma}

\newcommand{\mc}{\mathcal}
\newcommand{\mbb}{\mathbb}

\newcommand{\be}{\begin{equation}}
\newcommand{\ee}{\end{equation}}
\newcommand{\bi}{\begin{itemize}}
\newcommand{\ei}{\end{itemize}}

\newcommand{\ec}{\chi}							
\newcommand{\ind}{\mathrm{ind}}					
\newcommand{\res}[1]{\tilde{#1}}		    	
\newcommand{\nod}[1]{{\cal{#1}}}			    
\newcommand{\defm}[1]{{#1}}			            
\newcommand{\norm}[2]{\mc{N}_{#1}}				
\newcommand{\smct}{\pi}							
\newcommand{\pb}[1]{\pi^*({#1})}					
\newcommand{\pt}[1]{\mathrm{P}({#1})}				
\newcommand{\spec}[1]{{#1}_{\text{0}}}				

\def\fnote#1#2{\begingroup\def\thefootnote{#1}\footnote{#2}
     \addtocounter{footnote}{-1}\endgroup}

\begin{document}
\vspace{1cm}

\title{{\Large \bf Branes and Bundles through Conifold Transitions \\ and Dualities in Heterotic String Theory  }}

\vspace{2cm}

\author{
Lara~B.~Anderson,${}{}$
Callum~R.~Brodie${}{}$ and
James~Gray${}{}$
}
\date{}
\maketitle
\begin{center} {\small ${}${\it Department of Physics, 
Robeson Hall, Virginia Tech \\ Blacksburg, VA 24061, U.S.A.}}

\fnote{}{lara.anderson@vt.edu}
\fnote{}{callumb@vt.edu}
\fnote{}{jamesgray@vt.edu}

\end{center}

\begin{abstract}
\noindent
Geometric transitions between Calabi-Yau manifolds have proven to be a powerful tool in exploring the intricate and interconnected vacuum structure of string compactifications. However, their role in ${\cal N}=1$, 4-dimensional string compactifications remains relatively unexplored. In this work we present a novel proposal for transitioning the background geometry (including NS5-branes and holomorphic, slope-stable vector bundles) of 4-dimensional, ${\cal N}=1$ heterotic string compactifications through a conifold transition connecting Calabi-Yau threefolds. Our proposal is geometric in nature but informed by the heterotic effective theory. Central to this study is a description of how the cotangent bundles of the deformation and resolution manifolds in the conifold can be connected by an apparent small instanton transition with a 5-brane wrapping the small resolution curves. We show that by a ``pair creation" process 5-branes can be generated simultaneously in the gauge and gravitational sectors and used to describe a coupled minimal change in the manifold and gauge sector. This observation leads us to propose dualities for 5-branes and gauge bundles in heterotic conifolds which we then confirm at the level of spectrum in large classes of examples. While the 5-brane duality is novel, we observe that the bundle correspondence has appeared before in the Target Space Duality exhibited by $(0,2)$ GLSMs. Thus our work provides a geometric explanation of $(0,2)$ Target Space Duality.
\end{abstract}

\thispagestyle{empty}
\setcounter{page}{0}
\newpage
\tableofcontents

\section{Introduction}

Geometric transitions 
\cite{Candelas:1988di,Green:1988wa,Green:1988bp,Candelas:1989js,Candelas:1989ug,Aspinwall:1993yb,Aspinwall:1993nu} connecting topologically distinct background geometries of string theory have long played a key role in the study of the string landscape and string effective field theories in diverse dimensions. In the best understood examples of Calabi-Yau threefold (CY3) compactifications, such transitions consist of 1) conifold transitions -- in which the complex structure of one CY3 manifold is tuned to a singular limit and then a new smooth manifold can be obtained via a small resolution, and 2) flop transitions - in which two CY3s are different small resolutions of the same nodal variety. In the case of CY3 manifolds (and more generally manifolds of $\mathrm{SU}(3)$ structure), it has been conjectured that \emph{all such topologically distinct manifolds can be connected by geometric transitions} \cite{reid}.

The most robust examples of such transitions being understood field-theoretically have arisen in ${\cal N}=2$ theories in 4-dimensions (see e.g.\ \cite{Strominger:1995cz,Greene:1995hu}). In contrast, geometric transitions in 4-dimensional theories exhibiting ${\cal N}=1$ supersymmetry have proven much harder to study both field-theoretically and geometrically. For example, in heterotic string theory, the background consists not only of a compact manifold, $X$, but also a non-trivial gauge bundle over it (more precisely a holomorphic, slope-stable vector bundle $V$) and possibly other non-perturbative elements such as NS5-branes. Thus, in a conifold transition, not only must the singular geometry of $X$ be addressed, but also the question of what happens to the bundle $V$ (or 5-branes) on this singular geometry. Moreover, singularities arising in the bundle in heterotic theories are known to sometimes lie outside the limits of ordinary field theory. A key example of this are the so-called ``small instanton transitions" \cite{Witten:1995gx} involving NS5-branes and singular limits of the bundle which are known to lead to tensionless, non-critical strings \cite{Ganor:1996mu}. Despite some exploration \cite{Candelas:2007ac}, it remains an open question whether geometric transitions can controllably be described in heterotic theories at all, or whether the presence of gauge fields/5-branes in heterotic theories could physically obstruct such a topology changing transition from taking place.

In this paper we will explore these questions in the context of conifold transitions in heterotic theories. In particular, \emph{we will outline a novel proposal for how the full heterotic background can naturally and consistently be taken through a conifold transition, and provide substantial evidence that this proposal is correct}.

Our approach is for the most part geometric in nature, though guided/informed by field theory. It is important to note that some limitations are forced on any would-be, purely field-theoretic analysis in the ${\cal N}=1$ heterotic context, since key information about the heterotic moduli space remains unknown. These missing ingredients include an explicit/analytic description of the matter field K\"ahler potential, and hence the full moduli space metric (see \cite{Candelas:2016usb,Candelas:2018lib,McOrist:2019mxh,Ashmore:2019wzb,Douglas:2020hpv,Anderson:2020hux,Ashmore:2021ohf,Larfors:2022nep} for some recent progress), as well as the fact that field theory alone (as opposed to tensionless string limits, SCFTs, etc.) is likely not sufficient to describe the relevant singularities in the manifold/bundle (as described above for small instantons).

One powerful constraint in heterotic compactifications arises from the mixed gauge/gravitational anomalies whose cancellation requires that
\begin{equation}
c_2(T_X)=c_2(V)+\left[C\right]
\label{anomaly1}
\end{equation}
where $C$ is an effective curve in the CY3, wrapped by a 5-brane. Across a conifold transition $X \to \res{X}$, the second Chern character of the CY3 manifold changes as
\begin{equation}
\text{ch}_2(T_{\res{X}})=\text{ch}_2(T_X)+\left[\mathbb{P}^{1}s\right]
\label{ch2_change1}
\end{equation}
where $X$ and $\res{X}$ are respectively the deformation and resolution manifolds of the conifold transition, and $\left[\mathbb{P}^{1}s\right]$ is the class of the small resolution curves. From this formula and \eqref{anomaly1} it is clear that if anomalies are to be cancelled consistently on each side of a conifold transition, the bundle (or 5-brane) must also change and ``compensate" for the change in the second Chern character of the geometry seen in \eqref{ch2_change1}. That is, the bundle/brane background must dynamically play a very non-trivial role in a heterotic conifold transition. This is in contrast to some early studies of heterotic bundles in conifold transitions which focused on so-called ``spectator" bundles which changed as little as possible through the transition (and in particular had an unchanged second Chern character) \cite{Candelas:2007ac}.

In the following sections we will outline a proposal for how such a coupled change to manifold and bundle (or brane) occurs via a kind of pair creation process (in the singular limit) in which 5-branes are created in both the gauge and gravitational sectors of the theory simultaneously before being ``absorbed" back into the holomorphic cotangent bundle and the background gauge bundle (or brane configuration) respectively. Written on the resolution side of the conifold, this pair creation contributes to the anomaly cancellation condition as
\begin{equation} \label{intropair}
c_2(T_{\res{X}})+ \left[\mathbb{P}^{1}s\right]=c_2(\res{V})+[\tilde{C}]+ \left[\mathbb{P}^{1}s\right]~.
\end{equation}
where $\tilde{V},\tilde{C}$ are respectively a vector bundle and effective curve in $\tilde{X}$. Viewing this as an addition of charges, we see the pair creation process cancels out of the anomaly due to the opposite signs arising in the gravitational vs.\ gauge sectors of the theory. In order to connect this observation to conifold transitions, we make note of the fact that geometrically the cotangent bundles of the manifolds $X$ and $\res{X}$ are connected via the absorption of a 5-brane (i.e.\ a small instanton) wrapping the class $\left[\mathbb{P}^{1}s\right]$ as
\begin{equation}
0 \to \pb{\Omega_{\nod{X}}} \to \Omega_{\res{X}} \to \mc{O}_{\mbb{P}^1\mathrm{s}}(-2) \to 0
\label{eq:rel_cot_seq_intro}
\end{equation}
where $\nod{X}$ is the singular (i.e.\ nodal) variety, and $\smct \colon \res{X} \to \nod{X}$ the small contraction (see Figure~\ref{conifoldtrans}). As we will review in Section~\ref{sec:gravinst}, the form of this short exact sequence is exactly that of a so-called ``Hecke transform", which mirrors geometrically the way that 5-branes can be absorbed into bundles during a heterotic small instanton transition. This process effects a part of the conifold transition (bringing the resolution geometry to the nodal limit), and thus we can ask what happens in the gauge sector.

The answer to this latter question is a similar small instanton (i.e.\ 5-brane) absorption into the gauge sector, but here the geometry of the curves is more rich/subtle and involves curves which play a key role in the conifold geometry. In particular, curves in $X$ and $\res{X}$ which enhance to Weil non-Cartier divisors in the nodal limit. To elucidate the geometric details of this process we approach it in two steps.
\begin{enumerate}
\item We identify special curves from the point of view of a conifold pair of CY3 manifolds ($X,\res{X}$) which allow a 5-brane wrapped on them to move through the conifold transition in an anomaly-consistent manner (absorbing the required $\left[\mathbb{P}^1s\right]$ from \eqref{intropair}).
\item Once these curves are identified it is possible to merge them consistently into a variety of bundles via small instanton transitions (i.e.\ Hecke transforms). This allows us to extend the correspondence found for 5-branes to one of vector bundles.
\end{enumerate}

As part of the pure 5-brane study of the transition enumerated in (1) above, we find that the total degrees of freedom of the theory (including the vector and chiral multiplets) are preserved across the conifold transition. In particular the massless singlets of the 5-brane-only theory as counted by
\begin{equation}
h^{1,1}(X)+h^{2,1}(X)+h^0(C,{\cal N}_C) = h^{1,1}(\res{X})+h^{2,1}(\res{X})+h^0(\res{C},\res{{\cal N}}_{\res{C}})
\end{equation}
agree perfectly (despite the fact that the Hodge numbers of the CY3 are changing). In the above formula $C \subset X$ and $\res{C} \subset \res{X}$ and ${\cal N}_C$ and $\res{{\cal N}}_{\res{C}}$ are their respective normal bundles.
This complete matching of the low energy effective theory provides evidence that we have uncovered a new form of heterotic 5-brane duality. We study the geometry of these curves and their role in the conifold transition.

Furthermore, with this new heterotic 5-brane duality in hand, we can extend our observations back into the perturbative limit by performing small instanton transitions (i.e.\ Hecke transforms) to extrapolate a duality for heterotic gauge bundles. Once again we find pairs of theories for which the complete massless spectrum is identical. At this point, it is intriguing to note that the perturbative duality we find is not wholly new. Upon forming our geometric results on bundles/conifolds, we find that we are able to reproduce the geometry of another known (conjectural) duality of heterotic theories -- the ``Target Space Duality" (TSD) of $(0,2)$ Gauged Linear Sigma Models (GLSMs) \cite{Distler:1995bc,Blumenhagen:1997vt,Blumenhagen:1997cn,Blumenhagen:2011sq,Rahn:2011jw}. This matching provides a deep and non-trivial confirmation of the validity of our approach. In heterotic TSD two $(0,2)$ GLSMs share a non-geometric phase (i.e.\ a Landau-Ginzburg or hybrid phase made identical by a non-trivial relabeling of fields). Upon extending each GLSM back to a geometric phase it can be observed that two very different 2-dimensional theories appear to give rise to 4-dimensional ${\cal N}=1$ heterotic theories with identical charged and uncharged massless spectra. In particular, in terms of singlets,
\begin{equation}
h^{1,1}(X)+h^{2,1}(X)+h^1(X,\text{End}_0(V))=h^{1,1}(\res{X})+h^{2,1}(\res{X})+h^1(\res{X},\text{End}_0(\res{V})) \,.
\label{singlet_match1}
\end{equation}
Although target space duality has been observed in the GLSM literature for several decades, it was unclear why conifold singularities of the CY3 manifolds in the geometric phase were arising and an open question as to \emph{why} the target space spectrum (including \eqref{singlet_match1}) was identical. Our work provides the first answers to these questions from a geometric/target space point of view. Furthermore, we have found that {\it every} example of target space duality that we have studied consists of a single transition of the type we discuss in this paper, or a chain of such processes. This detailed structure, revealed to be underlying TSD, provides considerable evidence for our proposal.

Unlike in mirror symmetry where a pair of Type II theories lead to the same physics, in our examples and target space duality, whole chains of heterotic manifolds/bundles can be found connected by conifold transitions which lead to the same spectrum. Moreover, recent work has indicated that the form of non-trivial scalar potentials also match across such chains \cite{Anderson:2016byt}. This gives hope that this geometric correspondence may underlie some deeper true duality of heterotic theories. 

Importantly we find that not all bundles on CY3s can traverse a conifold transition (beginning from either the deformation or resolution side) in this manner. Instead, only those with special properties (which we outline) can be taken across consistently. In a heterotic theory it remains an open question just how much manifold/bundle topology determines the moduli space of the heterotic theory and this work could shed light on how such moduli spaces can be extended through the whole interconnected web of CY3s (or more generally $\mathrm{SU}(3)$ structure manifolds). We will return to such broader moduli space questions in future work \cite{us_to_appear}. 

It should be noted that the mathematical questions/results underpinning this analysis are by necessity intricate since we are studying not only singular limits of CY threefolds and holomorphic, slope-stable vector bundles over them, but also extrapolating such structures across conifold transitions. We have explored this geometry in a multitude of examples and have provided proofs in as much generality as possible. However, due to the complexity above, for some results it is beyond the scope of this work to prove them in complete generality for any threefold/bundle and we restrict ourselves to certain classes of examples (i.e.\ toric complete intersection threefolds, etc.). We have tried to be clear throughout this work about the level of generality of each result.

The structure of this paper is as follows. In Section~\ref{sec2} we review necessary background material on the geometry of conifold transitions in CY3s and then provide a novel interpretation of the change in the cotangent bundle in terms of gravitational/gauge sector ``pair creation" in the theory. In a series of Subsections we provide brief descriptions of the geometric ``rules" for carrying both 5-branes and bundles backgrounds through conifold transitions and illustrate these with a simple, explicit example. In Section~\ref{sec:5brane_duality} we explore the 5-brane duality in more detail including providing arguments for why the spectrum of the theory is preserved across the transition. In Section~\ref{sec:tsd} we detail the correspondence of the full theories for heterotic bundles across conifold transitions including spectrum/moduli matching. Moreover we explore in detail the relationship between our results and $(0,2)$ heterotic target space duality. The appendices provide a number of useful technical results. In particular, in the process of describing how bundles can be ``transitioned" through a conifold, we provide the first general description/count of how moduli change for a heterotic small instanton transition which is a useful addition to the literature (see Appendix~\ref{app:hecke}).

\section{Bundles and branes through conifold transitions} \label{sec2}

\subsection{Conifold transitions}
\label{sec:con_trans}

\begin{figure}[!t]\centering
\includegraphics[scale=0.65]{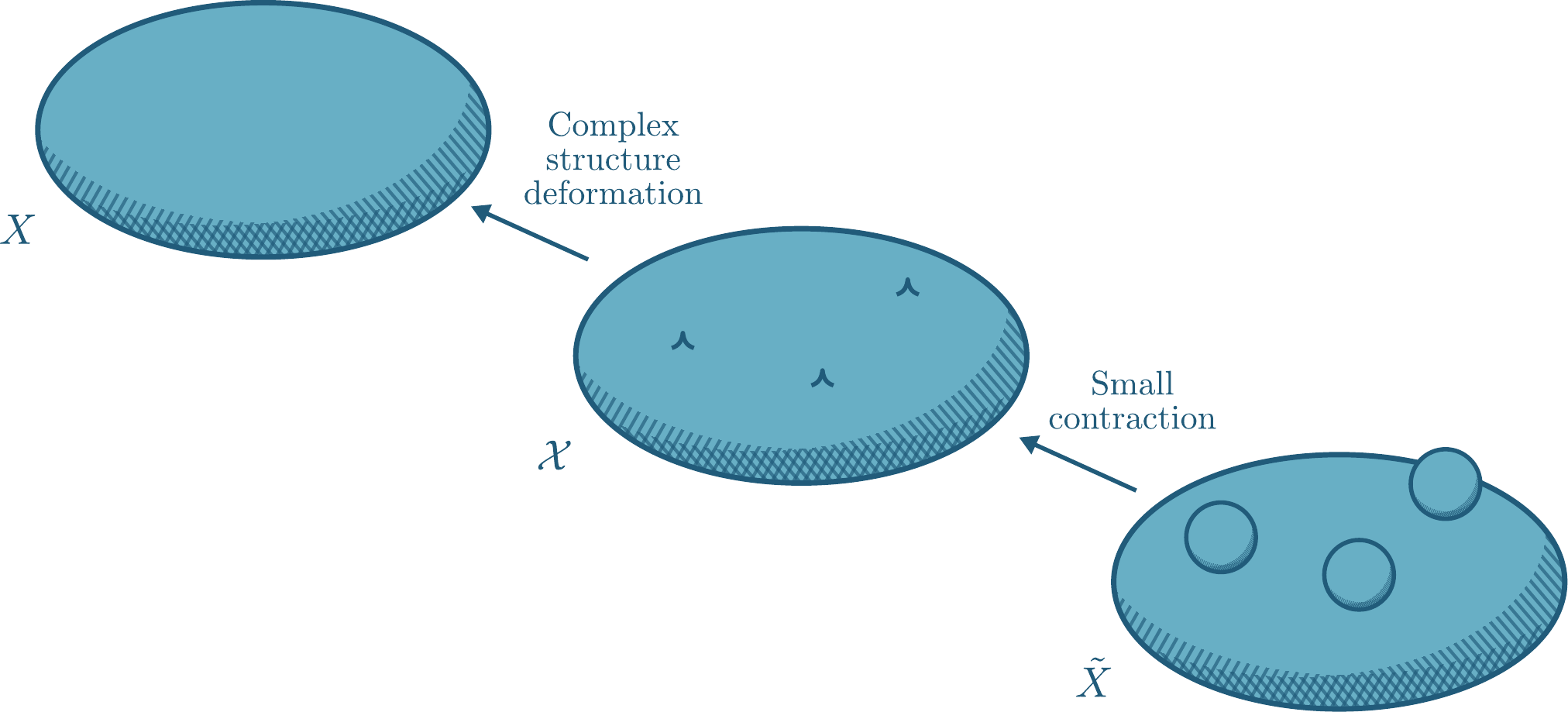}\caption{{\it A schematic depiction of a conifold transition between CY3s, as described in the text.}}
\label{conifoldtrans}
\end{figure}

Conifold transitions between smooth CY3s can be described in the following manner. Beginning with an initially smooth variety $\defm{X}$, referred to as the deformation side of the transition, the complex structure is tuned until singularities appear at a number of isolated points. We shall refer to the singular CY3 thus obtained as the nodal variety, $\nod{X}$. The singular points are then replaced with $\mbb{P}^1$s (in what is called a `small resolution') whose volumes are controlled by one or more new K\"aher moduli. Upon performing this small resolution one arrives at a new smooth CY3 which is referred to as the resolution side of the transition, $\res{X}$. One can also consider the transition in the other direction, performing first a small contraction on $\res{X}$ and then deforming the complex structure of the resulting $\nod{X}$ to generic values, thus arriving at $\defm{X}$. This process is depicted schematically in Figure~\ref{conifoldtrans}. The collection of $\mbb{P}^1$ curves which are involved in the small resolution are referred to as the exceptional locus. We note that in this paper we will only consider conifold transitions where the normal bundle to the exceptional locus, restricted to those curves, takes the form ${\cal O}(-1) \oplus {\cal O}(-1)$.

Despite being arguably the simplest example of a topological transition, a conifold transition nonetheless has rather drastic consequences for a number of topological invariants, including many which are important in the context of compactifications of the heterotic string. In the remainder of this subsection we discuss some of these changes, as well as their importance. 

\medskip

Perhaps the most obviously important topological quantities in the context of compactifying the heterotic string are the Hodge numbers and the second Chern character of the manifold. The change in the Hodge numbers is most clearly seen by first considering the simpler quantity of the Euler characteristic. Since the Euler characteristic $\ec(X)$ of a manifold $X$ is additive under surgery, and since $\ec(\mbb{P}^1) = 2$, we have 
\be \label{eulerp1s}
\ec(\res{X}) - \ec(\defm{X}) = 2\,\#(\mbb{P}^1\mathrm{s}) \,,
\ee
where $\#(\mbb{P}^1\mathrm{s})$ is the number of resolution $\mbb{P}^1$s in the transition, or equivalently the number of singular points on $\nod{X}$. The Euler characteristic of a smooth CY3 Y is related to the Hodge numbers by $\ec(Y) = 2\big( h^{1,1}(Y) - h^{2,1}(Y) \big)$. Moreover, since during a conifold transition a number of complex structure moduli become frozen at special values, while new K\"ahler moduli appear, the Hodge numbers $h^{2,1}$ and $h^{1,1}$ must be altered as
\be \label{defdeltas}
h^{1,1}(\res{X}) = h^{1,1}(\defm{X}) + \Delta(h^{1,1}) \,, \quad h^{2,1}(\res{X}) = h^{2,1}(\defm{X}) - \Delta(h^{2,1}) \,.
\ee
Hence, using (\ref{eulerp1s}), we have that
\be \label{preres1}
\Delta(h^{2,1}) =\#(\mbb{P}^1\mathrm{s})  -\Delta(h^{1,1})  \,.
\ee
In the context of compactifying the heterotic string, the importance of this change is that the number of K\"ahler and complex structure moduli in the theory is then altered, using (\ref{defdeltas}), as
\be \label{res1}
h^{1,1}(\res{X})+h^{2,1}(\res{X}) = h^{1,1}(\defm{X})+h^{2,1}(\defm{X}) - \#(\mbb{P}^1\mathrm{s}) + 2 \Delta(h^{1,1})\,.
\ee

\vspace{0.2cm}

In a conifold transition, the second Chern character $\mathrm{ch}_2(X)$ of the manifold grows a contribution exactly equal to the Poincar\'e dual of the curve class of the exceptional locus of the resolution. By a slight abuse of notation we can write this in the following manner\footnote{More precisely, we should say that $\mathrm{ch}_2(\res{X}) =\smct^*( \mathrm{ch}_2(\nod{X}) )+ [\mbb{P}^1\mathrm{s}]$ where $\smct$ is the small contraction map. This relation can easily be derived from the cotangent sequence (\ref{eq:rel_cot_seq}) that we will introduce in Section~\ref{sec:gravinst}, using the fact that $\mathrm{ch}_2(\mc{O}_{\mbb{P}^1\mathrm{s}}(-2)) = [\mbb{P}^1\mathrm{s}]$. For the Chern character of the nodal variety we should more properly refer to the relevant Chern-Schwarz-Macpherson (``CSM") class, but this subtlety does not affect the discussion of this paper. In addition, the `second Chern class' of $\nod{X}$ derived from the CSM class of the nodal variety is the same as the second Chern class $c_2(\defm{X})$ of the deformation geometry in every case we have checked. The code \cite{Paolo} was used in checking these examples.\label{fn:c2_rel}}.
\be
\mathrm{ch}_2(\res{X}) = \mathrm{ch}_2(\defm{X}) + [\mbb{P}^1\mathrm{s}] \,.
\label{eq:ch2_rel}
\ee
In terms of second Chern classes, this condition reads as follows.
\be 
c_2(\res{X}) = c_2(\defm{X}) - [\mbb{P}^1\mathrm{s}] \,.
\label{eq:ch2_rel2}
\ee
The importance of this change, in the context of compactifying the heterotic string, comes in considering the gravitational contribution to the anomaly cancellation condition,
\be \label{canom}
c_2(X) = c_2(V) + [C] \,.
\ee
In this expression, $V$ is the gauge bundle and $[C]$ is the Poincar\'e dual to the curve class wrapped by any 5-branes present in the background. Since the gravitational contribution to (\ref{canom}) is altered in the transition as in (\ref{eq:ch2_rel2}), the gauge sector of the theory will also have to be altered to counteract this new contribution and so preserve anomaly cancellation.

\subsubsection{Example}\label{firsteg}

Throughout this paper, we will illustrate our discussion with a simple and explicit example. In this section we will describe the CY3 geometries involved in this case, while the associated gauge and 5-brane structures will be presented as they are introduced in later subsections. To describe the conifold transition underlying our example, we will describe the smooth CY3s involved and then perform a small contraction and complex structure deformation respectively to illustrate how they meet at a nodal variety.

Let us start with the resolution variety. For this purpose, we will consider the following CY3, which is a complete intersection in a product of projective spaces, or ``CICY". 
\be
\res{X}=X_{7885} =
\left[
\begin{array}{c | c c}
\mbb{P}^1 & 1 & 1 \\
\mbb{P}^4 & 1 & 4 
\end{array}
\right] \,.
\ee
This description of the manifold, which has the identifier 7885 in the exhaustive list of CICY threefolds first described in \cite{Candelas:1987kf,ciCY3list}\footnote{Closely related data sets can be found here \cite{Gray:2013mja,Gray:2014fla,Anderson:2017aux,Gray:2021kax}.}, is called a  configuration matrix. Each row corresponds to one of the projective spaces in the product making up the ambient space, while each column contains the multi-degrees of one of the equations which describe the manifold as a complete intersection in that ambient space. In the present example, this means that the geometry is described by two equations inside $\mbb{P}^1 \times \mbb{P}^4$, namely:
\be \label{7885config}
X_{7885} \colon ~ \big\{ (x,y) \in \mbb{P}^1[x] \times \mbb{P}^4[y] ~\big|~x_0 l_0(y) + x_1 l_1(y) = 0 ~\mathrm{and}~ x_0 q_0(y) + x_1 q_1(y) = 0 \big\} \,.
\ee
Here, the $l_i$ and the $q_i$ are respectively arbitrary degree-one and degree-four polynomials in the homogeneous coordinates $y_0,\,\ldots\,,y_4$ of the ambient $\mbb{P}^4$, while $x_0$ and $x_1$ are the homogeneous coordinates of the ambient $\mbb{P}^1$.

Considered as equations in $x$ with coefficients which vary as one moves around in $y$, these two equations generically have no solution, except when the following determinant vanishes.
\be \label{nodeg}
l_0\,q_1 - l_1\,q_0 = 0 
\ee
Therefore, this equation is satisfied at all points on the CY3. Additionally, if all four polynomials vanish,
\be
l_0 = l_1 = q_0 = q_1 = 0 \,,
\ee
then $x$ is unconstrained. This means that over each such point in $\mathbb{P}^4$ there is an entire $\mbb{P}^1$. Hence, the geometry of $X_{7885}$ can be described as the hypersurface $\{ y \in \mbb{P}^4[y] ~|~ l_0\,q_1 - l_1\,q_0 = 0 \}$ inside $\mbb{P}^4$ with the addition of $\mbb{P}^1$s at the 16 points where $l_0 = l_1 = q_0 = q_1 = 0$. The hypersurface (\ref{nodeg}) in $\mbb{P}^4[y]$ is singular at precisely these points since all of the derivatives of the equation vanish there, and hence $X_{7885}$ is the small resolution $\res{X}$ of this nodal hypersurface, which we refer to as $\nod{X}$.

\vspace{0.1cm}

To obtain the deformation side of the transition one can simply deform the equation (\ref{nodeg}) of the nodal hypersurface in $\mbb{P}^4[y]$ to a generic polynomial of the same degree. This gives rise to a smooth manifold $\defm{X}$ described by a quintic polynomial inside $\mbb{P}^4$, which can also be described by a configuration matrix, having identifier 7890 in the list of CICY threefolds.
\be
X=X_{7890} =
\left[
\begin{array}{c | c}
\mbb{P}^4 & 5
\end{array}
\right] 
\ee

\vspace{0.1cm}

The process of shrinking the $\mbb{P}^1$s inside $X_{7885}$ to give a nodal hypersurface before deforming to give $X_{7890}$, or indeed the reverse process, is manifestly a conifold transition. Such a description of a conifold transition between CICYs is typically known as a `$\mbb{P}^1$-split' \cite{Candelas:1987kf,Candelas:1989ug}, since the ambient space in going from $\defm{X}$ to $\res{X}$ is altered by the introduction of a $\mbb{P}^1$ factor, and the degree of a defining polynomial of $\defm{X}$ is split across multiple defining polynomials of $\res{X}$.

\vspace{0.1cm}

It will turn out to be useful to describe the quintic $\defm{X}$ in a somewhat redundant fashion as follows.
\begin{eqnarray} \label{redunq}
\defm{X} = \left[ \begin{array}{c|cc} \mathbb{P}^1 & 1& 0 \\ \mathbb{P}^4&0&5\end{array} \right]
\end{eqnarray}
The linear equation associated to the first numerical column of this matrix can simply be solved to obtain a point in $\mathbb{P}^1$. Thus, this matrix describes the direct product of a point with the quintic, that is the quintic manifold itself. The advantage of this description is that (\ref{redunq}) and (\ref{7885config}) are now described in terms of the same ambient space. This will be practically expedient in future discussions.

\vspace{0.2cm}

Let us connect this example with the topological properties discussed earlier in this section. As noted above, this conifold transition involves 16 nodal points, or equivalently 16 exceptional $\mbb{P}^1$s. The Euler characteristics and Hodge numbers on either side of the transition are
\be
\begin{array}{c c}
\ec(\res{X}) = -168, & \ec(\defm{X}) = -200 \,, \vspace{.15cm} \\
h^{1,1}(\res{X}) = 2 \,, ~ h^{2,1}(\res{X}) = 86 \,, & h^{1,1}(\defm{X}) = 1 \,, ~ h^{2,1}(\defm{X}) = 101 \,. \\
\end{array}
\ee

\vspace{0.1cm}

The second Chern classes of the two geometries are as follows.
\be
c_2 (\res{X}) = 5 J_0 J_1 + 6 J_1^2 \,, \quad c_2 (\defm{X}) = 10 J_1^2 \,
\ee
Here,  $J_0$ and $J_1$ are respectively the K\"ahler forms of the ambient $\mbb{P}^1$ and $\mbb{P}^4$ factors, restricted to $\res{X}$, and in a slight abuse of notation we also write $J_1$ in the case of the restriction of the K\"ahler form of $\mbb{P}^4$ to $\defm{X}$ in the description (\ref{redunq}). We also note the class of the set of exceptional $\mbb{P}^1$s,
\be
[\mbb{P}^1\mathrm{s}] = -5 J_0 J_1 + 4 J_1^2 \,.
\ee
(This can be established directly from the normal bundle of the exceptional set, which we determine in \eqref{eq:gcicy_reps_ex} below.)

From the above results we see that the claimed general relationships (\ref{preres1}), (\ref{res1}) and (\ref{eq:ch2_rel2}) do indeed hold in this example.

\vspace{0.2cm}

Although the discussion of this section does indeed convey the structure of a conifold transition correctly, it is useful to view the process in a different manner. By viewing the transition as corresponding to a certain modification of the cotangent bundle of the variety, the way is opened to an understanding of how gauge bundles can be consistently taken through the conifold. It is to this reinterpretation of conifold transitions, as a small instanton transition in the cotangent bundle, that we now turn. 

\subsection{The conifold as a gravitational small instanton transition} \label{sec:gravinst}

In understanding how gauge field backgrounds behave during conifold transitions it will turn out to be useful to view the geometry of these processes in terms of the dynamics of the cotangent bundle of the manifolds. However, describing the change in the cotangent bundle is inherently difficult because not only the bundle but also the base geometry over which it is defined is altered during the transition. This is in contrast to the simpler case of describing changes in a gauge bundle in a standard small instanton transition or Higgsing process. In those cases, since the geometry on which the bundle lives is fixed, such a change can typically be described using the formalism of an exact sequence, in which two of the terms are the old and the new gauge bundles, while the other terms, as well as the maps between them, give a reasonably explicit and well-controlled description of how these two are related.

As an example of the simpler situation, consider the case of a small instanton transition in the gauge bundle \cite{Witten:1995gx,Ovrut:2000qi,Buchbinder:2002ji}. Here, beginning with some gauge bundle $V$, a 5-brane wrapping a curve locus $C$ is absorbed into the bundle, and as a result a new sheaf $\hat{V}$ is produced (which may then be smoothed to give a final gauge bundle). This absorption of a small instanton is an example of a change which is described by a short exact sequence, namely
\be \label{hecke1}
0 \to \hat{V} \to V \to \mc{F}_C \to 0 \,.
\ee
Here $\mc{F}_C$ is a sheaf with support precisely on the locus $C$ which the 5-brane wraps. This is the appropriate description of the 5-brane for this context. The short exact sequence (\ref{hecke1}) is referred to as a `Hecke transform' in \cite{Ovrut:2000qi}. It is worth noting, for later sections of this paper, that while the small instanton transition described by a Hecke transform of the form (\ref{hecke1}) will be rank-preserving: $\textnormal{rk}(V) =\textnormal{rk}(\hat{V})$, one can modify the short exact sequence to obtain more general results. For example, if one were to obtain $\hat{V}$ via the following Hecke transform instead \cite{Ovrut:2000qi}
\be \label{hecke2}
0 \to \hat{V} \to V \oplus {\cal O}\to \mc{F}_C \to 0 \,,
\ee
one would obtain a $\hat{V}$ whose rank is one greater than that of $V$. This will be important in later sections where we will indeed encounter such transitions.

Sequences such as (\ref{hecke1}) and (\ref{hecke2}) are defined over a single base geometry.  This is the origin of the difficulty of finding such a description of the change in the cotangent bundle during a conifold transition, since the cotangent bundles $\Omega_{\defm{X}}$ and $\Omega_{\res{X}}$ of the deformation and resolution manifolds respectively are defined over different spaces. One possible solution to this difficulty would be to find a way to capture $\Omega_{\defm{X}}$ as a bundle or sheaf on the resolution geometry $\res{X}$. The object obtained in this fashion would no longer bear the intimate connection with the geometry that a cotangent bundle would. However, it could still, in a precise way, encode the structure of $\defm{X}$ thus providing all of the information required. If such a sheaf could be found then one may hope to be able to write down a single exact sequence which describes the change in the cotangent bundle during the conifold transition.

\begin{figure}[!t]\centering
\includegraphics[scale=0.65]{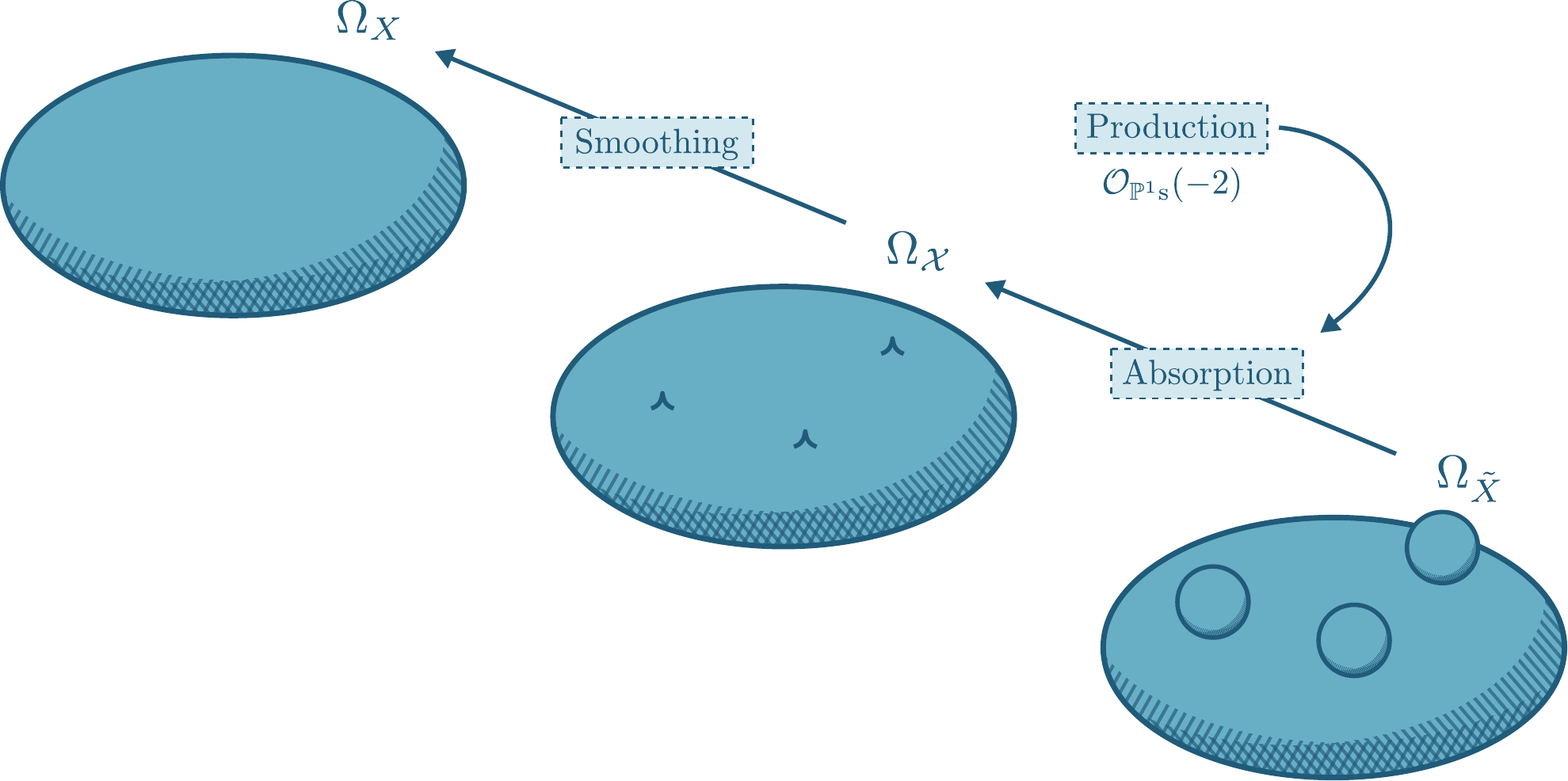}
\caption{{\it The conifold transition as a small instanton transition in the gravitational sector. As explained in the text, from the resolution side $\res{X}$, small instantons $\mc{O}_{\mbb{P}^1\mathrm{s}}(-2)$ with support on the exceptional $\mbb{P}^1$s are produced and absorbed into the cotangent bundle $\Omega_{\res{X}}$ to give the cotangent sheaf $\Omega_{\nod{X}}$ of the singular geometry, which is then smoothed to give the cotangent bundle $\Omega_{\defm{X}}$ of the deformation geometry.}}
\label{grav_inst_trans}
\end{figure}

In fact, such a description arises entirely naturally in the present case of the conifold transition. One can show that there exists the following `relative cotangent sequence' between the resolution and nodal geometries,\footnote{A sequence such as this always exists for any two varieties with a morphism between them, however, it is not always short exact on the left. While we have not been able to find a general proof of (\ref{eq:rel_cot_seq}) for an arbitrary conifold transition, we have been able to prove it for large classes of constructions, such as all $\mathbb{P}^1$-splits of CICYs for example. In addition, even in the general case, evidence can be provided for (\ref{eq:rel_cot_seq}) by showing that any extra term on the left of the sequence would have to consist of a sheaf which has both entirely vanishing cohomology and Chern classes.}
\be
0 \to \pb{\Omega_{\nod{X}}} \to \Omega_{\res{X}} \to \mc{O}_{\mbb{P}^1\mathrm{s}}(-2) \to 0 \,.
\label{eq:rel_cot_seq}
\ee
Here, $\smct \colon \res{X} \to \nod{X}$ is the small contraction map, $\Omega_{\nod{X}}$ is the cotangent sheaf of the nodal variety (a sheaf since $\nod{X}$ is singular), and $\mc{O}_{\mbb{P}^1\mathrm{s}}(-2)$ is a sheaf with support on the exceptional $\mbb{P}^1$s, given by taking the pushforward of the line bundle $\mc{O}(-2)$ under each embedding $\mbb{P}^1 \hookrightarrow{} \res{X}$. The first term in the relative cotangent sequence, being simply a pullback of the cotangent sheaf of the nodal variety, contains all of the geometric information about $\Omega_{\nod{X}}$, but represents this in an object on the resolution geometry. Hence, this sequence captures the relationship between the cotangent sheaf of the nodal variety and the cotangent bundle of the resolution manifold. 

The particularly striking feature of the short exact sequence (\ref{eq:rel_cot_seq}) is that the first two objects have support over the entire manifold, while the third object has support only over a curve. In other words, this short exact sequence is precisely what one would interpret as the Hecke transform describing a small instanton transition, were it to occur in the gauge sector. Although the sequence (\ref{eq:rel_cot_seq}) is defined entirely on $\res{X}$ the transition only really occurs when the system goes through the nodal point in moduli space. At this stage, the object $\pb{\Omega_{\nod{X}}} $ truly becomes the cotangent sheaf of the variety over which it is defined and the transition then completes via a smoothing of the nodal variety to obtain $\defm{X}$. This smoothing is exactly analogous to the manner in which $\hat{V}$ is smoothed out into a gauge bundle in a standard small instanton transition. The complete process is depicted schematically in Figure~\ref{grav_inst_trans}.

We see, therefore, that a conifold transition is precisely described by what we could call a small instanton transition in the gravitational sector. A different, less concise but somewhat more explicit, description of this transition to (\ref{eq:rel_cot_seq}) is detailed in Appendix~\ref{sec:sequencecomb_grav}. In the context of the present paper, this transition between cotangent bundles will be particularly important in that it will guide us towards a proposal for how the gauge bundle and 5-branes should behave during a conifold transition in order to correctly interact with the changes in the gravitational sector throughout the transition. It is to this topic that we will turn, after illustrating the preceding discussion with an example.

\subsubsection{Example}

In our example of a conifold transition, as described in Section~\ref{sec:con_trans}, we have the following two descriptions of the cotangent bundles of the resolution and nodal varieties.
\begin{eqnarray} \label{cotangtext1}
0 \to {\cal O}(0,-5) \to \Omega_{\mathbb{P}^4} \to  \smct^*(\Omega_{\nod{X}}) \to 0 \\  \label{cotangtext2}
0 \to {\cal O}(-1,-1) \oplus {\cal O}(-1,-4) \to \Omega_{\mathbb{P}^1} \oplus \Omega_{\mathbb{P}^4} \to \Omega_{\res{X}} \to 0
\end{eqnarray}
In these expressions $\Omega_{\mathbb{P}^1}$ and $\Omega_{\mathbb{P}^4}$ are the cotangent bundles of the indicated projective spaces, which in the notation of this example have the following Euler sequence resolutions.
\begin{eqnarray}
0 \to \Omega_{\mathbb{P}^1} \to {\cal O}(1,0)^{\oplus 2} \to {\cal O} \to 0 \\ \nonumber
0 \to \Omega_{\mathbb{P}^4} \to {\cal O}(0,1)^{\oplus 4} \to {\cal O} \to 0 
\end{eqnarray}

The normal bundle to the exceptional locus in $\res{X}$ is ${\cal O}(-1,1)\oplus {\cal O}(-1,4)$ in this example. To see this one may write out so-called `gCICY' representatives of general examples of the global sections of these line bundles \cite{Anderson:2015iia}. Given defining relations of the form given in (\ref{7885config}) these are as follows.
\begin{eqnarray} \label{eq:gcicy_reps_ex}
\frac{l_0}{x_1} \sim - \frac{l_1}{x_0} \in H^0(\res{X},{\cal O}(-1,1)) \\ \nonumber
\frac{q_0}{x_1} +c \frac{l_0}{x_1} \sim -\frac{q_1}{x_0} +c \frac{l_1}{x_0} \in H^1(\res{X},{\cal O}(-1,4))
\end{eqnarray}
In these expressions, $c$ is a general cubic in the homogeneous coordinates of $\mathbb{P}^4$. Clearly these sections vanish iff $l_0=l_1=q_0=q_1=0$, which is precisely the exceptional locus in $\res{X}$.

Given that the normal bundle takes this form, one can write the following twisted Koszul resolution of $\mc{O}_{\mbb{P}^1\mathrm{s}}(-2)$.
\begin{eqnarray} \label{exceprestext}
0 \to {\cal O}(0,-5) \to {\cal O}(-1,-1) \oplus {\cal O}(-1,-4) \to {\cal O}(-2,0) \to {\cal O}_{\mathbb{P}^1\mathrm{s}}(-2) \to 0
\end{eqnarray}

Given the above, one can form a commuting grid of sequences which has (\ref{eq:rel_cot_seq}) as a top row, and the resolutions of the objects in that sequence as given in (\ref{cotangtext1}),(\ref{cotangtext2}) and (\ref{exceprestext}) arrayed vertically underneath it. Diagram chasing this grid and using what is essentially the nine lemma \cite{ninelemma}, one can indeed prove that the sequence (\ref{eq:rel_cot_seq}) is well defined and short exact as claimed.

\subsection{The heterotic conifold as gauge-gravity pair creation}\label{secggpair}

In the previous subsection we have seen that the cotangent bundles of two varieties linked by a conifold transition are related by a specific small instanton transition. During this transition a sheaf, supported on the exceptional curves of the conifold, is absorbed into the cotangent bundle of the resolution side variety to form the cotangent bundle of the variety on the deformation side. This small instanton transition occurs as the manifold transitions through the nodal variety which is shared in the moduli space of the two geometries.

The obvious question that occurs in a physical setting is where did the curve-supported sheaf involved in this transition come from? Consider the heterotic anomaly cancellation condition. We start on the resolution geometry $\res{X}$ with a condition (\ref{canom}) that can be rewritten in the following form.
\begin{eqnarray} \label{anomschematic}
c_2(\Omega_{\res{X}}) = c_2(\res{V}) + [\res{C}]
\end{eqnarray}
In this expression $c_2(\Omega_{\res{X}})$ is more commonly rewritten as $c_2(T_{\res{X}})$. These two quantities are equal, however, and given that the small instanton transition just discussed is most naturally presented in terms of the cotangent bundle we chose to write (\ref{anomschematic}) in this manner. The other quantities in (\ref{anomschematic}) are a gauge bundle $\res{V}$ and a potentially non-trivial class $[\res{C}]$ which is wrapped by 5-branes.

\begin{figure}[!t]\centering
\includegraphics[scale=0.65]{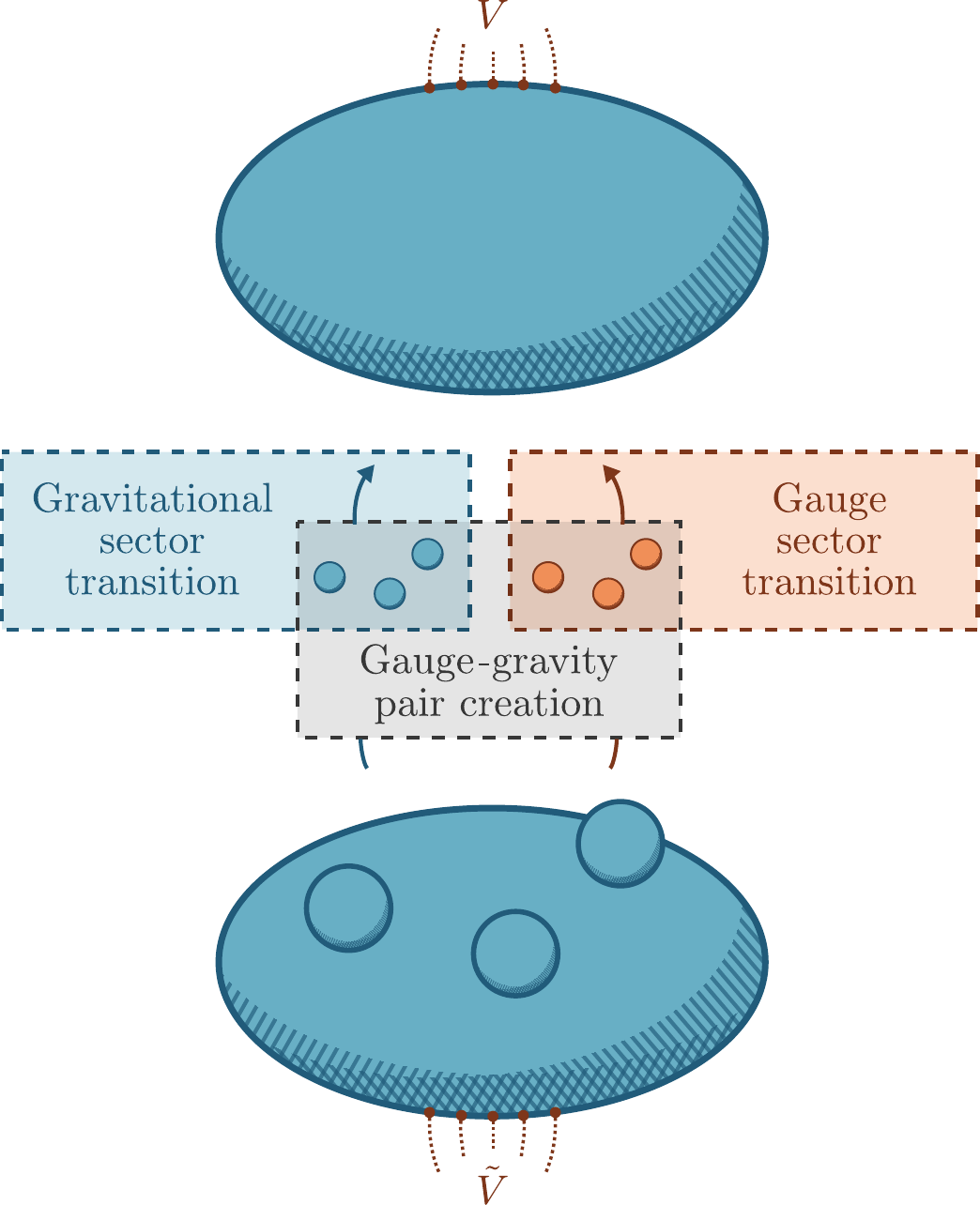}\caption{{\it The conjectured description of the heterotic conifold as a process of gauge-gravity pair creation of small instantons $\mc{O}_{\mbb{P}^1\mathrm{s}}(-2)$, which are depicted schematically by their support. \mbox{Absorption} and smoothing of the gravitational small instantons performs the geometric conifold transition ${\res{X} \to \defm{X}}$, while the gauge small instantons allow the gauge sector transition $\res{V} \to \defm{V}$, as we describe in the text below.}}
\label{paircreation}
\end{figure}

In order to perform the transition to the deformation side manifold, we must add a class $[\mathbb{P}^1\textnormal{s}]$ to the left hand side of (\ref{anomschematic}), so that the associated sheaf $\mc{O}_{\mbb{P}^1\mathrm{s}}(-2)$, whose second Chern character is given by this class, can then be absorbed via a small instanton transition to obtain the new cotangent bundle. If this process is to be consistent with anomaly cancellation, we must add the same class to the right hand side as well.
\begin{eqnarray} \label{anomschematic2}
c_2(\Omega_{\res{X}}) +[\mathbb{P}^1\textnormal{s}]= c_2(\res{V}) + [\res{C}]+[\mathbb{P}^1\textnormal{s}]
\end{eqnarray}
This would appear to be a kind of brane pair creation process, wherein 5-branes are created in both the gauge and gravitational sectors of the theory simultaneously, before being reabsorbed into other objects as we have already discussed for the cotangent bundle and will discuss shortly for the gauge sector of the transition. Since this transition happens at a singular point in the geometry of both sectors, it is hard to maintain calculational control to prove conclusively that such a process does take place. Nevertheless, in this paper we will provide a substantial amount of evidence that this pair creation process indeed does occur in heterotic string theory. A schematic depiction of this process is given in Figure~\ref{paircreation}.

As a first point, we should consider the usual rules for pair creation of branes. Such processes are usually thought to be present in a quantum theory if they are not explicitly forbidden by some selection effect such as charge conservation. For example, brane/anti-brane creation is well understood and occurs because the two objects have opposite charges, meaning that there is nothing to forbid the process. The same is true for the pair creation process we are proposing here. Indeed, this is precisely what (\ref{anomschematic2}) states.

A difference between standard brane/anti-brane creation and what we are proposing is seen when one considers supersymmetry. In order to ensure charge conservation, in a standard brane/anti-brane pair nucleation process the two extended objects involved must preserve complementary supersymmetries. Thus the process breaks supersymmetry completely. The same is not true for the process we are proposing here. Because supersymmetric objects in the gravitational and gauge sectors of heterotic string theory appear with opposite signs of charge in (\ref{anomschematic2}) the pair creation process allowed by charge conservation considerations preserves supersymmetry.

Although, given the above, a pair creation process such as that we propose here may seem reasonable, the real evidence for its existence will follow from the structure we present in the rest of this paper. There are a vast number of existing examples of `transitions', and indeed dual theories, where adding $[\mathbb{P}^1\textnormal{s}]$ to the gauge sector as in (\ref{anomschematic2}) does indeed lead precisely to known structure. For example, this sheaf can be combined, via a process we will describe in detail, with the gauge sector of the theory to yield exactly the gauge sector which is expected on the deformation side of the transition. The presence of this detailed structure, present across a huge number of known examples, would have to be pure coincidence if the pair creation process presented in this section is not physically realized. The authors find such a possibility, while logically possible, hard to believe.

\vspace{0.1cm}

As a final comment one could wonder why such pair creation processes don't simply continue, with more and more sheaves being nucleated. There are several effects which terminate this process. For example, it should be remembered that each pair creation event, and subsequent small instanton transition, is associated with a singular transition in the geometry such that the cotangent bundle can change topologically (e.g.\ $h^{1,1}$ goes down upon ``absorption" of the $\mathbb{P}^1s$). The set of such geometric transitions where the geometries on either side of the process preserve supersymmetry are, of course, extremely limited. If one nucleated too many sheaves, or indeed sheaves in the gravitational sector of the wrong form, then such a process would not be supersymmetric in nature.

\subsubsection{Absorption into the gauge sector?}

Above we have proposed that the correct understanding of a conifold transition in heterotic string theory is as a kind of pair creation process between the gravitational and gauge sectors, in which both a gravitational and a gauge small instanton are produced simultaneously on the same sublocus. In going from the resolution to the deformation side of a conifold the gravitational small instanton on the exceptional $\mbb{P}^1$s is absorbed into the cotangent bundle. The gauge small instanton on the exceptional $\mbb{P}^1$s allows the gauge sector to continue to balance the gravitational contribution to the 5-brane charge after the transition in order to maintain an anomaly-free theory.

While we have an explicit description in \eqref{eq:rel_cot_seq} of how the gravitational small instanton is absorbed, we still must describe the fate of the gauge small instanton. Of course, the very natural guess is that it is absorbed into the gauge bundle by an exactly analogous process, described by a Hecke transform given by a short exact sequence 
\be \label{notthisone}
0 \to \hat{V} \to \res{V} \to \mc{O}_{\mbb{P}^1\mathrm{s}}(-2) \to 0 \,.
\ee
Here, $\hat{V}$ is the bundle or sheaf produced as a result of this absorption (which it may subsequently be possible to smooth). However, what one finds is that whenever $\res{V}$ is a bundle which one might reasonably expect to be a candidate to be taken through a conifold transition, the map $\res{V} \to \mc{O}_{\mbb{P}^1\mathrm{s}}(-2)$ vanishes\footnote{Strictly speaking we should evaluate whether or not this map exists on the nodal variety $\nod{X}$ where this transition actually takes place. Doing so does not change any of the conclusions presented here.}. That is, \ $H^0(\res{X},{\res{V}}^\vee \otimes \mc{O}_{\mbb{P}^1\mathrm{s}}(-2)) = 0$, so that this absorption process does not exist. These examples of bundles which one `might expect' could be taken through a conifold transition will be associated to target space dual theories, and we will discuss these at length in Section~\ref{sec:tsd}. 

In fact, the impossibility of the absorption process (\ref{notthisone}) can also be seen from very general considerations. In the mathematics literature, it is expected (see e.g.\ \cite{Chuan:2010si,Collins:2021qqo} and references therein) that if the connection on a bundle is to have any chance of being taken smoothly through a conifold transition, that bundle should restrict to the exceptional $\mbb{P}^1$s to give $\mc{O}_{\mbb{P}^1\mathrm{s}}^{\oplus n}$, for some power $n$. Hence, for any such candidate bundle $\res{V}$, the map $\res{V} \to \mc{O}_{\mbb{P}^1\mathrm{s}}(-2)$ restricts on the $\mbb{P}^1$s to a map $\mc{O}_{\mbb{P}^1\mathrm{s}}^{\oplus n} \to \mc{O}_{\mbb{P}^1\mathrm{s}}(-2)$. However, such a map does not exist, due to the negative twist in the target line bundle, and thus the above Hecke transform does not exist either.

Given that this small instanton can seemingly not be absorbed into the gauge bundle in a simple fashion, a second natural approach would be to attempt to leave it as a 5-brane wrapped on the $\mbb{P}^1$s, and to carry this object through the conifold transition directly. However, in the small contraction limit, the exceptional $\mbb{P}^1$s shrink to zero volume, so that the volume of the wrapped 5-brane would also go to zero, giving rise in the limit to a tensionless spacetime-filling brane. What we will seek to show in what follows is that there exists a much smoother process by which the gauge sector can traverse the conifold transition, which will allow the compactified theory to pass through without any such drastic change.

We will see that an absorption process is possible for the above gauge small instanton, but it is more complicated than a process that is captured simply by a single Hecke transform. Essentially, a brane recombination process occurs, after which the desired small instanton transition does indeed exist. Alternatively, after this brane recombination process, it will be possible to leave the small instanton as a new 5-brane which is better behaved through the conifold. In terms of exposition, it will be most straightforward to consider this latter possibility first, and hence this will be the subject of the next subsection. This will ultimately also lead us quite directly to the correct description of the small instanton absorption process into the gauge bundle, which we will then treat in Section~\ref{sec:bund_thru_con}.

\subsection{Branes through the conifold transition}
\label{sec:brane_thru_con}

We would like to describe a way in which 5-branes wrapping the exceptional $\mbb{P}^1$s in the resolution side of a conifold transition might be combined with another 5-brane to give an object which traverses the conifold transition smoothly, unlike the 5-branes wrapped on the $\mbb{P}^1$s alone, which would produce in the contraction limit a tensionless spacetime-filling brane.

Describing such a process would provide a 5-brane theory on the resolution side and a 5-brane theory on the deformation side which are (according to our proposal of the description of a conifold transition as a gauge-gravity pair creation process) connected through the conifold transition. With this in mind, we will in fact find it most natural to begin the discussion by searching for such candidate pairs of 5-brane theories, and then subsequently showing that indeed these 5-brane theories are such that they can be matched on the nodal variety precisely through a recombining of the brane on the resolution side with branes wrapping the exceptional $\mbb{P}^1$s. Further evidence that these theories are indeed connected through the conifold transition, and hence also evidence for our general proposal of a gauge-gravity pair creation description of the conifold, will be provided in Section~\ref{sec:5brane_duality} below, where we will argue that these 5-brane theories are in fact dual theories, strongly suggesting that they are indeed connected by a smooth transition.

To collect the objects that we will need in order to describe the candidate pairs of 5-brane theories, and the brane recombination process through the conifold transition, it is necessary to pause to understand better the geometry of the conifold, and in particular the presence of certain curves and divisors whose existence is directly related to the nature of this transition.

\subsubsection{Objects canonically associated to a conifold transition}

\begin{figure}[!t]\centering
\includegraphics[scale=0.65]{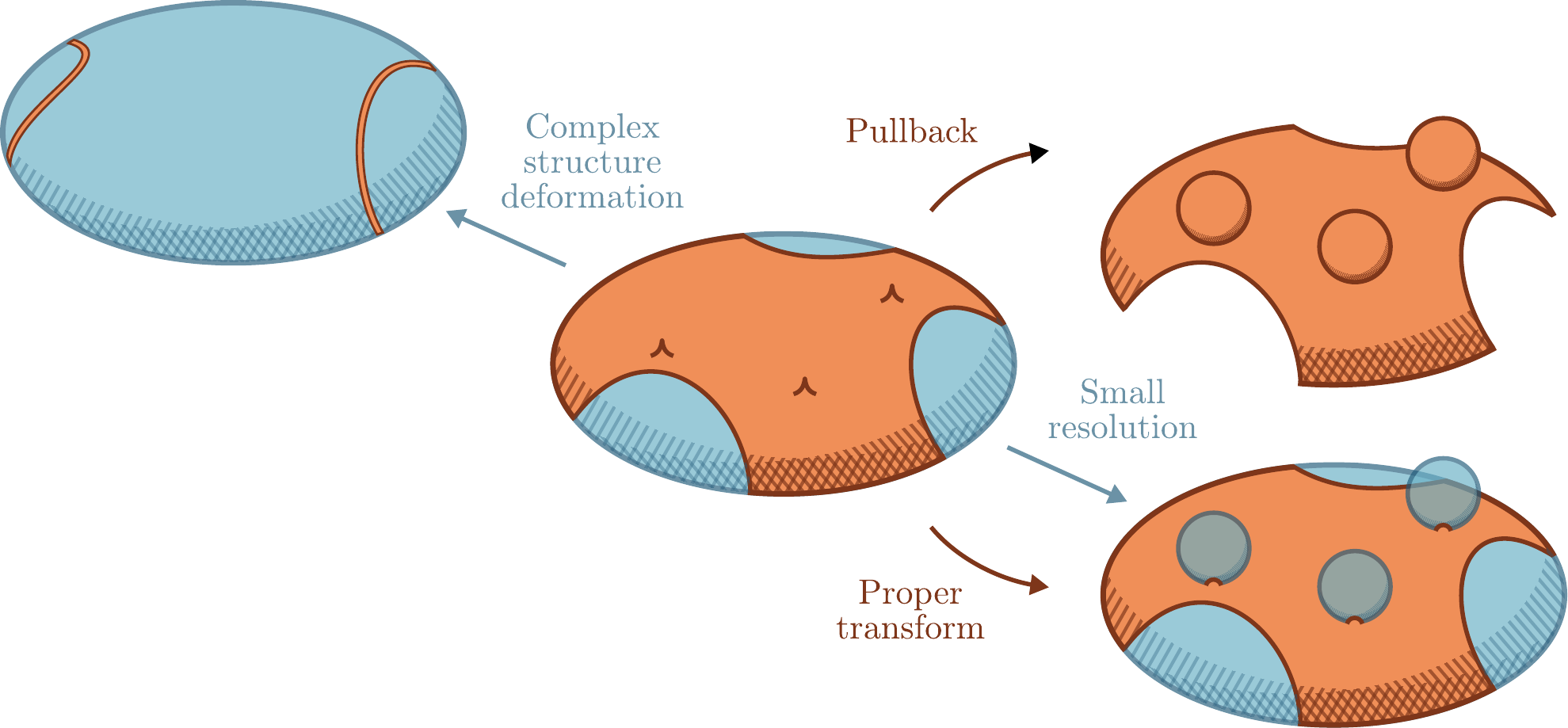}
\caption{{\it A Weil non-Cartier divisor on the nodal variety becomes a curve on the deformation branch, while on the resolution branch one can consider either the pullback or the proper transform divisor.}}
\label{fig3}
\end{figure}

A characteristic property of a conifold transition is that, as the deformation manifold is tuned to become a nodal variety, there are certain curves which jump in dimension to become divisors. This process of producing new divisors is directly linked to the fact that additional divisors must appear to generate the change in the Picard number, $h^{1,1}(\res{X}) = h^{1,1}(\defm{X})+1$. We will call the primitive divisor which appears in this fashion $\nod{D}$. This divisor in $\nod{X}$ can be lifted to two distinct divisors on the small resolution: its pullback $\smct^*(\nod{D})$ and its proper transform $\pt{\nod{D}}$. This situation is depicted schematically in Figure~\ref{fig3}.

Let us examine this structure in more concrete detail in terms of the illustrative example which we first introduced in Section~\ref{firsteg}. In this case, the deformation geometry $\defm{X}$ is described by a configuration matrix
\be
\defm{X} =
\left[
\begin{array}{c | c}
\mbb{P}^4 & 5
\end{array}
\right] \,,
\ee
i.e.\ by a generic quintic polynomial $Q(y)$ inside $\mbb{P}^4[y]$. The nodal geometry $\nod{X}$ is reached when this quintic is tuned to a form $l_0(y)\,q_1(y) - l_1(y)\,q_0(y) = 0$, where the $l_i$ and $q_i$ are respectively degree one and degree four polynomials. Consider here the example of a curve $\defm{C}$ which is described on $\defm{X}$ by
\be
\defm{C}\colon~ \{ l_0 = q_0 = 0 \} \cap \defm{X} ~\sim~ \{l_0 = q_0 = Q = 0 \} \subset \mbb{P}^4 \,.
\ee
As the generic quintic equation of $\defm{X}$ is tuned to the nodal one of $\defm{X}$, the two defining equations of this curve become no longer independent of the defining equation of the geometry, and in particular automatically satisfy the nodal quintic. Hence, this curve jumps in dimension to a divisor $\nod{D}$,
\be
\nod{D}\colon~ \{ l_0 = q_0 = 0 \} \cap \nod{X} ~\sim~ \{l_0 = q_0 = 0 \} \subset \mbb{P}^4 \,.
\ee
(One could also have considered for example the curve defined by $\{l_1 = q_1 = 0\} \cap \defm{X}$, or any linear combinations of these two curves.) We note that the divisors $\nod{D}$ which arise in this way are Weil but non-Cartier divisors, and it is clear that their existence is intimately linked to the geometry of the conifold transition.

\medskip

Under the small resolution along $\smct: \res{X} \to \nod{X}$, these objects $\nod{D}$ naturally remain divisors, since the resolution is an isomorphism except at the nodal points. However, as mentioned above, there are two distinct objects to which the divisor can be lifted: the pullback $\pb{\nod{D}}$ and the proper transform $\pt{\nod{D}}$. The loci of these two objects differ in that the pullback contains the exceptional $\mbb{P}^1$s, while the proper transform $\pt{\nod{D}}$ intersects these transversely and in a single point. 

In our example, the resolution geometry $\res{X}$ is described by a configuration matrix
\be
\res{X} =
\left[
\begin{array}{c | c c}
\mbb{P}^1 & 1 & 1 \\
\mbb{P}^4 & 1 & 4 
\end{array}
\right] \,,
\label{cicy_res}
\ee
i.e.\ by the following two generic equations of multi-degrees $(1,1)$ and $(1,4)$ inside $\mbb{P}^1 \times \mbb{P}^4$,
\be
\res{X}: ~ \big\{ (x,y) \in \mbb{P}^1[x] \times \mbb{P}^4[y] ~\big|~x_0 l_0(y) + x_1 l_1(y) = 0 ~\mathrm{and}~ x_0 q_0(y) + x_1 q_1(y) = 0 \big\} \,.
\ee
The pullback divisor $\pb{\nod{D}}$ is described by
\be
\pb{\nod{D}}\colon~ \{ l_0 = q_0 = 0 \} \cap \res{X} \,,
\ee
which is a locus that manifestly contains the exceptional $\mbb{P}^1$s, since these sit over the points $l_0 = l_1 = q_0 = q_1 = 0$. The proper transform divisor is described by
\be \label{egpt}
\pt{\nod{D}}\colon~ \{ x_1 = 0 \} \cap \res{X} \,,
\ee
and manifestly intersects each exceptional $\mbb{P}^1$ transversely in a point. Looking at the defining equations of $\res{X}$, it is clear that the locus of $\pb{\nod{D}}$ is indeed simply that of $\pt{\nod{D}}$ plus the exceptional $\mbb{P}^1$s , since imposing $l_0 = q_0 = 0$ in those equations gives $x_0 l_0(y) = x_0 q_0(y) = 0$, which has these two components as solutions.

\subsubsection{Candidates for 5-brane transition}

The reason we have introduced the above objects - the Weil non-Cartier divisors on the nodal variety - is that they lead almost directly to the description of a pair of a 5-brane theory on the resolution side and a 5-brane theory on the deformation side which are clear candidates to be continuously connected through the conifold transition. Let us make this explicit.

We have already seen that a Weil non-Cartier divisor gives rise immediately to a curve $\defm{C}$ on the deformation side. The 5-brane theory we will define on the deformation geometry is of a 5-brane wrapped on this curve, plus a set of additional essentially arbitrary 5-branes which saturate the remainder of the anomaly cancellation condition. These will turn out to have unimportant, trivial behavior as mere `spectators', as we will discuss in detail below.

In contrast we have seen that in passing to the resolution geometry the Weil non-Cartier divisor gives rise only to a divisor, which we will take to be the proper transform divisor $\pt{\nod{D}}$\footnote{It is this choice, not $\smct^*(\nod{D})$ that will turn out to correctly link two well defined 5-brane theories across the transition as we will show in the following}. We now take an intersection of this divisor with an additional hypersurface to define a curve. Specifically, we intersect with a hypersurface of the same form, denoted by $\defm{H}$, as that involved in the deformation of the geometry on the other side of the conifold transition. In our canonical $\mbb{P}^1$-split example, this means intersecting with the zero locus of a generic quintic polynomial for example.

The reason to define curves $\defm{C}$ and $\res{C}$ in this way is that the `difference' between the classes of $\res{C}$ and $\defm{C}$ is exactly the same as the `difference' between the second Chern classes of the deformation and resolution geometries. That is, upon wrapping 5-branes on these curves, the remaining parts of the anomaly cancellation conditions, $c_2(\res{X})-[\res{C}]$ and $c_2(\defm{X})-[\defm{C}]$, are `identical', in the precise sense that this remaining discrepancy can now be trivially made up by the addition of the `spectator' branes which we mentioned above. Hence, the curve pairing of $\defm{C}$ and $\res{C}$ does the `hard work' in allowing the pair of 5-brane theories to both be anomaly consistent.

Below we illustrate this in our specific example. Additionally, we collect in Appendix~\ref{sec:exp_for_pn_split} the analogous explicit results for the much more general case of any $\mbb{P}^n$-split of a toric complete intersection.

\medskip

Recall the $\mbb{P}^1$-split conifold transition that we have been using as our canonical example. In this case, the two curves $\defm{C}$ and $\res{C}$ are described by
\be
\begin{aligned} \label{mrcdef}
\defm{C}\colon& \{ l_0 = q_0 = 0 \} \cap \defm{X}  \,, \\
\res{C}\colon& \{ x_0 = Q = 0 \} \cap \res{X}  \,,
\end{aligned}
\ee
and hence the classes of the two curves within the two geometries $\defm{X}$ and $\res{X}$ are
\be
\begin{aligned}
[\defm{C}] &= J_1 \times 4J_1 = 4 J_1^2\,, \\
[\res{C}] &= J_0 \times 5J_1 = 5J_0J_1  \,.
\end{aligned}
\ee
Recalling the class of the exceptional $\mbb{P}^1$s inside the resolution geometry $\res{X}$ in this conifold example,
\be
[\mbb{P}^1\mathrm{s}] = -5 J_0 J_1 + 4 J_1^2 \,,
\ee
we see manifestly that the difference between the two curve classes above is indeed `identical'\footnote{We recall that by abuse of notation `$J_1$' means the restriction of the K\"ahler form of the common ambient $\mbb{P}^4$ to either of $\defm{X}$ or $\res{X}$. Hence, it is only as the two geometries limit to the nodal variety at the middle of the conifold transition that these objects become identical (and indeed comparable at all). } to this class which, as seen in (\ref{eq:ch2_rel}), is the difference in the Chern characters of the two manifolds on either side of the conifold transition. Hence, more explicitly, if we wrap 5-branes on these curves on the geometry on each side of the conifold transition, the piece left in the anomaly cancellation condition is exactly `identical',
\be
c_2(\defm{X}) - [\defm{C}] = 6J_1^2 \,, \quad c_2(\res{X}) - [\res{C}] = 6J_1^2 \,.
\ee

If one includes an additional `spectator' brane, meaning a brane which trivially traverses the conifold transition by simply remaining far from the singular points that appear on the intermediate nodal variety, its contributions to the anomaly cancellation condition on the two geometries are also naturally `identical'. Hence, if we are able to show the continuous matching across the conifold transition of the above pair of 5-branes, this remaining part of the story of the 5-brane theory traversing the transition is trivial to complete.

We will describe in detail below these spectator branes, and show that they indeed behave and contribute to the anomaly condition as just described. However, first we consider the more critical question of whether the above pair of 5-branes are indeed connected continuously across the conifold transition.

\subsubsection{Brane recombination and the transition}

We wish to show that the 5-branes just described on either side of the conifold transition can be made to continuously meet in some specific sense at the nodal variety. Said differently, we want to show that each of these 5-branes can be taken through the conifold transition, becoming the other during the process.

By construction of the curve $\defm{C}$, when the deformation geometry is tuned to become the nodal variety, this curve enhances into a Weil non-Cartier divisor $\nod{D}$. In contrast, the curve $\res{C}$ simply remains a curve as the resolution geometry shrinks to the nodal variety. This behavior can be modified, however. The curve $\res{C}$ is defined as an intersection of two divisors, specifically the intersection between the proper transform $\pt{\nod{D}}$ and the zero locus of a divisor in some class $\defm{H}$. But the nodal geometry too is described by the vanishing of a series of polynomials, one of which is also the zero locus of a divisor in class $\defm{H}$. Hence, if we tune this defining relation of the curve so that in the nodal limit it becomes to equal the defining equation of the nodal geometry, this equation in the curve's definition will be trivially satisfied and the curve will jump to become the proper transform divisor $\pt{\nod{D}}$. In the nodal limit, the locus of this divisor becomes precisely that of the Weil non-Cartier divisor $\nod{D}$. Hence, there exists a special tuning of the curve $\res{C}$ which, if performed at the same time as the resolution geometry shrinks to the nodal variety, allows this curve to limit to the same locus as the curve $\defm{C}$ limits to from the deformation side.\footnote{It is notable that the meeting of the two 5-brane theories from the two sides of the conifold transition requires a geometric tuning in coming from the deformation side and a tuning in the gauge/5-brane theory coming from the resolution side. Moreover, despite these tunings having drastically different physical interpretations, they are of precisely the same mathematical form, both involving tuning a quintic in our canonical case for example. It is hence natural to guess that these two theories are not only connected, but in fact dual. This will be the subject of a detailed discussion in Section~\ref{sec:5brane_duality} below. Further, this exchange of geometric and gauge degrees of freedom is precisely what is seen in examples of $(0,2)$ target space duality of the heterotic string, and we will discuss the very concrete connections between this and the present discussion in Section~\ref{sec:tsd} below.} We will return to discuss the curious fact that the curves that the 5-branes are wrapping limit to higher-dimensional cycles on the nodal geometry at the end of this subsection.

\vspace{0.1cm}

We have now seen that (with a bit of tuning) the curves that the 5-branes on the deformation and resolution sides wrap become coincident objects when those two geometries meet as the intermediate singular variety. However, the proper mathematical description of these 5-branes that we are employing is in terms of sheaves. The curves we have been discussing are just those cycles over which these sheaves have support. The 5-brane on the resolution side of the transition is described by a sheaf ${\cal O}_{\res{C}}$. This becomes $\mc{O}_{\pt{\nod{D}}}$ in the limit which has support only over the limit of $\pt{\nod{D}}$. The 5-brane on the deformation side of the transition is described by a sheaf ${\cal O}_{\defm{C}}$. This becomes $ \mc{O}_{\pb{\nod{D}}}$ in the nodal limit which only has support on $\nod{D}$. These two sheaves do {\it not} match in the nodal limit. Rather, it is at this stage that the gauge small instanton $\mc{O}_{\mbb{P}^1\mathrm{s}}(-2)$ that was created during the gravitational/gauge pair creation process comes into play.

Naively, it would be desirable if a brane recombination process could occur to take the sheaf ${\cal O}_{\mbb{P}^1\mathrm{s}}(-2)$ that was generated during the pair creation process and combine this with ${\cal O}_{\res{C}}$ to produce the sheaf ${\cal O}_{\defm{C}}$ that is expected after the transition. If we first consider a case where we do not tune $\defm{C}$ such that is becomes a divisor in the nodal limit (simply by taking a general element of its curve class rather than a tuned example such as (\ref{mrcdef})) a natural way in which one might try to combine two sheaves in this way would be via a short exact extension sequence, of the following form\footnote{We have shown explicitly that the statements we make about the following sequences hold in the classes of examples we discuss in this paper. While we are not aware of a general proof, we expect this to hold much more widely.}.
\begin{eqnarray}
0 \to \mc{O}_{\mbb{P}^1\mathrm{s}}(-2) \to \mc{O}_{\smct^* ( \defm{C})} \to \mc{O}_{\res{C}} \to 0
\end{eqnarray}
This sequence is not correct however, as it suffers from several problems. First, this sequence as stated leads to an incorrect relationship between the Chern classes involved. To obtain the correct relationship, the central object of the extension has to be twisted as follows.
\begin{eqnarray} \label{seqtwo}
0 \to \mc{O}_{\mbb{P}^1\mathrm{s}}(-2) \to \mc{O}_{\smct^* ( \defm{C})} \otimes \mc{O}_{\res{X}}(-\pt{\nod{D}}) \to \mc{O}_{\res{C}} \to 0
\end{eqnarray}
We will address the meaning of the twisting of the central term above by $\mc{O}_{\res{X}}(-\pt{\nod{D}})$ shortly. Before addressing that feature however, there is another problem that means that the sequence (\ref{seqtwo}) is not correct. The issue is that the extension class associated to that sequence, written as it is on the resolution manifold, vanishes. No such non-trivial recombination of the sheaves involved exists. This, however, is an artifact of our trick of describing the physics of the transition in terms of objects pulled back to the resolution variety. If one now takes $\defm{C}$ to be the tuned curve which becomes a divisor in the nodal limit (and similarly for $\res{C}$) and then tries to form such an extension of divisor supported sheaves on the nodal variety, the relevant extension class {\it does} exist. This requirement of properly going to the nodal geometry is perhaps not surprising at this stage as, after all, the entire process really takes place as the singular geometry is traversed.

The easiest way in which to see that the extension class does indeed become non-vanishing in this limit where the curves become divisors is to still use the trick of working on the resolution manifold, but to take $\defm{C}$ to be the divisor $\nod{D}$ and to tune the hypersurface $\defm{H}$  appearing in the definition of $\res{C}$ to be the relevant defining equation of the nodal variety. This mimics the structure of the relevant curves blowing up into divisors and leads to the following, now finally correct, short exact sequence of brane recombination.
\be \label{rightseq}
0 \to \mc{O}_{\mbb{P}^1\mathrm{s}}(-2) \to \mc{O}_{\pb{\nod{D}}} \otimes \mc{O}_{\res{X}}(-\pt{\nod{D}}) \to \mc{O}_{\pt{\nod{D}}} \to 0 \,.
\ee
This sequence essentially says that the two sheaves, $\mc{O}_{\mbb{P}^1\mathrm{s}}(-2)$ describing the pair-created brane and $\mc{O}_{\pt{\nod{D}}}$ describing the limit of the 5-brane on the resolution manifold, recombine to a sheaf $\mc{O}_{\pb{\nod{D}}} \otimes \mc{O}_{\res{X}}(-\pt{\nod{D}})$ describing the limit of the 5-brane from the deformation variety. Since this brane recombination process only produces the 5-brane configuration we require from the deformation side up to a twist by $ \mc{O}_{\res{X}}(-\pt{\nod{D}})$, we must now discuss the origin of this seemingly additional structure.

\vspace{0.1cm}

For any two varieties $X_1$ and $X_2$ the following short exact sequence holds.
\begin{eqnarray} \label{seqgen}
0 \to {\cal O}_{X_1 \cup X_2} \to {\cal O}_{X_1} \oplus {\cal O}_{X_2} \to {\cal O}_{X_1 \cap X_2} \to 0
\end{eqnarray}
For the case at hand we shall take $X_1=\pt{\nod{D}}$ and $X_2=\mbb{P}^1\mathrm{s}$. We then have that $X_1 \cup X_2=\smct^*(\nod{D})$ and $X_1\cap X_2 = \text{pts}$, a set of points where the proper transform divisor intersects the exceptional locus. The sequence (\ref{seqgen}) then becomes the following.
\begin{eqnarray} 
0 \to {\cal O}_{\smct^*(\nod{D})} \to {\cal O}_{\pt{\nod{D}}} \oplus {\cal O}_{\mbb{P}^1\mathrm{s}} \to {\cal O}_{\text{pts}} \to 0
\end{eqnarray}
If we twist this sequence up by $ \mc{O}_{\res{X}}(-\pt{\nod{D}})$ we obtain the following.
\begin{eqnarray}
0 \to {\cal O}_{\smct^*(\nod{D})}\otimes \mc{O}_{\res{X}}(-\pt{\nod{D}}) \to {\cal O}_{\pt{\nod{D}}}  \oplus {\cal O}_{\mbb{P}^1\mathrm{s}} \otimes \mc{O}_{\res{X}}(-\pt{\nod{D}}) \to {\cal O}_{\text{pts}} \to 0
\end{eqnarray}
Here we have used the fact that the twisting does not affect the sheaves that are supported only over points or the sheaf whose support strikes the exceptional locus at points. Pushing this sequence forward to the nodal variety and using the fact that $\smct_*\left( {\cal O}_{\mbb{P}^1\mathrm{s}} \otimes \mc{O}_{\res{X}}(-\pt{\nod{D}})\right)=0$ we arrive at the following.
\begin{eqnarray} \label{lastseq}
0 \to \smct_*\left({\cal O}_{\smct^*(\nod{D})}\otimes \mc{O}_{\res{X}}(-\pt{\nod{D}})\right) \stackrel{g}{\longrightarrow} {\cal O}_{\nod{D}}  \stackrel{f}{\longrightarrow} {\cal O}_{\text{pts}} 
\end{eqnarray}
The map $f$ in the above sequence is non-zero precisely because the points over which the last sheaf has support lie within $\nod{D}$.  Now consider deforming the nodal Calabi-Yau manifold to return to the smooth manifold $\defm{X}$. Such a deformation removes the singular points from $\nod{D}$, rendering the map $f$ vanishing. The map $g$ in (\ref{lastseq}) then becomes an isomorphism. This shows that upon deforming to the smooth deformation manifold, the unwanted twist in the central sheaf of (\ref{rightseq}) goes away. Indeed, the Weil non-Cartier divisor $\nod{D}$ even becomes an (untwisted) curve under this deformation.

\vspace{0.2cm}

{\it The final upshot of the lengthy preceding discussion is that the gauge small instanton $\mc{O}_{\mbb{P}^1\mathrm{s}}(-2)$, produced in our conjectured gauge-gravity pair creation process is precisely what is required to combine with a 5-brane wrapping $\res{C}$ on the resolution geometry to become the limit of the 5-brane wrapping $\defm{C}$ from the deformation geometry when the common point in moduli space is approached.}

\vspace{0.2cm}

Let us look at all of this structure in the context of the illustrative example we have been employing throughout this section. In this case we have, from (\ref{egpt}) that ${\cal O}_{\res{X}}(\pt{\nod{D}})= {\cal O}_{\res{X}}(1,0)$. Given this, the sequence (\ref{rightseq}) becomes the following in this example.
\be \label{rightseqeg}
0 \to \mc{O}_{\mbb{P}^1\mathrm{s}}(-2) \to \mc{O}_{\pb{\nod{D}}}(-1,0) \to \mc{O}_{\pt{\nod{D}}} \to 0 
\ee
It is easy to show that this sequence is indeed correct in this case. The extension class \newline $\text{Ext}^1( \mc{O}_{\pt{\nod{D}}},\mc{O}_{\mbb{P}^1\mathrm{s}}(-2))=\mathbb{C}^{\#(\mbb{P}^1\mathrm{s})}$, which is not vanishing and thus there is some object appearing in the central position in (\ref{rightseqeg}) which is not just a direct sum. This object has support over $\pb{\nod{D}}$. That it is $\mc{O}_{\pb{\nod{D}}}(-1,0)$  can then be ascertained by demanding that the Chern classes and line bundle cohomologies of the central object agree with what is implied by the short exact sequence.
 
The sequence (\ref{lastseq}) becomes the following in this example.
\begin{eqnarray}
0 \to {\cal O}_{\nod{D}}(-1,0) \to  {\cal O}_{\nod{D}} \to {\cal O}_{\text{pts}}
\end{eqnarray}
 In the nodal limit ${\cal O}_{\nod{D}}(-1,0)$ and ${\cal O}_{\nod{D}}$ are indeed different and the above sequence is non-trivial. However, as we deform away from the nodal point, the divisor by which one twists ${\cal O}_{\nod{D}}$ to obtain ${\cal O}_{\nod{D}}(-1,0)$ disappears, and so the two sheaves indeed become the same object (in addition to $\nod{D}$ transitioning to become $\defm{C}$).
 
 \vspace{0.2cm}

Finally, we return to the fact that the objects to which the 5-branes limit on the nodal variety appear, somewhat surprisingly, to be described by divisors, rather than curves. It is certainly clear that the sheaf which describes the small instanton in the gauge theory undergoes this radical change as the geometry limits from either side to the nodal variety. However, it is not clear whether this change is purely an effect in the small instanton limit, or a physically important change in the case of true 5-branes as well. The correct conditions to impose on the dimensionality of extended objects in such a singular limit of heterotic string theory, where those objects intersect the singularities, is not known and thus such a ``dimension jumping" effect could be real. It would certainly be interesting to investigate this effect, in addition to the pair creation process we have described earlier, in a simpler, non-compact, setting where one might have more direct control of such a process. In this paper, however, we will concentrate on the compact setting where one obtains a vast amount of indirect evidence that the process we have described here does occur from the highly constrained structure of heterotic compactifications.

\subsubsection{Spectator branes}\label{specsec}

It remains to complete the above pair of 5-brane theories with the addition of spectator branes, which, unlike the special 5-branes above, traverse the conifold transition essentially trivially, staying far from the singularities or equivalently the exceptional $\mbb{P}^1$s, and hence remaining `ignorant' of the transition, so that they simply make up the `identical' remaining part of the anomaly cancellation condition on each side of the transition.

Note that any generic curve will be an example of a spectator, since any generic curve as it passes through the nodal variety will miss the singular points. Hence, the existence of spectator branes and thus the final piece required to complete the connection of a pair of 5-brane theories across a conifold transition is guaranteed.

We illustrate this in an example. Recall our standard conifold transition example, in which the deformation $\defm{X}$ and resolution geometries $\res{X}$ are CICYs described by the configuration matrices
\be
\defm{X} =
\left[
\begin{array}{c | c}
\mbb{P}^4 & 5
\end{array}
\right] \,, \quad 
\res{X} =
\left[
\begin{array}{c | c c}
\mbb{P}^1 & 1 & 1 \\
\mbb{P}^4 & 1 & 4 
\end{array}
\right] \,,
\ee
meaning they are described by complete intersections inside $\mbb{P}^4$ and $\mbb{P}^1 \times \mbb{P}^4$ as
\be
\begin{gathered}
\defm{X} \colon \big\{ (y) \in \mbb{P}^4[y] ~\big|~ Q(y) = 0 \big\} \,, \\
\res{X} \colon \big\{ (x,y) \in \mbb{P}^1[x] \times \mbb{P}^4[y] ~\big|~ x_0 l_0(y) + x_1 l_1(y) = 0 ~\mathrm{and}~ x_0 q_0(y) + x_1 q_1(y) = 0 \big\} \,,
\end{gathered}
\ee
where $Q$, the $l_i$, and the $q_i$ are (generic) polynomials of degrees 1, 4, and 5 respectively, and for which the nodal geometry $\nod{X}$ is given by tuning the quintic polynomial $Q$ to the special choice
\be
\nod{X} \colon \big\{ (y) \in \mbb{P}^4[y] ~\big|~ l_0(y)q_1(y) - l_1(y) q_0(y) = 0 \big\} \,.
\ee
This example provides a simple case in which we can track spectator curves through the transition and compute the curve classes on each geometry. In particular, it is easy to track a curve through the transition due to the fact that these $\mbb{P}^1$-split (or more generally $\mbb{P}^n$-split) examples of conifold transitions have the convenient property that the ambient space of the deformation geometry $\defm{X}$ continues on to naturally form part of the ambient space of the resolution geometry $\res{X}$. Hence, we can define a spectator curve by using only the coordinates of the ambient space of $\defm{X}$. Explicitly in our example, we can define a curve in the geometry at any point during the transition by taking the intersection of the geometry with the common zero locus of two generic polynomials $P_1(y)$ and $P_2(y)$, and then we can track how the curve behaves through the transition simply by continuing to take this intersection with the geometry at each stage. That is, explicitly, the curve at each stage is described by
\be
\left\{
\begin{array}{l l l}
\defm{\spec{C}} \\
\nod{\spec{C}} \\
\res{\spec{C}} \\
\end{array}
\right\}
= 
\{ P_1(y) = P_2(y) = 0 \} \cap
\left\{
\begin{array}{l l l}
\defm{X}  & \subset \mbb{P}^4[y]  \\
\nod{X} & \subset \mbb{P}^4[y]  \\
\res{X} & \subset \mbb{P}^1[x] \times \mbb{P}^4[y] \\
\end{array}
\right\} \,.
\ee
We note that since $P_1$ and $P_2$ are generic the curve $\defm{\spec{C}}$ misses the singularities on the nodal geometry so that this is indeed a spectator curve. The classes of the curves $\defm{\spec{C}} \subset \defm{X}$ and $\res{\spec{C}} \subset \res{X}$ are now simple to compute. Setting the degrees of the polynomials $P_1$ and $P_2$ to be $d_1$ and $d_2$, these classes are simply
\be
[\defm{\spec{C}}] = (d_1 \, d_2) J_1^2 \,, \quad [\res{\spec{C}}] = (d_1 \, d_2) J_1^2 \,,
\ee
i.e.\ they are `identical'. Hence, by wrapping an additional 5-brane on such a spectator curve, the above pair of theories on the deformation and resolution geometries can be made anomaly-consistent, specifically in this example by any choice of spectator curve for which $d_1 d_2 = 6$.

\subsection{Bundles through the conifold transition}
\label{sec:bund_thru_con}

In this section we will demonstrate that the process of mapping a 5-brane through a conifold transition (as outlined above) leads naturally to a way to follow a vector bundle through the transition via a heterotic small instanton transition (see e.g.\ \cite{Witten:1995gx,Ovrut:2000qi,Buchbinder:2002ji}).

As first described in \cite{Ovrut:2000qi,Buchbinder:2002ji}, and reviewed in Section~\ref{sec:gravinst}, the mathematical process of ``absorbing" a 5-brane into a vector bundle proceeds in several steps. Consider a 5-brane wrapping a curve $C$ inside a CY 3-fold $X$. In the limit that the 5-brane starts to dissolve onto an $E_8$ fixed plane (in the language of heterotic M-theory), the subsequent ``small instanton" can be described via a skyscraper sheaf supported over $C$ \cite{Ovrut:2000qi,Buchbinder:2002ji} or equivalently via an ideal sheaf of $C$ \cite{Aspinwall:1998he}\footnote{The relationship between these two descriptions in the context of a small instanton transition is rather evident in the case of a Hecke transform of the form given in (\ref{idealsky}). In this case, before deformation to a smooth bundle, one can either think of the process as a Hecke transform involving the skyscraper sheaf ${\cal O}_{\res{C}}$ as written, or simply as adding the ideal sheaf ${\cal I}_{\res{C}}$ to the bundle $\res{V}_0$}. Then as outlined already in Section~\ref{sec:gravinst}, the correct description of its ``merging" into a pre-existing vector bundle $V_0$ is given by a so-called Hecke transform \cite{Ovrut:2000qi}
\begin{equation}
0 \to \hat{V} \to V_0 \stackrel{f}{\longrightarrow} {\cal F}_C \to 0 ~
\label{hecke_def}
\end{equation}
where ${\cal F}_C$ is a rank 1 vector bundle\footnote{Note that ${\cal F}_C$ is physically determined by the cycle which the 5-brane wraps and the choice of form-field backgrounds on that extended object.} on $C$. Because $C$ is a co-dimension 2 object in $X$, $\hat{V}$ is in general singular (and hence a sheaf rather than a vector bundle) as it appears in \eqref{hecke_def} and must be further deformed into a smooth bundle. The key operation in \eqref{hecke_def} is the surjective morphism denoted by $f$ which ``weaves" together the fibers of $V_0$ with those of ${\cal F}$ over the locus where they overlap. It is straightforward to show that
\begin{equation}
c_2(\hat{V})=c_2(V_0) + [C]
\end{equation}
(for any choice of ${\cal F}$) as expected by a 5-brane ``absorption". 

As noted in Section~\ref{sec:gravinst}, it should be observed that this type of process can occur either as a rank-changing transition for the bundle or as a rank-preserving one. In the former case, if the initial bundle $V_0$ is of the form $V_0=U \oplus {\cal O}_X$ then $\text{rk}(\hat{V})=\text{rk}(U)+1$. The fact that $\text{rk}(\hat{V}) > \text{rk}(U)$ implies that the gauge group of the 4-dimensional theory is broken to a subgroup. In contrast, for generic $V_0$, the transition will be rank-preserving (and hence generically gauge-group preserving). Such transitions can also change the chiral index of the theory \cite{Kachru:1997rs,Ovrut:2000qi,Anderson:2019agu}.

In the present context, we can extend the discussion of 5-brane transitions from the previous section to bundles by considering a bundle $V$ which can ``emit" a 5-brane in the special classes defined in Section~\ref{sec:brane_thru_con} (i.e.\ $C$ and $\res{C}$ as in \eqref{mrcdef}) and leave the remaining part of the bundle as a ``spectator" in the conifold process (akin to the spectator branes of Section~\ref{specsec}). 

Before outlining this process in more detail, it is important to clarify in what way we expect a bundle to be a ``spectator" to a conifold transition. As in the case of 5-branes described above, intuitively it makes sense that if the crucial defining data of the bundle misses the conifold points (respectively $\mathbb{P}^1$s), then this stands a chance of leading to a good bundle on both sides of the transition. More precisely, we wish to parametrically define a pair $(X,V_0)$ denoting a vector bundle on the deformation side of the conifold which can be tuned to the nodal limit of the CY3 geometry and then pulled back to the resolution manifold $\res{X}$ to produce a smooth vector bundle $\res{V}_0$. These rather rough intuitive notions can be made somewhat more precise by studying the form of the connection at the singularities (resp.\ the $\mathbb{P}^1$s) and interesting work has been done on special aspects of this question (see \cite{Candelas:2007ac}). As mentioned previously, one simple condition placed on bundles defined over the resolution geometry $\res{X}$ that can possibly be extended onto the deformation manifold $X$ is that the restriction of $\res{V}_0$ to the $\mathbb{P}^1s$ is trivial (i.e.\ $\res{V}_0|_{\mathbb{P}^1}= {{\cal O}_{\mathbb{P}^1}}^{\oplus \mathrm{rk}(\res{V})}$)\cite{Chuan:2010si}.

As a toy example of a spectator bundle, suppose that ${\cal O}_{X}(D)$ is a line bundle on $X$. Then as proved in \cite{Anderson:2013qca}, ${\cal O}_{\res{X}}(D)$ is a line bundle on $\res{X}$ with the same cohomology and Chern classes as it had on $X$ (since the K\"ahler cone of $X$ is simply extended, never reduced, in moving to that of $\res{X}$). Thus, bundles built as kernels, cokernels or extensions of line bundles of this form, can all potentially serve as spectators. Note that line bundles of the form shown above have the property that ${\cal O}_{\res{X}}(D)|_{\mathbb{P}^1}= {\cal O}_{\mathbb{P}^1}$ as described above. 

It should be noted however, that even if a spectator bundle can be carried simply through the conifold transition, spectators do not in general have the right behavior for a \emph{physical heterotic theory} to pass through the transition, in that the end result of a bundle on $X$ trivially extended onto $\res{X}$ will in general not satisfy anomalies (or slope-stability) on $\res{X}$. To have a chance at a physically consistent transition, we need to employ the ideas of the previous subsections. Returning to this goal, suppose that $V_0$ constitutes a vector bundle on $X$ whose second Chern class is of the form $c_2(V_0)=c_2(T_X) - \left[C\right]$. Then consider a rank-changing small instanton transition of the form
\begin{equation} \label{laraseq1}
0 \to {\hat V} \to V_0 \oplus {\cal O}_X \to {\cal O}_C \to 0~.
\end{equation}
Next, let $\res{V}_0$ be the trivial extension of $V_0$ to $\res{X}$ and on this manifold, consider absorbing the small instanton associated to $\res{C}$  into $\res{V}_0 \oplus {\cal O}_{\res{X}}$ as
\begin{equation}
0 \to \hat{\res{V}} \to \res{V}_0 \oplus {\cal O}_X \to {\cal O}_{\res{C}} \to 0~.
\end{equation}
In this process we have absorbed the 5-branes paired by the transition described earlier in this Section into effectively the ``same" spectator bundle on both sides of the transition. If both $\hat{\defm{V}}$ and $\hat{\res{V}}$ can be deformed to smooth bundles, $\defm{V}$, $\res{V}$, then we have a pair for which  $c_2(\defm{V})=c_2(T_{\defm{X}})$ and $c_2(\res{V})=c_2(T_{\res{X}})$. Thus, we would have extended the 5-brane transition of the previous section into a transition of vector bundles! As we will demonstrate in future sections, both the 5-brane and bundle transitions detailed here lead to a matching of the charged and uncharged massless matter spectra of the resulting ${\cal N}=1$ 4-dimensional theories. In other words, they lead to apparently dual theories. In the case of the bundles, this will be a known duality (arising from $(0,2)$ GLSMs).

To conclude this section, it should be noted that while the small instanton process above was described as the \emph{absorption} of the special 5-branes in Section~\ref{sec:brane_thru_con}, it is equally natural to describe this process as an \emph{emission} of that 5-brane, which then can traverse the conifold. For example, beginning with a smooth bundle $V$ which can be tuned to a singular limit of the form 
\begin{equation}
V \to \hat{V}=V_0 \oplus {\cal I}_C \;,
\end{equation}
it then naturally fits into the short exact sequence
\begin{equation}
0 \to {\hat V} \to V_0 \oplus {\cal O}_X \to {\cal O}_C \to 0~,
\end{equation}
where the exactness follows from the Koszul sequence of ${\cal I}_C$. This sequence implies that via a small instanton transition such a bundle $V$ can emit an instanton supported on the curve $C$. Then if we allow the 5-brane wrapping $C$ to transition through the conifold to the dual 5-brane/curve $\res{C}$, we can ``reabsorb" the new curve into $\res{V}_0$ (if we bring the spectator bundle $V_0$ passively through the conifold transition) via
\begin{equation} \label{idealsky}
0 \to \res{V}_0 \oplus {\cal I}_{\res{C}} \to \res{V}_0 \oplus {\cal O}_X \to {\cal O}_{\res{C}} \to 0~.
\end{equation}
As a final step, this singular bundle $\res{V}_0  \oplus {\cal I}_{\res{C}}$ can be deformed back into a smooth vector bundle $\res{V}$ on $\res{X}$. Thus, the pair $V$, $\res{V}$ has been linked across the conifold transition.

\begin{figure}[!t]\centering
\includegraphics[scale=0.65]{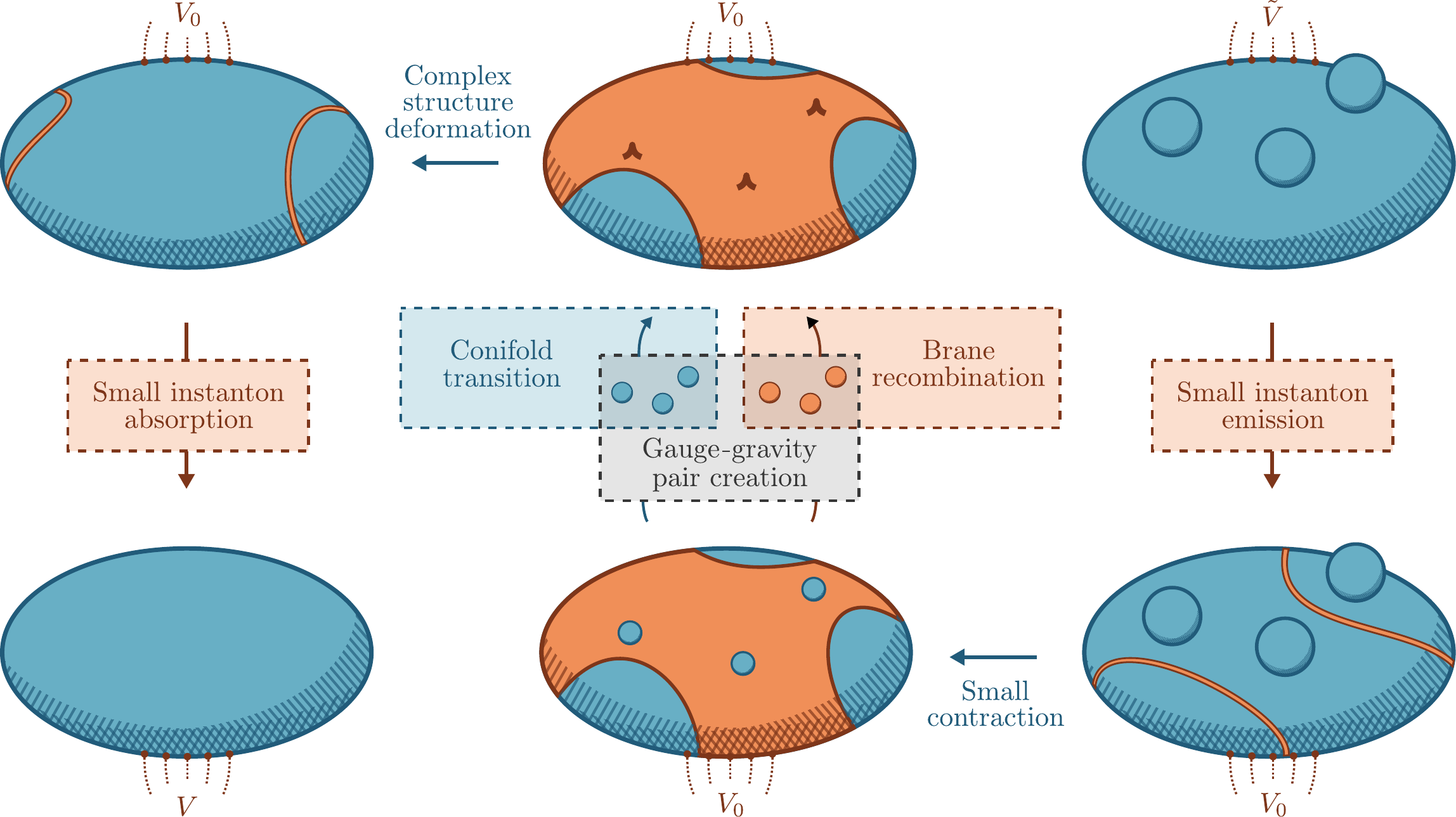}
\caption{{\it Beginning at the top-right: the path of the heterotic gauge bundle through a conifold transition, as described in the text. The result is the target space dual of the original theory, as we discuss in Section~\ref{sec:tsd}.}}
\end{figure}

\subsubsection{Example}
As an example of the phenomenon above, let us consider the 5-brane transition involving a conifold of the quintic used previously in this section. In this case, the relevant curve class on the quintic was $\left[C\right]=4D^2$ with explicit polynomial description given by $l_0=q_0=0$ as in \eqref{mrcdef}. Recall that the skyscraper (${\cal O}_C$) and ideal sheaves (${\cal I}_C$) of such a curve on the quintic are defined by the Koszul sequence as
\begin{align} \label{ideal_def}
&0 \to {\cal O}(-5) \to {\cal O}(-1) \oplus {\cal O}(-4) \to {\cal I}_C \to 0 \\
& 0 \to {\cal I}_C \to {\cal O}_X \to {\cal O}_C \to 0
\end{align}

 Let us define the following bundle on the quintic threefold as a spectator (in the sense above) to this conifold transition:
\begin{equation}
0 \to {\cal O}(-4) \to {\cal O}(-1)^{\oplus 4} \to V_0 \to 0
\label{spec_mon}
\end{equation}
Over a smooth quintic threefold, this bundle can be combined with a 5-brane wrapping $C$ via a Hecke transform as in \eqref{laraseq1} as a rank-changing transition
\begin{equation}
0 \to \hat{V} \to V_0 \oplus {\cal O} \to {\cal O}_C \to 0\;.
\end{equation}
The bundle $V_0$ has $c_2(V_0)=6D^2$ and as a result, can correctly pair with $C$ given above to exactly saturate the anomaly cancellation condition of $c_2(T_X)=10D^2= c_2(V) + \left[C\right]$. 

In this case, by direct computation, we find that the space of morphisms, $\text{Hom}(V_0, {\cal O}_C)$ is trivial and as a result $\hat{V}= V_0 \oplus {\cal I}_C$. Fortunately however, the cokernel descriptions of both our spectator bundle in \eqref{spec_mon} as well as the ideal sheaf in \eqref{ideal_def} lead to a natural addition of the short exact sequences as
\begin{equation}
0 \to {\cal O}(-5)+{\cal O}(-4) \stackrel{f^\mathrm{T}}{\longrightarrow} {\cal O}(-4) \oplus {\cal O}(-1) \oplus {\cal O}(-1)^{\oplus 4} \to V_0 \oplus {\cal I}_C \to 0
\end{equation}
which, for a block-diagonal map, $f^\mathrm{T}$, leads to a unified cokernel description of $V_0 \oplus {\cal I}_C$. If we deform the map $f^\mathrm{T}$ away from a block diagonal form\footnote{In the bundle moduli space of the Hecke transform $\hat{V}$, such a deformation is of so-called ``non-Hecke" type. See Appendix~\ref{app:hecke}.} this amounts to an appropriate smoothing of the sheaf ${\hat V}$ back to a smooth bundle\footnote{Note that in this case the repeated entries of ${\cal O}(-4)$ in the first and second terms in the sequence can be eliminated without changing the cokernel.}. The resulting bundle is a familiar one on the quintic
\begin{equation}
0 \to {\cal O}(-5) \to {\cal O}(-1)^5 \to V \to 0~.
\label{vdef}
\end{equation}
$V$ is a rank $4$ holomorphic deformation of the cotangent bundle of the quintic (i.e.\ essentially a deformation of the so-called ``standard embedding" on the quintic). 

On the resolution side of the conifold given by the CICY threefold in \eqref{cicy_res} likewise we begin with the spectator
\begin{equation}
0 \to {\cal O}(0,-4) \to {\cal O}(0,-1)^{\oplus 4} \to \res{V}_0 \to 0~,
\end{equation}
on this CY3 geometry and combine this object with $\res{C}$ from \eqref{mrcdef} after the 5-brane transition. The curve $\res{C}$ is in the class $5D_1D_2$ and its ideal sheaf is given by
\begin{align}
&0 \to {\cal O}(-1,-5) \to {\cal O}(-1,0) \oplus {\cal O}(0,-5) \to {\cal I}_{\res{C}} \to 0 \\
& 0 \to {\cal I}_{\res{C}} \to {\cal O}_X \to {\cal O}_{\res{C}} \to 0
\end{align} 
As above, we can combine the sequences for $\res{V}_0$ and ${\cal I}_{\res{C}}$ to obtain
\begin{equation}
0 \to {\cal O}(0,-4) \oplus {\cal O}(-1,-5) \stackrel{\tilde{f}^{\,\mathrm{T}}}{\longrightarrow} {\cal O}(0,-1)^{\oplus 4} \oplus {\cal O}(-1,0) \oplus {\cal O}(0,-5) \to \res{V}_0 \oplus {\cal I}_{\res{C}} \to 0 
\label{vres}
\end{equation}
This singular sheaf once again can be deformed into a smooth bundle $\res{V}$ on $\res{X}$ from \eqref{cicy_res} by tuning the map $\tilde{f}^{\,\mathrm{T}}$ away from block-diagonal form. Thus, we have constructed a \emph{dual pair of bundles} in the sense of Section~\ref{sec:brane_thru_con} in $V$ and $\res{V}$ in \eqref{vdef} and \eqref{vres} with general maps. As we will see in later sections, the language of ``duality" is justified in describing these connected bundles as they lead to apparently identical 4-dimensional theories (with perfect matching of their massless spectra) across the conifold transition.

\section{5-brane duality}
\label{sec:5brane_duality}

It transpires that the 5-brane transitions described in the previous section link compactified theories which appear to be dual. In this section, we provide proofs and general arguments of the matching for the various parts of the spectrum, and exhibit this 5-brane duality through examples.
Indeed, in this section, we move beyond the single example with which we have illustrated our discussion so far and in Section~\ref{sec:simple_class} we provide a large class of pairs of 5-brane theories connected by the transitions of the previous section. We then turn to the discussion of duality in Section~\ref{m5modmatch}.

\subsection{A simple class of 5-brane theory pairs}
\label{sec:simple_class}

The construction of Section~\ref{sec:brane_thru_con} details a prescription for continuously taking a 5-brane theory across a conifold transition. A very large class of examples in which this can be be conveniently and explicitly described is that of conifold transitions described by $\mbb{P}^n$-splits of toric complete intersections, and for any such case we have collected all of the relevant expressions in Appendix~\ref{sec:exp_for_pn_split}. A subclass of these for which it is less cumbersome to illustrate the construction is that of conifold transitions described by $\mbb{P}^1$-splits between CICYs, and hence we consider these below.

\medskip

Consider a pair of CICY manifolds $\defm{X}$ and $\res{X}$ which are related by a conifold associated to a $\mathbb{P}^1$-split.
\begin{eqnarray} \label{m5egsplit}
\defm{X} =\left[ \begin{array}{c|cc} {\cal A} & {\bf v}_{\hspace{.4pt}0}+{\bf v}_1 &{\bf R} \end{array} \right] \longleftrightarrow  \left[ \begin{array}{c|ccc} \mathbb{P}^1 & 1&1&0 \\ {\cal A} &{\bf v}_{\hspace{.4pt}0}&{\bf v}_1&{\bf R}\end{array} \right] =\res{X}
\end{eqnarray}
Here $\mc{A}$ is a product of $N$ projective spaces, ${\bf v}_{\hspace{.4pt}0}$ and ${\bf v}_1$ are vectors of length $N$ and ${\bf R}$ is an $N \times (K+1)$ matrix where $K = \mathrm{dim}({\cal A})-5$. It will be useful in what follows to use the following, equivalent, description of $\defm{X}$.
\begin{eqnarray} \label{redunXD}
\defm{X}= \left[ \begin{array}{c|ccc} \mathbb{P}^1&1&0&0\\ {\cal A} &{\bf 0}& {\bf v}_{\hspace{.4pt}0}+{\bf v}_1 &{\bf R} \end{array} \right] 
\end{eqnarray}

\begin{figure}[!t]\centering
\includegraphics[scale=0.65]{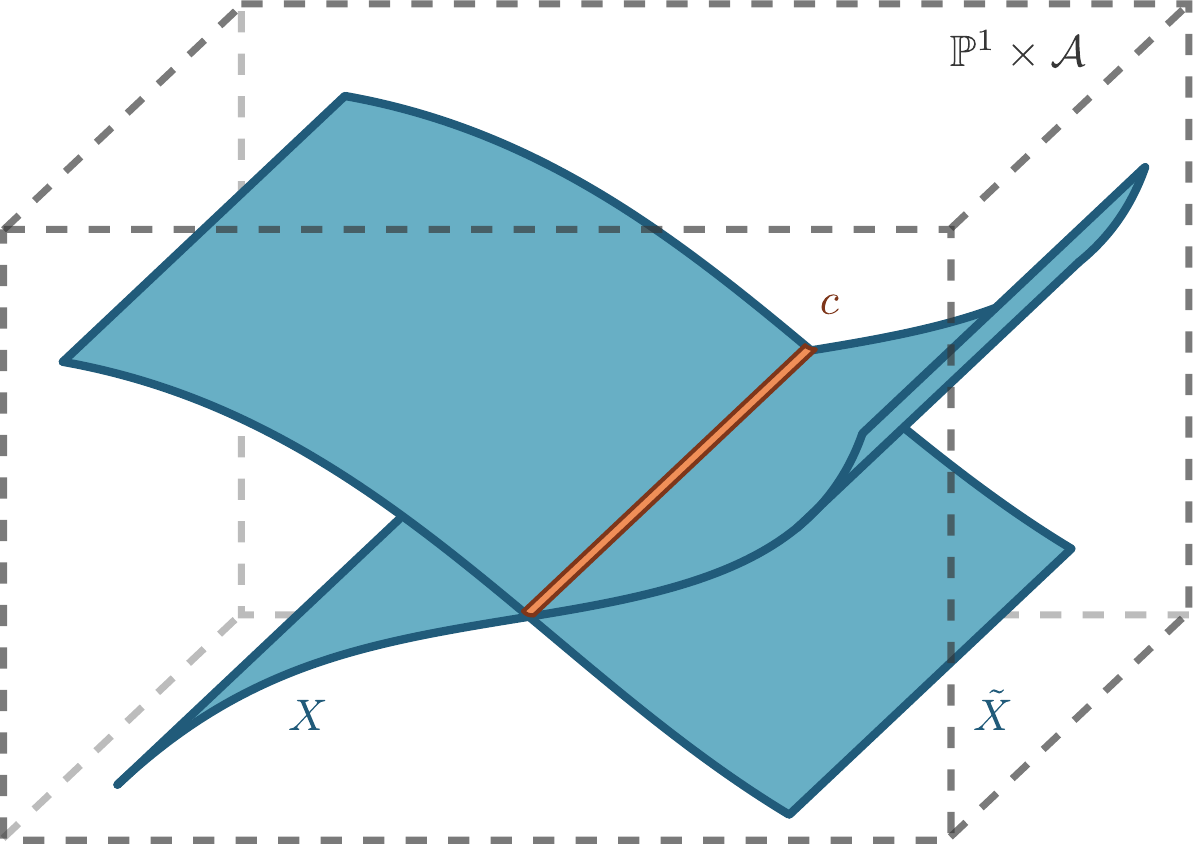}
\caption{{\it A graphical depiction of the construction of dual 5-brane theories for the special case of a $\mbb{P}^1$-split. The CICYs related by the $\mathbb{P}^1$-split, $\defm{X}$ and $\res{X}$, can be embedded in the same ambient space, and the 5-branes playing a crucial role in the construction are then given by the curve $c$ on which these two manifolds intersect.}}
\label{5dualpic}
\end{figure}

On these two manifolds we will consider a 5-brane stack wrapping a curve $c$ which is the intersection of $\res{X}$ in (\ref{m5egsplit}) and $\defm{X}$ in (\ref{redunXD}) in their shared ambient space.
\begin{eqnarray}
c=  \left[ \begin{array}{c|ccccc} \mathbb{P}^1&1&0&0&1&1\\ {\cal A} &{\bf 0}& {\bf v}_{\hspace{.4pt}0}+{\bf v}_1 &{\bf R} &{\bf v}_{\hspace{.4pt}0} & {\bf v}_1\end{array} \right] 
\end{eqnarray}
This situation is depicted schematically in Figure~\ref{5dualpic}. One can then complete these constructions by adding a `spectator' 5-brane stack, or indeed a spectator bundle, to each configuration in order to saturate the heterotic anomaly cancellation condition. Our statement is that these two theories are dual to each other\footnote{There is one small class of apparent exceptions to this duality, namely those where the resolution manifold is isomorphic to the Sch\"{o}n manifold. We discuss this in some detail below.}: they give the same low energy spectrum.

\vspace{0.2cm}

The curve $c$ has a different normal bundle considered as a complete intersection, $\defm{C}$ or $\res{C}$ respectively, in the two manifolds $\defm{X}$ and $\res{X}$.
\begin{eqnarray} \label{nbs}
{\cal N}_{\defm{C}} = {\cal O}({\bf v}_{\hspace{.4pt}0})\oplus {\cal O}({\bf v}_1) \;\;\; ,\;\;\; {\cal N}_{\res{C}} = {\cal O}(1,{\bf 0}) \oplus {\cal O}(0,{\bf v}_{\hspace{.4pt}0}+{\bf v}_1)
\end{eqnarray}
The classes of these 5-branes, as appearing in the anomaly cancellation condition, are as follows.
\begin{eqnarray}
\left[ \defm{C}\right] &=& ({\bf v}_{\hspace{.4pt}0} \cdot {\bf J})({\bf v}_1 \cdot {\bf J}) \\ \nonumber
[ \res{C} ] &=& J_0 ({\bf v}_{\hspace{.4pt}0}+ {\bf v}_1) \cdot {\bf J} 
\end{eqnarray}
In the above, $J_0$ is the K\"ahler form of the $\mathbb{P}^1$ involved in the split, restricted to the CY3 $\res{X}$ and ${\bf J}$ is a vector of the K\"ahler forms of the projective factors in ${\cal A}$ restricted to the relevant CY3 manifolds. Note that these objects are precisely the distinguished curves associated to the transition that were discussed in Section~\ref{sec:brane_thru_con}.

We can compute the second Chern characters of $\defm{X}$ and $\res{X}$ to find the following.
\begin{eqnarray}
\text{ch}_2 (T_{\defm{X}}) &=& \text{ch}_2(T_Y) - \frac{1}{2} \left( ({\bf v}_{\hspace{.4pt}0}+ {\bf v}_1) \cdot {\bf J}\right)^2 \\ \nonumber
\text{ch}_2(T_{\res{X}}) &=& \text{ch}_2(T_Y) -J_0 ({\bf v}_{\hspace{.4pt}0}+{\bf v}_1) \cdot {\bf J} - \frac{1}{2} ({\bf v}_{\hspace{.4pt}0} \cdot {\bf J})^2- \frac{1}{2} ({\bf v}_1 \cdot {\bf J})^2
\end{eqnarray}
In these expressions we have defined $Y = \left[{\cal A} \,|\, {\bf R}\right]$. Thus, we see that,
\begin{eqnarray}
\text{ch}_2 (T_{\defm{X}})-\text{ch}_2 (T_{\res{X}})= [ \res{C}] - \left[ \defm{C} \right]\;.
\end{eqnarray}

Given that these two 5-branes account for the difference in second Chern class between the two manifolds, a spectator bundle or 5-brane stack can be added to complete the model and make it anomaly free as claimed above. This spectator 5-brane stack wraps a curve in the class %
\begin{eqnarray} \label{specclass}
[C_0]= [\tilde{C}_0] = \frac{1}{2}\left(({\bf v}_{\hspace{.4pt}0}\cdot {\bf J})^2+({\bf v}_1\cdot {\bf J})^2\right)-\textnormal{ch}_2(T_Y)
\end{eqnarray}
on both sides of the duality. Notice that this class has no contributions involving $J_0$. In the case of completion by a spectator bundle, (\ref{specclass}) is the second Chern class of that object.

\subsection{Moduli matching across 5-brane duality}
\label{m5modmatch}

In the compactified theory of a Calabi-Yau threefold $X$ containing only a 5-brane wrapped on a single curve $C$, there are the following massless multiplets (see for example \cite{Lukas:1998hk}):
\bi
\item $h^{1,1}(X)+h^{2,1}(X)$ chiral multiplets
\item $g_C$ vector multiplets, where $g_C$ is the genus of the curve $C$
\item $h^0(C,{\cal N}_C)$ chiral multiplets\footnote{A derivation that this is indeed the correct enumeration of brane moduli is given in Appendix~\ref{appbranemod}.}, where ${\cal N}_C$ is the normal bundle of $C$ inside $X$
\ei
There is also an additional universal chiral multiplet for each 5-brane. If there are multiple 5-branes wrapped on various curves $C_i$ inside $X$, then one has a contribution $g_{C_i}$ and a contribution $h^0(C_i,{\cal N}_{C_i})$ from each. In our construction above, in the theory on $X$ (respectively $\tilde{X}$) there is a 5-brane wrapped on the curve $C$ ($\tilde{C}$), as well as a spectator 5-brane wrapped on a curve $C_0$ ($\tilde{C}_0$), so there are two contributions to consider from the 5-brane sector.

For these pairs of 5-brane theories resulting from our construction the above massless spectra match, and in this section we provide general arguments for this.

\medskip

Consider first the massless vector multiplets. Since these are given by the genera of the curves on which the 5-branes are wrapped, the matching of the vector multiplets across the pair of 5-brane theories will follow immediately if one can establish the isomorphisms $C_0 \cong \res{C}_0$ and $C \cong \res{C}$.

In the case of the spectator curves $C_0$ and $\res{C}_0$, this is clear. This is because these curves are related by smooth passage through the conifold transition, staying far from the singular/exceptional loci. That this is also true for the curves $C \cong \res{C}$ can be seen manifestly in the $\mbb{P}^1$-split class of examples in Section~\ref{sec:simple_class} above, since the curves $C$ and $\res{C}$ are simply different embeddings of the same curve $c$ into the two geometries $X$ and $\res{X}$. For a more general argument of the isomorphism $C \cong \res{C}$ in our construction, we refer the reader to Appendix~\ref{curvemoduliproof}. 

\medskip

Consider next the chiral multiplet moduli. In the case of the spectator branes, we expect that the contributions precisely match, $h^0(C_0,{\cal N}_{C_0}) = h^0(\res{C},{\cal N}_{\res{C}_0})$, and indeed, we provide in Appendix~\ref{curvemoduliproof} a proof that this is the case for any $\mbb{P}^n$-split between CICYs. By contrast, the moduli of the 5-brane stacks wrapping $\defm{C}$ and $\res{C}$ do differ in the two theories. Indeed, these contributions must compensate for the difference in geometric moduli across the conifold transition, i.e.\
\begin{eqnarray} \label{matchingresult}
h^0(\defm{C},{\cal N}_{\defm{C}})+h^{1,1}(\defm{X}) + h^{2,1}(\defm{X}) = h^0(\res{C},{\cal N}_{{\res{C}}})+h^{1,1}(\res{X})+h^{2,1}(\res{X}) \,.
\end{eqnarray}

For the large class of examples of conifold transitions described by $\mbb{P}^n$-splits of toric complete intersections, for which we explicitly perform the construction of the 5-brane theory pairs in Appendix~\ref{sec:exp_for_pn_split}, the descriptions of the geometry and the 5-branes are sufficiently explicit that one can determine in any particular case the above cohomologies, and hence verify that the above equality indeed holds.

We have performed this check in many explicit examples, including many cases where $h^{1,1}$ changes by more than one, or where the complete intersection description is non-favorable, providing a wealth of evidence for the general result.\footnote{There is however one small class of exceptions to this. In the few cases in which the resolution geometry is isomorphic to the Sch\"{o}n manifold, which appear also to be precisely the cases where the Weil non-Cartier divisor $\nod{D}$ on the singular variety is isomorphic to $T^4$, the moduli matching fails. The most obvious qualitative difference in these cases is the appearance of a non-zero $h^1(\nod{D},\mc{O}_{\nod{D}})$. An understanding of the role of this contribution may provide an explanation for why the moduli fail to match in these cases, and it would be interesting to understand the unique features of this special case further. However if one considers cases where the Weil non-Cartier divisor is not isomorphic to $T^4$, then one may hope for a general proof of moduli matching, along the lines of the argument in Appendix~\ref{curvemoduliproof}.} Moreover, in Appendix~\ref{curvemoduliproof}, we outline a proof of this in the most tractable case, of a conifold transition which changes $h^{1,1}$ by only one, and which can be described as a $\mbb{P}^1$-split between CICYs. Given that the result has been verified in a wealth of examples we also expect a more general proof to be possible along similar lines.

Clearly, when coupled with the preceding result concerning the moduli of the spectator 5-brane stacks, such a result guarantees that the total number of moduli, coming from 5-branes, complex structure moduli, K\"ahler moduli and the dilaton, will always match across the dual theories.

\vspace{0.2cm}

As a final comment for this section,  we believe that the 5-brane duality described here is a true duality, and not just an accidental matching at the level of spectrum. In particular, we expect other properties of the dual configurations, such as the potentials in their low energy effective theories, should also match. To see why this is expected to be so, we must discuss the relationship between the physical process we have been describing in this paper and the phenomenon of $(0,2)$ target space duality. It is to this topic that we turn in Section~\ref{sec:tsd}.

\subsubsection{Example}
\label{eg1}

Let us consider the case of the following $\mathbb{P}^1$-split of the quintic.
\begin{eqnarray} \label{spliteg1}
\defm{X} = \left[\begin{array}{c|c} \mathbb{P}^4 & 5 \end{array}\right] \longleftrightarrow \left[\begin{array}{c|cc} \mathbb{P}^1&1&1 \\ \mathbb{P}^4 &1&4  \end{array} \right] =\res{X}
\end{eqnarray}
In this case we have
\begin{eqnarray}
{\cal N}_{\defm{C}} = {\cal O}(1) \oplus {\cal O}(4) \;\;\;,\;\;\; {\cal N}_{\res{C}} = {\cal O}(1,0) \oplus {\cal O}(0,5)
\end{eqnarray}
One should of course complete this example by providing either a spectator bundle or a spectator 5-brane which will saturate the anomaly cancellation condition. If we opt for the latter possibility one can choose to include such objects on curves with the following normal bundles on the two sides.
\begin{eqnarray}
{\cal N}_{C_0} = {\cal O}(2) \oplus {\cal O}(3) \;\;\;,\;\;\; {\cal N}_{\res{C}_0} = {\cal O}(0,2) \oplus {\cal O}(0,3)
\end{eqnarray}
Counting the chiral multiplet moduli on both sides of the transition, we arrive at the following.
\begin{eqnarray} \label{matchingeg}
\begin{array}{|c|c|c|c|c|c|c|c|} \hline &&&&& \\[-1em] & h^0({\cal N}_{\defm{C}/\res{C}})& h^0({\cal N}_{C_0/\tilde{C}_0}) & h^{1,1}(X/\res{X}) & h^{2,1}(X/\res{X})  & \text{Total} \\ \hline \text{Deformation} & 38&30&1&101&170 \\ \text{Resolution} & 52&30&2&86& 170\\\hline \end{array}
\end{eqnarray}
Thus we see that these moduli match as claimed. One also finds that
$g_{C} = g_{\tilde{C}} = 51$ and ${g_{C_0} = g_{\tilde{C}_0} = 76}$, so that the vector multiplet moduli also match.

\vspace{0.1cm}

As was pointed out above, one could have completed this model with a spectator bundle rather than a spectator 5-brane stack. The following bundles are suitable, having the correct Chern classes to saturate anomaly cancellation.
\begin{eqnarray} \label{speceg}
0 \to {\cal O}(-4) \to {\cal O}(-1)^4 \to \defm{V}_0 \to 0 \\ \nonumber
0 \to {\cal O}(0,-4) \to {\cal O}(0,-1)^4 \to \res{V}_0 \to 0 
\end{eqnarray}
A short computation shows that the spectra of these two bundles, both singlet and charged, match on the two sides of the transition.

As a side note, we also note that $\res{V}_0$ restricts trivially to the exceptional locus in $\res{X}$ as was predicted in Section~\ref{sec:bund_thru_con}. A simple Koszul sequence computation, using the fact that the normal bundle of the exceptional locus is ${\cal O}(-1,1)\oplus {\cal O}(-1,4)$, reveals that ${\cal O}(0,-1)$ and ${\cal O}(0,-4)$ restrict to the trivial bundle on those $\mathbb{P}^1$s. The restriction of the sequence defining $\res{V}_0$ then immediately tells us that this does indeed restrict to a trivial bundle.

\section{Heterotic conifold transitions and target space duality}
\label{sec:tsd}

In this section we consider the simple extension of the duality outlined above for 5-branes to a duality involving gauge bundles (via heterotic small instanton transitions \cite{Witten:1995gx,Ovrut:2000qi,Buchbinder:2002ji}). As we will demonstrate below, in terms of the effective 4-dimensional ${\cal N}=1$ theory, the correspondence we derive is not new, but rather provides a geometric explanation of a known phenomenon arising in heterotic $(0,2)$ GLSMs -- so-called $(0,2)$ target space duality \cite{Distler:1995bc,Blumenhagen:1997vt,Blumenhagen:1997cn,Blumenhagen:2011sq,Rahn:2011jw}.

\subsection{Moduli matching for transitioning bundles} \label{bunmodmatch}

Let us describe how the degrees of freedom match for bundles connected across the conifolds as described in Section~\ref{sec:bund_thru_con}. We begin by considering the moduli of the sheaves described by two Hecke transforms of the form (\ref{idealsky}), one on each side of the transition, which describe 5-branes and spectator bundles on the point of being recombined into a smooth higher rank object.

The degrees of freedom of these Hecke transform sheaves, while they are still in the form (\ref{idealsky}) with a separated ideal sheaf, are derived in Appendix~\ref{app:hecke}. There it is shown that the moduli are given by\footnote{Note that, although here we have given the expressions for the resolution side, an identical form would hold for the deformation geometry.}
\begin{eqnarray} \label{heckemoduli}
H^1(\res{V}_0^{\vee} \otimes \res{V}_0)  \oplus  \textnormal{Ext}^1(\res{V}_0,{\cal I}_{\res{C}})   \oplus \textnormal{Ext}^1({\cal I}_{\res{C}},\res{V}_0)  \oplus  H^0({\cal N}_{\res{C}}|_{\res{C}})\;.
\end{eqnarray}
Although it may not be apparent at first, the first three of these terms are the same on both sides of the transition. From the discussions of Section~\ref{m5modmatch} this is clear for the spectator bundle moduli $H^1(\res{V}_0^{\vee} \otimes \res{V}_0)$. For the extension groups, this follows from the following two expressions, which hold for the case at hand.
\begin{eqnarray}
\text{Ext}^1(\res{V}_0,{\cal I}_{\res{C}})&=&\ker\left(H^1(\res{V}_0^{\vee} )\to H^1(\res{V}_0^{\vee}|_{\res{C}})\right) \\ 
\text{Ext}^1({\cal I}_{\res{C}},\res{V}_0)&=&H^1(\res{V}_0)\oplus \ker\left(\text{Ext}^2({\cal O}_{\res{C}},\res{V}_0) \to \text{Ext}^2({\cal O},\res{V}_0) \right)
\end{eqnarray}
These quantities only involve $\res{V}_0$ and quantities intrinsic to the curves, which are identical on the two sides of the duality.

The last term in (\ref{heckemoduli}) does not match on the two sides of the duality but is exactly the 5-brane moduli from Section~\ref{m5modmatch}. As described in that section, these moduli differ by exactly the same number of degrees of freedom required to account for the difference in Hodge numbers and so geometrical moduli of the underlying manifolds.

Given the above discussion, we see that the moduli on the resolution and deformation sides of the duality match before the smoothing is performed to turn the Hecke transforms of the form (\ref{idealsky}) into smooth bundles. In deforming to the smooth bundle situation, moduli are lost. In every case we have examined the two bundles change their moduli by the same number. As might be expected, this is often a change of a single modulus, as is the case in our canonical example that was used throughout Section~\ref{sec2}.  

A similar, but simpler analysis holds for the charged matter of the system. Starting with the Hecke sequence (\ref{idealsky}) one can work out the charged matter, associated to cohomology groups such as $H^1(\defm{V})$ and $H^1(\res{V})$, in terms of properties of the spectator bundles and properties intrinsic to the curves $\defm{C}$ and $\res{C}$. These match on the two sides of the transition and, in this case, are generically unchanged in deforming to the smooth point in bundle moduli space in examples we have seen.

The real evidence that the final moduli and matter counts do always match on the two sides of the transition, however, is given by linking the process we are describing here to a well known duality that has already been discussed at length in the literature. It is to this that we turn in the next subsection.

\subsection{Connections to $(0,2)$ GLSM target space duality}
In $(0,2)$ Target Space Duality (TSD), two $(0,2)$ heterotic GLSMs are found to share a non-geometric branch of their vacuum space (either a Landau-Ginzburg phase or a more general hybrid phase) and the subsequent pair of GLSMs reveal 4-dimensional ${\cal N}=1$ theories that appear to be ``dual" in the sense that their total massless spectrum for both charged and uncharged fields is identical. In particular, the number of uncharged singlets, as counted by
\begin{equation}
h^{1,1}(X)+h^{2,1}(X) + h^1(X,\text{End}_0(V))\;,
\end{equation}
is preserved across the pair. This is true despite the fact that the underlying CY3 manifolds are topologically distinct (with different Hodge numbers), as are the vector bundles over them. Moreover, subsequent work \cite{Anderson:2016byt} demonstrated that even when D- or F-term contributions to the scalar potential ``lift" some of these flat directions in the vacuum space, the true number of singlets remaining matches across the TSD paired theories.

Since the primary focus of the present paper is on the geometry of the heterotic manifold/bundle and the associated 4-dimensional field theory, we will not provide a detailed review of target space duality as it arises in 2-dimensional $(0,2)$ GLSMs here, but instead summarize its effective action on a monad bundle over a complete intersection CY3 manifold inside a toric variety (the geometry that naturally arises in $(0,2$) GLSMs).

\vspace{0.1cm}

In the context of a $(0,2)$ GLSM, we are given a bundle $V$ defined as the kernel\footnote{In the interests of simplicity, we will for now exclude fermionic gauge symmetries which can lead to more general monads.} ($V=\ker(F)$) of a morphism between sums of line bundles
\begin{equation}\label{mon_def}
0 \to V \to \bigoplus_a {\cal O}({\bf b}_a) \stackrel{F}{\longrightarrow}\bigoplus_l {\cal O}({\bf c}_l) \to 0
\end{equation}
over a CY3 manifold defined as a complete intersection (of polynomials $G_j(x_\alpha)=0$) with normal bundle ${\cal N}=\bigoplus_j {\cal O}({\bf s}_i)$ and a set of homogeneous coordinates $x_i$ with weights ${\bf q}_i$ (where the boldface quantities are vectors running over $h^{1,1}(X)$ components). In this notation, the Calabi-Yau condition is satisfied if $\sum_i {\bf q}_{i}=\sum_j {\bf s}_j$, and $c_1(V)=0$ leads to $\sum _a {\bf b}_a=\sum_l {\bf c}_l$ (for each component of the vectors).

In this notation, the monad is defined as the kernel of a holomorphic map,
\begin{equation}
F_a^l(x_\alpha) \in H^0(X, {\cal O}({\bf c}_l - {\bf b}_a))\;,
\end{equation}
while the manifold is defined by the vanishing of a set of holomorphic functions,
\begin{equation}
G_j(x_\alpha) \in H^0(X, {\cal O}({\bf s}_j))~.
\end{equation}

In the appropriate circumstances, target space duality simply involves the observation that in a non-geometric vacuum of the $(0,2)$ GLSM, a pair (or more) of functions $F_i^j(x_\alpha)$ and $G_r(x_\alpha)$ can be interchanged \emph{without changing the Landau-Ginzburg or hybrid theory in that phase.} If this apparent symmetry is used to relabel GLSM fields in that limit and then then one moves back to a geometric phase of the theory, this interchange of $F \leftrightarrow G$ has effectively defined a new monad bundle and CY3 manifold. 

In the present discussion we will consider pairs of $F \leftrightarrow G$ interchanges which will be labeled without loss of generality by fixing $l=1$ and considering $a=1,2$, thus focusing on the bundle maps $F_1^1$ and $F_2^1$. If the multi-degrees of the polynomials match such that
\begin{equation}
2{\bf c}_1-{\bf b}_1-{\bf b}_2={\bf s}_1 +{\bf s}_2
\end{equation}
then a target space dual geometry can exist\footnote{Subject to verifying that an appropriate hybrid phase vacuum actually exists with the appropriate vev, $\langle p_1 \rangle \neq 0$.} in which 
\begin{equation}
F_1^1 \leftrightarrow G_1~~~~\text{and}~~~~~F_2^1 \leftrightarrow G_2
\label{switch_eg}
\end{equation}
leading to a new $(0,2)$ GLSM, i.e.\ a new manifold/bundle pair.

To make contact with previous literature involving bundles constructed as monads (see for example \cite{Distler:1987ee,Kachru:1995em,Douglas:2004yv,Anderson:2008uw,Anderson:2009mh}) and the $(0,2)$ GLSM literature on target space duality \cite{Distler:1995bc,Blumenhagen:1997vt,Blumenhagen:1997cn,Blumenhagen:2011sq,Rahn:2011jw} we will allow the GLSM charge matrix data to determine the multi-degrees of the line bundles \eqref{mon_def} via the following dictionary:
\begin{equation}
{\bf b}_a={\bf \Lambda}_a~~~,~~\;{\bf c}_l=|{\bf p}|_l
\end{equation}
while the normal bundle of the CY3 manifold is determined by ${\bf s}_i=|{\bf \Gamma}|_j$. 

We turn now to an example and consider the following manifold and bundle pair (given in terms of GLSM charge data), originally presented in \cite{Rahn:2011jw}. An $\mathrm{SU}(3)$ bundle $V$, given in monad form, over a manifold $X$ is presented as follows.
\begin{eqnarray}\begin{aligned}&\begin{array}{|c|c||c|c|}
\hline
x_i & \Gamma^j & \Lambda^a & p_l \\ \noalign{\hrule height 1pt}\begin{array}{ccccccc}
0& 0& 0& 1& 1& 1 & 1  \\
1& 1& 1& 2& 2& 2 &0
\end{array}&\begin{array}{cc}
-2 &-2 \\
-4 & -5
\end{array}&\begin{array}{cccc}
0 & 2 & 1& 0  \\
 1& 6 & 0& 1
\end{array}&\begin{array}{c}
 -3 \\
 -8
\end{array}\\
\hline
\end{array} \label{tsd_eq1}\end{aligned}\end{eqnarray}
The massless singlet spectrum of this theory is counted by
\begin{equation}
h^{1,1}(X)+h^{2,1}(X)+h^1(X,\text{End}_0(V))=2+68+322=392 \label{sing_count1}
\end{equation}
and the charged matter is given by $n_{{\bf 27}}=120$, $n_{{\bf\bar{27}}}=0$.

As shown in \cite{Rahn:2011jw}, \eqref{tsd_eq1} is linked by target space duality to the following manifold bundle pair $(\res{X},\res{V})$:

\begin{eqnarray}\begin{aligned}&\begin{array}{|c|c||c|c|}
\hline
x_i & \Gamma^j & \Lambda^a & p_l \\ \noalign{\hrule height 1pt}\begin{array}{ccccccc}
0& 0& 0& 1& 1& 1 & 1  \\
1& 1& 1& 2& 2& 2 &0
\end{array}&\begin{array}{cc}
-3 &-1 \\
-7 & -2
\end{array}&\begin{array}{cccc}
 1& 1& 1& 0  \\
 4& 3& 0& 1 
\end{array}&\begin{array}{c}
 -3 \\
 -8
\end{array}\\
\hline
\end{array} \label{tsd_eq2}\end{aligned}\end{eqnarray}
where the maps that were interchanged as in \eqref{switch_eg} to produce \eqref{tsd_eq2} are here of multi-degree
\begin{align}
&G_1=(\tilde{F}^1_1)_{(2,4)} && G_2=(\tilde{F}^1_2)_{(2,5)} \\
&\tilde{G}_1=(F^1_1)_{(3,7)} && \tilde{G}_2=(F^1_2)_{(1,2)}
\end{align}
Interestingly, we may without loss of generality, choose the bi-degree $(1,2)$ defining equation of $\res{X}$ to consist of a single weight $(1,2)$ coordinate $x_i$ (by choice of coordinates) and hence, the description can be reduced to a single hypersurface.
\begin{eqnarray}\begin{aligned}&\begin{array}{|c|c||c|c|}
\hline
x_i & \Gamma^j & \Lambda^a & p_l \\ \noalign{\hrule height 1pt}\begin{array}{cccccc}
0& 0& 0& 1& 1& 1  \\
1& 1& 1& 2& 2 &0
\end{array}&\begin{array}{c}
-3  \\
-7 
\end{array}&\begin{array}{cccc}
 1& 1& 0& 1  \\
 4& 3& 1& 0 
\end{array}&\begin{array}{c}
 -3 \\
 -8
\end{array}\\
\hline
\end{array} \label{tsd_eq2a}\end{aligned}\end{eqnarray}
This removal of a ``redundant" constraint equation (with the same multi-degree as a coordinate) can be consistently realized in the GLSM by integrating out a massive pair of fields (i.e.\ $x_4$ and $\Gamma^2$).

As expected of target space duality, the massless spectrum remains the same, though distributed differently,
\begin{equation}
h^{1,1}(\res{X})+h^{2,1}(\res{X})+h^1(X,\text{End}_0(\res{V}))=2+95+295=392 \label{sing_count2}
\end{equation}
with the same $E_6$ gauge group and charged matter spectrum as before. Note that in this example, in moving from $(X,V)$ to $(\res{X},\res{V})$ some of the complex structure and bundle moduli were interchanged, while the number of K\"ahler moduli remained the same. In general, it was observed in \cite{Blumenhagen:1997vt,Blumenhagen:2011sq} that target space dual pairs can involve a mixing of \emph{all three} types of geometric moduli by using similar ``redundancies" to the one observed in the defining equations of the example above, only using them in reverse. In particular, by introducing a ``redundant" description of $X$ which involves more $\mathbb{C}^*$ actions, the K\"ahler moduli can be non-trivially included in the process. The general procedure for this redundancy and then subsequent construction of the target space dual is laid out in detail in \cite{Blumenhagen:2011sq}. Here we will simply summarize the approach by means of an example. 

Consider the quintic hypersurface $\mathbb{P}^4[5]$. A simple redundant description of this Calabi-Yau threefold is given by 
\begin{eqnarray} \label{redund_quintic}
 \left[\begin{array}{c|cc} \mathbb{P}^1&1&0 \\ \mathbb{P}^4 &0&5  \end{array} \right] 
\end{eqnarray}
The two manifolds are equivalent for the same reasons as in the example above. Here the geometry of redundancy is especially simple as the linear constraint picks out a single point in the $\mathbb{P}^1$ ambient space (and clearly as a manifold, $X=X \times \{pt\}$). However, beginning with this redundant description as a starting point leads to novel target space dual pairs. 

As an example, we will take the same bundle which has appeared in prior sections of this work, namely the rank 4 deformation of the tangent bundle of the quintic CY3:
\begin{equation}
0 \to V \to {\cal O}(1)^{\oplus 5} \to {\cal O}(5) \to 0
\label{quintic_tan_def}
\end{equation}
The manifold redundancy mentioned above can be extended in a similar manner to the bundle (as first noted in \cite{Anderson:2016byt}) and we will choose here to add a repeated entry to the second and third terms of this sequence as
\begin{equation}
0 \to V \to {\cal O}(1)^{\oplus 5} \oplus {\cal O}(4) \to {\cal O}(5)\oplus {\cal O}(4) \to 0~~.
\end{equation}
Presenting this bundle and the redundant quintic in \eqref{redund_quintic} in GLSM charge matrix notation we find
\begin{eqnarray}
\begin{array}{|ccccccc|cc||cccccc|cc|}
\hline
y_0 & y_1 & y_2& y_3& y_4&x_0 &x_1& \Gamma^1 & \Gamma^2 & \Lambda^1 & \Lambda^2& \Lambda^3 & \Lambda^4 & \Lambda^5 & \Lambda^6 & p_1 & p_2 \\
\hline
0& 0& 0& 0& 0& 1 & 1 &-1 & 0 &0 & 0 & 0& 0  & 0 &0  & -1 & 0 \\
1& 1& 1& 1& 1& 0 & 0& -0 & -5 &  4& 1 & 1& 1 & 1 & 1 &  -5 & -4 \\
\hline
\end{array} \label{tsd_cano_eg}~.\end{eqnarray}
Note that in this redundant description it naively seems that $c_1(X) \neq 0$ and $c_1(V) \neq 0$. However due to the simple geometric nature of the redundancy this is not actually the case. In the GLSM the anomalies are cancelled by the condition that the net sum of charges is vanishing (i.e.\ $c_1(X)+c_1(V)=0$) which still holds. Explicitly we choose defining equations in \eqref{tsd_cano_eg} to be
\begin{align}
&x_0=0 \\
&p_5(y)=0 \label{quin_def_eq}
\end{align}

For the geometry described by \eqref{tsd_cano_eg}, the algorithm of \cite{Anderson:2016byt} leads us to a new manifold/bundle pair
\begin{eqnarray}
\begin{array}{|ccccccc|cc||cccccc|cc|}
\hline
y_0 & y_1 & y_2& y_3& y_4&x_0 &x_1& \Gamma^1 & \Gamma^2 & \Lambda^1 & \Lambda^2& \Lambda^3 & \Lambda^4 & \Lambda^5 & \Lambda^6 & p_1 & p_2 \\
\hline
0& 0& 0& 0& 0& 1 & 1 &-1 &-1 &1 & 0 & 0& 0  & 0 &0  & -1 & 0 \\
1& 1& 1& 1& 1& 0 & 0& -1 & -4 &  0& 5 & 1& 1 & 1 & 1 &  -5 & -4 \\
\hline
\end{array} \label{tsd_cano_eg_dual}~.\end{eqnarray}
Here the defining equations are given by
\begin{align}
&l_0(y)x_0+ l_1(y)x_1=0 \label{cicy_def_eq1} \\
&q_0(y)x_0+q_1(y)x_1=0 \label{cicy_def_eq2}
\end{align}

These manifold/bundle pairs are of course the canonical example that we have studied throughout Section~\ref{sec2}. In examples of this kind in target space duality, the fact that the base CY3 manifolds, $X, \res{X}$, are related by a conifold transition is a consequence of the redundant description used. Here a linear hypersurface constraint in $\mathbb{P}^1$ as in \eqref{redund_quintic} led to a $\mathbb{P}^1$-split, while in general $n$ linear constraints in $\mathbb{P}^n$ used as a redundancy leads to a conifold realized as a $\mathbb{P}^n$-split.

In this case the singlet spectrum of $(X,V)$ is given by
\begin{equation}
h^{1,1}(X)+h^{2,1}(X)+h^1(X,\text{End}_0(V))=1+101+325=427 \label{sing_count_quin}
\end{equation}
while in the target space dual geometry
\begin{equation}
h^{1,1}(\res{X})+h^{2,1}(\res{X})+h^1(\res{X},\text{End}_0(\res{V}))=2+86+339=427 \label{sing_count_quin2}
\end{equation}
and for both theories $n_{{\bf 27}}=100$ and $n_{\bar{{\bf 27}}}=0$ as expected. In addition, it was shown in \cite{Blumenhagen:2011sq} that for every anomaly-consistent geometry $(X,V)$ that generates a target space dual, $(\res{X},\res{V})$, via this redundant ambient space procedure, the dual geometry is guaranteed to also satisfy anomalies.

Note that in this case, the interchange of CY3 defining equations and monad maps takes the form
\begin{align}
&G_1= \tilde{F}^1_1=x_0 &&G_2=\tilde{F}^1_2=p_5 \\
&\tilde{G}_1=F^1_1=l_0(y)x_0+l_1(y)x_1 && \tilde{G}_2=F^1_2=q_0(y)x_0+q_1(y)x_1
\label{switcheroo}
\end{align}
and for TSD to hold, these defining equations/polynomial maps must be held equal (and all other bundle maps which remain unchanged are also chosen to agree). Note that this effectively provides a map from a point in the moduli space of $(X,V)$ to a point in the moduli space of $(\res{X},\res{V})$.

At a naive first pass, the TSD procedure implemented above seems to indicate that in some sense in the complete geometry, components of a manifold/bundle (i.e.\ a pair $G({\bf x}_i),F({\bf x}_i)$ as in \eqref{switcheroo}) have been interchanged in order to construct a new stable bundle/CY3 manifold.  The exact \emph{geometric} nature of this interchange and any direct links to the conifold transition connecting the CY3 manifolds has remained a mystery from the point of view of the heterotic backgrounds\footnote{The link to conifold transitions are also mysterious from the point of view of the GLSM since the matching of vacuum spaces typically happens deep in a non-geometric phase.}  $(X,V)$ and $(\res{X},\res{V})$. In the remainder of this section, we will argue that at least in the case of $X$ and $\res{X}$ connected by conifold transitions, the gauge/gravitational instanton transition described in previous sections provides such an explanation.

To begin this exploration, note that for the given polynomials exchanged in \eqref{switcheroo}, much of the bundle effectively carries through the transition trivially. We can exploit this fact by moving to a point in moduli space where the monad map becomes block diagonal (and hence the bundle itself becomes a direct sum). This allows us to divide the bundle into two pieces -- one that changes  and one that doesn't (we'll refer to this latter piece as $V_0$). In the dual (cokernel) bundle description we can write each bundle as $V_0\oplus {\cal I}$ (respectively $\res{V}_0 \oplus {\cal \res{I}}$). Here the unchanging parts (i.e.\ the ``spectators") are given as
\begin{align}
&0 \to {\cal O}(-4) \to {\cal O}(-1)^{\oplus 4} \to V_0 \to 0 \\
&0 \to {\cal O}(0,-4) \to {\cal O}(0,-1)^{\oplus 4} \to \res{V}_0 \to 0
\end{align}
which are familiar from \eqref{speceg} in Section~\ref{sec:bund_thru_con}, while the pieces of the bundles that actually change under the target space duality procedure are
\begin{align}
&0 \to {\cal O}(-5) \to {\cal O}(-1) \oplus {\cal O}(-4) \to {\cal I} \to 0 \label{def_ideal}\\
&0 \to {\cal O}(-1,-5) \to {\cal O}(0,-5)\oplus {\cal O}(-1,0) \to {\cal \res{I}} \to 0
\end{align}
Of course the suggestively named objects ${\cal I},{\cal \res{I}}$ are ideal sheaves and the ideal sheaves of very special curves that we have seen already arising in previous sections! In particular ${\cal I}$ is the ideal sheaf of a curve $\defm{C}$ (familiar from Section~\ref{sec:brane_thru_con}) in the class $4D^2$ in the quintic manifold $X$ defined by the vanishing of the polynomials
\begin{equation}
l_1(y)=q_1(y)=0
\label{def_curve}
\end{equation}
while ${\cal \res{I}}$ is the ideal sheaf of a curve in the class $5D_1D_2$ in $\res{X}$ given by
\begin{equation}
x_0=p_5(y)=0
\end{equation}
where $l_1(x),q_1(x)$ and $p_5(x)$ are defined as in \eqref{quin_def_eq}, \eqref{cicy_def_eq1} and \eqref{cicy_def_eq2}. Note that in order to maintain target space duality, the defining equations of these curves are toggled to the defining equations of the dual manifold as in \eqref{tsd_cano_eg_dual}. For generic choices of the quintic defining equation, the loci supporting the sheaves ${\cal I}$, ${\cal \res{I}}$ are co-dimension 2 (i.e.\ curves). However, for this correlated system of manifolds/curves in the limit that the quintic is tuned to the conifold (i.e.\ nodal) point
\begin{equation}
p_5=l_0q_1-l_1q_0
\label{nodal_quintic}
\end{equation}
they are precisely the Weil non-Cartier divisors described in Section~\ref{sec:brane_thru_con} and whose role in 5-brane physics was explored in Sections~\ref{sec2} and~\ref{sec:5brane_duality}. More precisely, as described in Section~\ref{sec:bund_thru_con}, the ideal sheaf given in \eqref{ideal_def} can be removed from the bundle $V$ via a small instanton transition (described by a Hecke transform of the form given in \eqref{idealsky} in Section~\ref{sec:bund_thru_con})
\begin{equation}
0 \to V_0 \oplus {\cal I} \to V_0 \oplus {\cal O} \to {\cal O}_C \to 0
\label{hecke_v2}
\end{equation}
Finally, and most importantly, as described in Section~\ref{sec:brane_thru_con}, the fact that the bundle decompositions exist of the form $V_0 \oplus {\cal I}$ above means that the arguments of Sections~\ref{sec:brane_thru_con} and \ref{sec:5brane_duality} guarantee that the observed matching of the massless moduli and charged across this TSD pair follows from the discussions of Sections~\ref{sec:bund_thru_con} and \ref{bunmodmatch} and Appendix~\ref{app:hecke}. \emph{Thus, we have understood the moduli matching of TSD from a geometric point of view!}

The results provided above are for a single pair of manifolds/bundles. However we expect these arguments to hold for all TSD pairs involving conifolds (and all toric $\mathbb{P}^n$-splits) and have verified this in a large number of examples. Indeed, as can be noted from previous sections the majority of our results hold for generic conifold transitions in toric complete intersections. Moreover, although required from the GLSM viewpoint our proofs do not rely on the monad construction of vector bundles and hence, in that sense (in addition to the sense in which they include the purely 5-brane duality) are more general than the setting of TSD.

\vspace{0.1cm}

One exhaustive playground in which to test the ubiquity of the correspondences above -- i.e.\ the explanation of target space duality via gauge/gravitational pair creation -- is to consider all stable monad bundles on the quintic with $c_2(V)=c_2(TX)$. The list of such bundles was first found in \cite{Douglas:2004yv} (see also \cite{Anderson:2008uw,Anderson:2009mh} for a description of systematic enumerations of monads with particular $c_2(V),c_3(V)$, etc.). For each vector bundle in the list with $c_2(V)=10D^2$ we can ask the following questions:
\begin{itemize}
\item Can this bundle be linked to some other vector bundle $\res{V}$ on the manifold in \eqref{tsd_cano_eg_dual} by TSD?
\item Does this bundle admit a non-trivial Hecke transform surjection $V \to {\cal O}_C \to 0$ for $C$ the curve defined in \eqref{def_curve}? (That is, can the necessary small instanton transition be performed that effectively partitions the dual monad bundle into $V_0$ and ${\cal I}_C$ as above?)\end{itemize}
In each case we find that the answer to the first question is positive if and only if the second is also true. That is, the existence of a target space dual pair and an appropriate gauge/gravitational instanton transition across the conifold is one-to-one for this set. Moreover, the questions posed above for the single conifold transition linking the quintic to the CICY threefold given in \eqref{tsd_cano_eg_dual} can be repeated for \emph{every} conifold transition beginning on the quintic that increases $h^{1,1}$ by $1$. There are 18 such manifolds whose second Chern classes take the form 
\begin{equation}
c_2=(10-n)D_2^2+\ldots
\label{c2_change}
\end{equation} 
where $D_2^2$ is the direction in the Mori cone of the resolved CY3 manifold that ``carries through the conifold" from the original quintic threefold. Conifold transitions can be found for each integer value of $n$ in \eqref{c2_change} (note that in the example presented above, $n=4$) and these come in pairs consisting of a CY3 threefold and its flop in the new (i.e.\ $\mathbb{P}^1$) direction associated with the small resolution. In each case, we find that the target space duality can be performed if and only if the appropriate ideal sheaf to a special curve (in the sense of Section~\ref{sec:brane_thru_con}) can be identified inside $V$.

To conclude this section we note that in the arguments above and those regarding Hecke transforms and small instantons given in Section~\ref{sec:bund_thru_con} we considered the fully decomposed, direct sum limit of the (dual) vector bundle into ``spectator + ideal sheaf" in order to explain the transition and continuity of moduli. However, in some examples of target space duality, it seems that such a complete direct sum decomposition is more than is required to follow the bundle through the conifold transition (or equivalently perform the TSD). Instead, in some cases only the weaker condition itemized above that there exists a map $V \to {\cal O}_C \to 0$ seems to be required. As an example of this we can consider the flop of the CY3 geometry given in \eqref{tsd_cano_eg_dual}. The same deformation of the tangent bundle of the quintic in \eqref{quintic_tan_def} can be written with a different redundant description as
\begin{eqnarray}
\begin{array}{|ccccccc|cc||ccccc|c|}
\hline
y_0 & y_1 & y_2& y_3& y_4&x_0 &x_1& \Gamma^1 & \Gamma^2 & \Lambda^1 & \Lambda^2& \Lambda^3 & \Lambda^4 & \Lambda^5  & p_1  \\
\hline
0& 0& 0& 0& 0& 1 & 1 &-1 & 0 &0 & 0 & 0& 0  & 0  & -1  \\
1& 1& 1& 1& 1& 3 & 0& -3 & -5 &  1& 1 & 1& 1 & 1  &  -5 \\
\hline
\end{array} \label{tsd_cano_egt}~,\end{eqnarray}
which is TSD to this manifold/bundle pair,
\begin{eqnarray}
\begin{array}{|ccccccc|cc||ccccc|c|}
\hline
y_0 & y_1 & y_2& y_3& y_4&x_0 &x_1& \Gamma^1 & \Gamma^2 & \Lambda^1 & \Lambda^2& \Lambda^3 & \Lambda^4 & \Lambda^5  & p_1  \\
\hline
0& 0& 0& 0& 0& 1 & 1 &-1 & -1 &1 & 0 & 0& 0  & 0  & -1  \\
1& 1& 1& 1& 1& 3 & 0& -4 & -4 &  0& 2 & 1& 1 & 1  &  -5 \\
\hline
\end{array} \label{tsd_cano_eg_flop}~.\end{eqnarray}
The toric complete intersection 3-fold is the flop of that given in \eqref{tsd_cano_eg_dual}. Note that in this case, the relevant curve in $X$ which controls the transition (in the sense of \eqref{def_ideal} and \eqref{def_curve}) is given by
\begin{equation}
q_0=q_1=0
\end{equation}
which lies in the class $16D^2$. This class is manifestly too large to support a 5-brane in an anomaly-consistent manner. As a result, the picture of a 5-brane-through-conifold-transition followed by a Hecke transform as outlined in Sections~\ref{sec:brane_thru_con} and \ref{sec:bund_thru_con} is unclear. However, without fully removing this 5-brane from the bundle, but instead moving to a tuned limit where a Hecke sequence such as \eqref{hecke2} can be defined, the general process can still be completed and a bundle transitioned along with the manifold through the conifold transition. Finally it is worth noting that in this particular example, the same manifold and bundle can be found via target space duality beginning with a \emph{different} redundant description, namely
\begin{eqnarray}
\begin{array}{|ccccccc|cc||cccccc|cc|}
\hline
y_0 & y_1 & y_2& y_3& y_4&x_0 &x_1& \Gamma^1 & \Gamma^2 & \Lambda^1 & \Lambda^2& \Lambda^3 & \Lambda^4 & \Lambda^5& \Lambda^6  & p_1& p_2  \\
\hline
0& 0& 0& 0& 0& 1 & 1 &-1 & 0 &0 & 0 & 0& 0  & 0 & 0  & -1 & 0  \\
1& 1& 1& 1& 1& 3 & 0&  0 & -5 &  2& 1 & 1& 1 & 1  & 1&  -5 & -5 \\
\hline
\end{array} \label{tsd_cano_egt_v2}~.\end{eqnarray}
which is associated to the key curve in the class $D^2$ given by
\begin{equation}
l_0=l_1=0
\end{equation}
which \emph{can} be fully removed from $V$ as a 5-brane in an anomaly-consistent way. In the case of this particular conifold transition there are 4 Weil non-Cartier divisors in the nodal limit given by \eqref{nodal_quintic}. These consist of two curves in the class $4D^2$ (given by $l_0=q_0=0$ and $l_1=q_1$ respectively) which both connect the quintic to the threefold given in \eqref{tsd_cano_eg_dual} and those described above in the classes $D^2$ and $16D^2$ (both of which lead to the CY3 manifold in \eqref{tsd_cano_eg_flop}). Of these only the one class ($16D^2$) is incompatible with a complete 5-brane transition in $X$. We leave as an open question whether every conifold pair has at least one anomaly-consistent curve connecting the CY3s in the sense of Section~\ref{sec:5brane_duality}. This has certainly been the case for every example we have studied.

\section{Discussion and outlook}
A key motivation of this work is the question of whether and how a compactification of heterotic string theory on a Calabi-Yau threefold may be able to consistently traverse a topological transition of the compactification geometry. That is, we have aimed to explore in the heterotic case an analogue of the story which is well known for the Type II string (e.g.\ \cite{Strominger:1995cz}). This question has remained open in the ${\cal N}=1$ heterotic case due (in part) to the added complication of a gauge sector background, whose behavior across the topological transition has historically presented a stumbling block. A broader goal for this undertaking is to determine which theories on distinct compactification topologies might secretly be smoothly connected, to hence illuminate the true structure of the moduli space of heterotic compactifications.

Separately, we have also been motivated by the phenomenon of heterotic (0,2) target space duality. While still at the level of an intriguing observation, through the rich structure of gauged linear sigma models there is by now significant evidence of pairs, or even whole chains of heterotic compactifications, which have distinct topologies and distinct gauge sector backgrounds, but which nonetheless appear to give rise to the same physical 4-dimensional ${\cal N}=1$ theory. In our context, we have in particular considered this as suggestive of the existence of consistent physical transitions between distinct compactification backgrounds, so that $(0,2)$ target space duality would be merely a symptom of the possibility of this traversal process. 

In summary then, this work has attempted to unite the two themes above and has provided a \emph{geometric process} by which a compactification of the heterotic string can traverse a conifold transition. Moreover, we find that this procedure reproduces the known ``dual" geometries connected by $(0,2)$ target space duality in GLSMs (and indeed pairs connected by a new 5-brane duality as discussed in Section~\ref{sec:5brane_duality}). As mentioned in previous sections, our results are primarily a geometric prescription (albeit heavily informed by heterotic effective theories). It remains an open question exactly how ``smooth" these conifold transitions are in the full heterotic moduli space or whether the tools of ordinary field theory are sufficient to describe them. The development of the conjectural dual pairs outlined in previous sections has involved combining a number of disparate elements, necessarily leaving open a number of intriguing questions. The  answers to these questions will be the subject of future work, and in particular, this includes the following important tasks.

First, it is natural to ask whether there exists a simple field-theoretic description of the gauge-gravity pair creation process outlined in Section~\ref{secggpair}. This process has arisen as a key component in our conifold traversal proposal, and while the compactified context has the advantage of providing significant non-trivial consistency checks, it also has arisen in an intrinsically intricate setting involving multiple aspects of heterotic bundles/branes. Hence, an interesting area of further investigation would be to study simple ``toy models" of this process in isolation and try to provide more detailed field-theoretic descriptions.

A related question would be to more deeply understand the significance of the apparent jump in dimension of the essential curves (wrapped by 5-branes) in the nodal limit described in Section~\ref{sec:brane_thru_con}. We have seen that this jump in dimension is crucial for the brane recombination process which facilitates the traversal of the gauge sector across the conifold transition. However, a physical interpretation is difficult, because this effect occurs only in the singular limit, and while it is clear that the supporting loci of the skyscraper sheaves in the gauge sector jump, it is not clear whether an interpretation exists as a genuine extended object in string theory, or whether this is only an effect arising in a small instanton limit. One avenue which may provide hints for the appropriate description is a detailed comparison with the data of the corresponding hybrid phase of the gauged linear sigma model (see e.g.\cite{Hubsch:2002st} for an analysis similar in spirit). It would also be interesting to see if realizations of 5-brane limits in GLSMs similar to those in \cite{Blaszczyk:2011ib,Quigley:2011pv} could make contact with our proposed 5-brane duality.

A further important point to note is that this work appears to hint at some deeper duality of the heterotic string. It would be interesting to pursue this more directly from a heterotic NLSM viewpoint and to also ask what its consequences might be for other theories under string dualities. Some initial steps in the latter direction were taken in \cite{Anderson:2019axt} in the context of heterotic/F-theory duality. The analysis undertaken there however was complicated by the fact that all known examples of $(0,2)$ target space duality involved the monad construction of vector bundles. In \cite{Bershadsky:1997zv,Anderson:2019axt} it was shown that under a Fourier-Mukai transform such bundles lead to reducible/non-reduced spectral covers and hence lead to inclusion of T-brane solutions \cite{Cecotti:2010bp,Anderson:2013rka,Anderson:2017rpr} in the dual F-theory compactification, which are necessarily complicated in nature. However, in the present work we have outlined a geometric prescription that is independent of GLSMs/the monad construction. As a result, it would be interesting to revisit the question of F-theory duals in simpler contexts and to understand the nature of these conjectural dualities for such theories (including any links to the more general heterotic/F-theory dual pairs in \cite{Anderson:2021oth}).

Finally, we hope to use this work as a starting point to develop a clearer picture of how a heterotic compactification might traverse other topological transitions more generally (including flop transitions). In particular, natural questions arise as to whether or not portions of the full heterotic moduli space (defined by a particular manifold/bundle as background) can be ``extended" into another that is connected by a topological transition. It is believed that geometric transitions can connect all known Calabi-Yau threefolds \cite{reid}, and in particular the manifold resulting from a flop may be reached instead through a sequence of two conifold transitions. However, analogous to the Type II story, we expect that it is possible to pass directly through the flop without ever moving to the deformation branch. In this case, we can expect, and indeed have seen evidence that, the description of the traversal of the gauge sector is qualitatively different to the conifold case. As a particular example, one can expect that the phenomenon of a jumping dimension of the gauge sector objects will no longer be present, since it arises chiefly from the behavior of curves which become divisors in going from the deformation to the resolution branch. Hence the understanding of the traversal process in the flop case may be expected to involve qualitatively different phenomena and so to provide distinct insights. We hope to return to these questions in future work.

\section*{Acknowledgements}

The work of LA, CB and JG is supported by the NSF grant PHY-2014086. 

\appendix{}

\section{Gravitational small instanton transitions from sequence \newline recombination} \label{sec:sequencecomb_grav}

There are a variety of different ways to describe the recombination of cotangent and skyscraper sheaves, embodied by the Hecke transform (\ref{eq:rel_cot_seq}), which underlies the gravitational small instanton transition described in Section~\ref{sec:gravinst}. As one example of this, if one has appropriate resolutions of the sheaves involved, one can study this process explicitly in terms of manipulations of these sequences. Here we will illustrate this using the canonical example that we have used throughout the main text.

For our example, the cotangent sheaf of the nodal variety admits a resolution of the following form.%
\begin{eqnarray} \label{cotangres}
0 \to {\cal O}(0,-5) \to \Omega_{\mathbb{P}^4} \to  \smct^*(\Omega_{\nod{X}}) \to 0
\end{eqnarray}
Note that while the tangent bundle is not described as a short exact sequence in this limit, the cotangent sheaf is. This sequence is the dual of the, non-short exact, adjunction sequence associated to $\nod{X}$. In (\ref{cotangres}) we have used the fact that the normal bundle to $\defm{X}$ is ${\cal O}(5)$.

The sheaf ${\cal O}_{\mathbb{P}^1\mathrm{s}}(-2,0)$ admits the following free Koszul resolution, given that it is a complete intersection with normal bundle ${\cal O}(-1,4)\oplus {\cal O}(-1,1)$.
\begin{eqnarray} \label{excepres}
0 \to {\cal O}(0,-5) \to {\cal O}(-1,-1) \oplus {\cal O}(-1,-4) \to {\cal O}(-2,0) \to {\cal O}_{\mathbb{P}^1\mathrm{s}}(-2,0) \to 0
\end{eqnarray}

The sequences (\ref{cotangres}) and (\ref{excepres}) can be combined, simply by adding their entries together. In doing so, we keep the maps to be the same as in the original two sequences with no additional components added. In other words, the maps are `block diagonal' and descend precisely from the structures of (\ref{cotangres}) and (\ref{excepres}).

\begin{eqnarray} \label{combined1}
0&\to& {\cal O}(0,-5) \to {\cal O}(-1,-1) \oplus {\cal O}(-1,-4) \oplus  {\cal O}(0,-5)  \\ \nonumber   &\to&{\cal O}(-2,0) \oplus \Omega_{\mathbb{P}^4} \to {\cal O}_{\mathbb{P}^1\mathrm{s}}(-2,0) \oplus  \pi^*(\Omega_{\nod{X}})\to 0
\end{eqnarray}
This sequence is one description of the split locus in moduli space of the central object of the Hecke transform (\ref{eq:rel_cot_seq}), which we reproduce here:
\begin{eqnarray}
0 \to \smct^*(\Omega_{\nod{X}}) \to \Omega_{\res{X}} \to {\cal O}_{\mathbb{P}^1\mathrm{s}}(-2) \to 0 \,,
\end{eqnarray}
viewed as an extension.

By generalizing the maps in (\ref{combined1}) away from the block diagonal structure inherited from (\ref{cotangres}) and (\ref{excepres}), we can obtain an explicit description of how the two sheaves recombine into $\Omega_{\res{X}}$. The first thing to note is that, once the maps are generalized, the two copies of ${\cal O}(0,-5)$ in consecutive terms in the sequence can be canceled with out changing the object being resolved.
\begin{eqnarray} \label{combined2}
0 &\to& \cancel{{\cal O}(0,-5)} \to {\cal O}(-1,-1) \oplus {\cal O}(-1,-4) \oplus  \cancel{{\cal O}(0,-5)}  \\ \nonumber &\to&{\cal O}(-2,0) \oplus \Omega_{\mathbb{P}^4} \to {\cal F}\to 0
\end{eqnarray}
Note that here we have renamed ${\cal O}_{\mathbb{P}^1\mathrm{s}}(-2,0) \oplus  \smct^*(\Omega_{\nod{X}})$ to indicate that we are no longer describing the direct sum but rather some sheaf ${\cal F}$ whose nature we wish to elucidate. 

Next we note that, because for the $\mathbb{P}^1$ factor in the resolution ambient space $\Omega_{\mathbb{P}^1}= {\cal O}(-2,0)$, the sequence can be written as follows.
\begin{eqnarray}
0 \to {\cal O}(-1,-1) \oplus {\cal O}(-1,-4) \to \Omega_{\mathbb{P}^1} \oplus \Omega_{\mathbb{P}^4} \to {\cal F}\to 0
\end{eqnarray}
This sequence is simply the dual of the adjunction sequence associated to the description of $\res{X}$ as a complete intersection in $\mathbb{P}^1 \times \mathbb{P}^4$.
\begin{eqnarray}
 0 \to T\res{X} \to T\mathbb{P}^1 \times T\mathbb{P}^4 \to {\cal O}(1,1) \oplus {\cal O}(1,4) \to 0
\end{eqnarray}
We can thus identify ${\cal F}= \Omega_{\res{X}}$ as expected. This analysis therefore gives a different, and in some senses more explicit, description of the small instanton in the gravitational sector that connects $\Omega_{\nod{X}}$ and $\Omega_{\res{X}}$.

\section{Brane moduli} \label{appbranemod}

For a 5-brane to preserve supersymmetry it must wrap a holomorphic curve in the Calabi-Yau threefold. In terms of an embedding, the holomorphic spacetime coordinates $X^a(y^i , \overline{y}^{\overline{i}})$ of points on the brane must be a holomorphic functions of the world volume coordinates $(y^i,\overline{y}^{\overline{i}})$.
\begin{eqnarray}
\overline{\partial}_{\overline{i}} X^a =0
\end{eqnarray}

By using projectors we can rewrite this condition in terms of real coordinates
\begin{eqnarray} \label{start}
\overline{\smct}_I^{\; J} \partial_J \left( \Pi_B^{\;A} X^B\right)=0 \;.
\end{eqnarray}
In this expression we have,
\begin{eqnarray} 
\overline{\pi}_I^{\;J} = \frac{1}{2} \left(1_I^{\;J} + i {\cal J}_I^{(C) J}\right)\;,
\end{eqnarray}
where ${\cal J}^{(C)}$ is the complex structure tensor on the world volume of the brane, and
\begin{eqnarray}
\Pi_I^{\;J} = \frac{1}{2} \left(1_I^{\;J} - i {\cal J}_I^{(X) J}\right)\;,
\end{eqnarray}
where ${\cal J}^{(X)}$ is the complex structure tensor on the spacetime manifold.

Starting by assuming that we have a solution to (\ref{start}) we can vary the embedding $X^A$ and the two complex structure tensors. Substituting such a variation into (\ref{start}) and using that the unperturbed configuration is a solution, we then arrive at the following constraint on the fluctuations if they are to preserve supersymmetry.
\begin{eqnarray} \label{final2}
\frac{1}{2} i \delta {\cal J}_I^{(C)J} \partial_J \left( \Pi_B^{\;A} X^B \right)+ \overline{\pi}_I^{\;J} \partial_J \left( -\frac{1}{2}i \delta {\cal J}_B^{(X)A} X^B \right)+ \overline{\pi}_I^J \partial_J \left( \Pi_B^{\;A} \delta X^B\right) =0
\end{eqnarray}
To analyze things further it is useful to consider  (\ref{final2}), component by component, in terms of the original, unperturbed, complex coordinates. In doing so we find that the $(I,A)=(i,a)$ and $(I,A)=(i,\overline{a})$ components of the equation are trivially satisfied. The $(I,A)=(\overline{i},a)$ and $(I,A)=(\overline{i},\overline{a})$ components however are not. The first of these reduces to the following condition. 
\begin{eqnarray} \label{importantone}
\frac{i}{2} \delta {\cal J}_{\overline{i}}^{(C)j} \partial_jX^a = - \partial_{\,\overline{i}} \left( \delta X^a - \frac{i}{2} \delta {\cal J}_{\overline{b}}^{(X)a} X^{\overline{b}}\right)
\end{eqnarray}
From this expression we see that $\delta {\cal J}_{\overline{i}}^{(C)j}$ corresponds to a modulus iff $\delta {\cal J}_{\overline{i}}^{(C)j} \partial_jX^a $ is exact. In other words, the subset of $\delta {\cal J}_{\overline{i}}^{(C)j}$ that correspond to moduli are those in the kernel of the map $H^1(TC) \to H^1(TX|_C)$. There are two types of fluctuation $\delta X^a$ which can solve (\ref{importantone}). First, any $\delta X^a \in H^0(TX|_C)$ satisfies the equation in isolation and so is a modulus. Second there are fluctuations which are paired to fluctuations of the threefold complex structure which ensure that the bracket on the right hand side of (\ref{importantone}) remains closed. Note that such a compensating $\delta X^a$ exists for any possible fluctuation $\delta {\cal J}_{\overline{b}}^{(X)a}$.

How do these allowed fluctuations of $\delta {\cal J}_{\overline{i}}^{(C)j}$ and $\delta X^a$ combine into something familiar. Recall the following short exact sequence.
\begin{eqnarray}
0 \to TC \to TX|_C \to N_C \to 0
\end{eqnarray}
Taking the associated long exact sequence in cohomology we arrive at the following result if $H^0(TC)=0$.
\begin{eqnarray}
H^0(N_C) = H^0(TX|_C) \oplus \ker \left( H^1(TC) \to H^1(TX|_C)\right)
\end{eqnarray}
This is exactly the set of moduli we obtained from the differential analysis above.  In all of our examples $H^0(TC)=0$ and indeed this holds for any curve with genus $g>1$.

The above analysis leaves us with just the $(I,A)=(\overline{i},\overline{a})$ component of (\ref{final}) to examine. This component takes the following form.
\begin{eqnarray}
\left(\partial_{\,\overline{i}} \delta {\cal J}_b^{(X)\overline{a}} \right) X^b =0
\end{eqnarray}
This is a constraint on the complex structure variation of the threefold which is necessary if the cycle the brane is wrapping is to be able to deform in order to remain supersymmetric.

\section{Moduli matching for the 5-brane theories}
\label{curvemoduliproof}

In Section~\ref{sec:brane_thru_con} we specified a special pair of heterotic 5-brane theories, one on the resolution and one on the deformation side of conifold transitions between CY3s. We have argued that these 5-brane theories are continuously connected across the transition. We have also illustrated the construction explicitly in a simple example. (Additionally we summarise in Appendix~\ref{sec:exp_for_pn_split} the explicit construction for any of the large class of (effective) $\mbb{P}^n$-splits of toric complete intersections.) 

Further, we have argued that this construction produces two theories which are not only continuously connected but are in fact dual. In particular, this is strongly evidenced by the direct connection between this construction and target space duality, as discussed in detail in Section~\ref{sec:tsd}.

In the simple example treated in the main text we also showed, in Section~\ref{sec:5brane_duality}, that the two 5-brane theories have matching numbers of moduli, providing further evidence for the duality in this example. In this appendix, we consider more general cases and attempt to provide proofs of the matching.

We first briefly consider vector multiplet moduli, before turning to the matching of the chiral multiplet moduli. In this latter case, to make the computations tractable, we focus on conifold transitions described by $\mbb{P}^n$-splits between CICYs. We first provide a proof that the chiral multiplets from the `spectator' 5-branes match across the pair of 5-brane theories. By contrast, the contributions from the non-trivial part of the 5-brane pairing, of the 5-branes wrapped on the special curves $C$ and $\tilde{C}$ which are intimately linked to the conifold geometry, are more difficult, and indeed should differ precisely by the difference in geometric moduli across the conifold transition. Below we outline a proof in the simplest case, of conifold transitions in which $h^{1,1}$ changes by one and which are described by a $\mbb{P}^1$-split between CICYs.

\subsection*{Vector multiplet moduli}

As discussed in Section~\ref{sec:5brane_duality}, there are $g_C$ vector multiplet moduli coming from a 5-brane wrapped on a curve $C$ with genus $g_C$. Hence, the matching of the vector multiplet moduli across the pair of 5-brane theories depends on the isomorphisms of the spectator curves, $C_0 \cong \res{C}_0$, and of the non-trivially transitioning curves, $C \cong \res{C}$. The spectator curves were discussed in Section~\ref{sec:5brane_duality}, but we now give a general argument for the isomorphism $C \cong \res{C}$.

Recall that both curves are defined beginning from the same Weil non-Cartier divisor $\nod{D}$ on the nodal variety. When the nodal variety is deformed $\nod{X} \to \defm{X}$, the equation describing $\nod{D}$ becomes independent of the equation describing the geometry, and hence the curve $\defm{C}$ arises from the intersection of $\nod{D}$ with the zero locus of the equation describing the deformation. On the resolution side on the other hand, the curve $\res{C}$ is explicitly defined by intersecting the proper transform $\pt{\nod{D}}$ with the zero locus of an equation of the same form as describes the deformation $\nod{X} \to \defm{X}$. Notably, this intersection takes place far from the nodal points, so the small resolution taking $\nod{D}$ to $\pt{\nod{D}}$ is irrelevant. Hence, the descriptions of the two curves are identical, so that they are indeed isomorphic.

\subsection*{Chiral multiplet moduli: spectator part}

We now consider the chiral multiplets coming from the spectator brane, in the case of a $\mbb{P}^1$-split between CICYs. In analyzing this situation the following lemma, derived in \cite{Anderson:2013qca}, is key.
\begin{UnLemma} \label{lemma_again}
Let $\defm{X}$ and $\res{X}$ be two CICY threefolds related by a ``splitting transition" of the type described in (\ref{m5egsplit})\footnote{In fact the obvious generalization of this lemma holds for CICYs related by an arbitrary number of general $\mathbb{P}^n$-splits.}. Suppose that ${\cal L}={\cal O}({\bf a})$ is a line bundle corresponding to a divisor $D \subset X$ such that $D$ is the restriction of a divisor in the ambient space ${\cal A}$ (a ``favorable" line bundle on $\defm{X}$). If we define $\tilde{{\cal L}}={\cal O}(0,{\bf a})$ then $h^i(\res{X},\tilde{{\cal L}})=h^i(\defm{X},{\cal L})\; \forall \,i$, on the common ``determinantal locus" in moduli space.
\end{UnLemma}

\vspace{0.2cm}

To apply this lemma to the case at hand, let us assume that the spectator 5-brane stacks on $\defm{X}$ and $\res{X}$ are described as complete intersections. Then their normal bundles, thanks to the class $[\defm{C}_0]=[\res{C}_0]$ having no contributions involving $J_0$ (see (\ref{specclass}) or more generally \eqref{eq:gen_spec_class}), are sums of line bundles of the form ${\cal N}_{C_0}={\cal O}({\bf a}) \oplus {\cal O}({\bf b})$ and ${\cal N}_{\res{C}_0}={\cal O}(0,{\bf a}) \oplus {\cal O} (0,{\bf b})$ on $\defm{X}$ and $\res{X}$ respectively.

The cohomologies of these normal bundles evaluated on the respective 5-brane curves can be obtained by using their associated Koszul sequences. For the curve $C_0 \subset X$ this is
\begin{eqnarray}
0 \to \wedge^2 {\cal N}_{C_0}^{\vee} \otimes {\cal N}_{C_0} \to {\cal N}_{C_0}^{\vee} \otimes  {\cal N}_{C_0} \to {\cal N}_{C_0} \to {\cal N}_{C_0}|_{C_0} \to 0 \,,
\end{eqnarray} 
and analogously for $\res{C}_0 \subset \res{X}$. Decomposing these long exact sequences into two short exact sequences and taking the associated long exact sequences in cohomology one can compute the cohomologies $h^0(C_0,{\cal N}_{C_0})$ and $h^0(\res{C}_0,{\cal N}_{\res{C}_0})$ of interest (where $h^0(C_0,{\cal N}_{C_0}) \equiv h^0(C_0,{\cal N}_{C_0}|_{C_0})$ etc.). The above lemma shows that all of the cohomologies on the CY3 that will be involved in this computation will be the same on the deformation and resolution side of the transition, at the common determinantal locus in moduli space. As such the two cohomologies will agree in this limit. As long as the tuning to the determinantal locus is not too special, this limit will share its cohomology with the generic point in moduli space. Indeed, we find this to be the case in every example we have examined. 

An analogous argument shows that the spectrum of spectator bundles, constructed using line bundles of the form given in the above Lemma, will also match on the two sides of the duality. For example consider spectator bundles $V_0$ which are two term dual monads of the form,
\begin{eqnarray}
0 \to C\to B \to V_0 \to 0\;,
\end{eqnarray}
where $B$ and $C$ are sums of line bundles of the form given in the Lemma. We see using the lemma that the cohomologies of $B$ and $C$ will match in the singular limit, and so too, therefore, will the cohomology of $V_0$ and various associated bundles. As in the 5-brane case, as long as the tuning to the determinantal locus is not too special, the cohomology of $V_0$ will be the same at a generic point in moduli space as it is in that limit, leading to a matching on the two sides of the duality.

\subsection*{Chiral multiplet moduli: non-trivial part}

The matching of the total number of chiral multiplet moduli requires that the change in 5-brane deformation moduli across the transition must balance the change in geometric deformation moduli. Since the deformation moduli match for the spectator 5-branes, this compensating change must come from the 5-branes wrapped on the curves $C$ and $\res{C}$, 
\be
h^{1,1}(\res{X})+h^{2,1}(\res{X})+h^0(\res{C},\norm{\res{C}}{\res{X}}) \stackrel{?}{=} h^{1,1}(\defm{X})+h^{2,1}(\defm{X})+h^0(\defm{C},\norm{\defm{C}}{\defm{X}}) \,,
\ee
where $\norm{\res{C}}{\res{X}}$ and $\norm{C}{X}$ are the normal bundles of $\res{C} \subset \res{X}$ and $C \subset X$. Recalling from Section~\ref{sec:con_trans} that in a conifold transition the Hodge numbers change as
\be
h^{1,1}(\res{X}) = h^{1,1}(\defm{X}) + \Delta(h^{1,1}) \,, \quad h^{2,1}(\res{X}) = h^{2,1}(\defm{X}) - \#(\mbb{P}^1\mathrm{s})+\Delta(h^{1,1}) \,,
\ee
for some $\Delta(h^{1,1}) > 0$, we see that the relation that is required to hold amongst the 5-brane moduli is
\be \label{eq:mod_match_req_gen}
h^0(\res{C},\norm{\res{C}}{\res{X}}) - h^0(\defm{C},\norm{\defm{C}}{\defm{X}}) + 2\Delta(h^{1,1}) - \#(\mbb{P}^1\mathrm{s}) \stackrel{?}{=} 0 \,.
\ee

\subsubsection*{Proof outline in a tractable case}

While we do not have a proof of the above relation in the general case, we here outline a proof in a particularly tractable case, namely of conifold transitions for which $h^{1,1}$ changes only by one, and which can be described as $\mbb{P}^1$-splits between CICYs.

Before assuming that $\Delta(h^{1,1})=1$, consider a general $\mbb{P}^1$-split. In this case the resolution geometry $\res{X}$ has a configuration matrix of the form
\be
\res{X} = \left[ \begin{array}{c | c c c c c}
\mbb{P}^1 & 1 & 1 & 0 & \ldots & 0 \\
\mc{A} & {\bf v}_0 & {\bf v}_1 & {\bf r}_0 & \ldots & {\bf r}_K
\end{array} \right] \,,
\ee
and the deformation geometry $\defm{X}$ has configuration matrix
\be
\defm{X} = \left[ \begin{array}{c | c c c c}
\mc{A} & ({\bf v}_0 + {\bf v}_1) & {\bf r}_0 & \ldots & {\bf r}_K
\end{array} \right] \,.
\ee
Here, $\mc{A}$ is a product of $N$ projective spaces, and ${\bf v}_{\hspace{.4pt}0}$, ${\bf v}_1$, and ${\bf r}_0$ $\ldots$ ${\bf r}_K$ are vectors of length $N$, where $K = \mathrm{dim}({\cal A})-5$.

Recalling the discussion in Section~\ref{sec:brane_thru_con}, the curves $\res{C} \subset \res{X}$ and $\defm{C} \subset \defm{X}$ on which we wish to wrap 5-branes can be described by configuration matrices
\be
\begin{aligned}
\res{X} &\supset \res{C} =
\left[ \begin{array}{c | c c c c c ;{2pt/2.5pt} c c}
\mbb{P}^1 & 1 & 1 & 0 & \ldots & 0 & 1 & 0\\
\mc{A} & {\bf v}_0 & {\bf v}_1 & {\bf r}_0 & \ldots & {\bf r}_K & \vec{0} & ({\bf v}_0+{\bf v}_1)
\end{array} \right] \,,
\\
\defm{X} &\supset \defm{C} =
\left[ \begin{array}{c | c c c c ;{2pt/2.5pt} c c}
\mc{A} &  ({\bf v}_0 + {\bf v}_1) & {\bf r}_0 & \ldots & {\bf r}_K & {\bf v}_0 & {\bf v}_1
\end{array} \right] \,,
\end{aligned}
\ee
where the dashed lines separate the equations defining the geometry from the additional equations describing the curve. Now note that the three equations of bidegree $(1,{\bf v}_0)$, $(1,{\bf v}_1)$, $(1,\vec{0})$ in the definition of $\res{C}$ are straightforwardly equivalent to ones of bidegree $(0,{\bf v}_0)$, $(0,{\bf v}_1)$, $(1,\vec{0})$. With this rewriting, the descriptions of the two curves become essentially identical (which also makes clear their isomorphism, $\defm{C} \cong \res{C}$),
\be
\begin{aligned}
\res{C} =
\left[ \begin{array}{c | c c c c c c c}
\mbb{P}^1 & 0 & 0 & 0 & \ldots & 0 & 1 & 0\\
\mc{A} & {\bf v}_0 & {\bf v}_1 & {\bf r}_0 & \ldots & {\bf r}_K & \vec{0} & ({\bf v}_0+{\bf v}_1)
\end{array} \right] \cong 
\left[ \begin{array}{c | c c c c c c}
\mc{A} &  ({\bf v}_0 + {\bf v}_1) & {\bf r}_0 & \ldots & {\bf r}_K & {\bf v}_0 & {\bf v}_1
\end{array} \right]
\end{aligned}
\ni \defm{C} \,,
\ee
so that, defining $\mc{O}_C(\vec{w}) := \mc{O}_{\mc{A}}(\vec{w})|_C$, the normal bundles of $\res{C}$ and $\defm{C}$ are clearly given by
\be
\norm{\res{C}}{\res{X}}|_{\res{C}} \cong \mc{O}_C \oplus \mc{O}_C({\bf v}_0+{\bf v}_1) \,, \quad \norm{\defm{C}}{\defm{X}}|_{\defm{C}} \cong \mc{O}_C({\bf v}_0) \oplus \mc{O}_C({\bf v}_1) \,.
\ee

We wish to compute the zeroth cohomology of each of the above bundles. In the case of $h^0(C,\norm{\res{C}}{\res{X}})$ this is straightforward. For example, noting from the configuration matrix describing $\defm{C}$ that the canonical bundle of $C$ is $K_C = \mc{O}_C({\bf v}_0+{\bf v}_1)$, and hence that the two line bundles in $\norm{\res{C}}{\res{X}}|_{\res{C}}$ are Serre dual, and also noting that $h^0(C,\mc{O}_C)=1$, it follows after a small amount of algebra that
\be
h^0(C,\norm{\res{C}}{\res{X}}) = - \ind(C,\mc{O}_C) + 2 = 1 + g_C \,,
\ee
where $g_C$ is the genus of the curve.

In contrast, it is more difficult to give a general expression for $h^0(C,\norm{\defm{C}}{\defm{X}})$, and indeed we expect this to differ from the cohomology $h^0(C,\norm{\res{C}}{\res{X}})$ above, which depends only on the intrinsic topology of the curve, by a piece that depends on the change in Hodge numbers across the conifold transition. To proceed, we make two assumptions.
\begin{enumerate}
\item Assume that $\nod{D}$ is not isomorphic to $T^4$.
\item Restrict to cases for which $\Delta(h^{1,1})=1$, and for which $\mc{O}_{\nod{D}}({\bf v}_0)$, $\mc{O}_{\nod{D}}({\bf v}_1)$, and $\mc{O}_{\nod{D}}({\bf v}_0+{\bf v}_1)$ have vanishing higher cohomologies.
\end{enumerate}
We explain these assumptions below. Here $\nod{D}$ is the Weil non-Cartier divisor on the nodal variety which we used to construct the two 5-brane curves, as discussed in Section~\ref{sec:brane_thru_con}, and which in the present case is given by
\be
\nod{D} = \left[ \begin{array}{c | c c c c c} \mc{A} & {\bf v}_0 & {\bf v}_1 & {\bf r}_0 & \ldots & {\bf r}_K \end{array} \right] \,.
\ee
Noting that from the degrees of the defining equations this surface is CY, we see that $\nod{D}$ is hence isomorphic to either a $T^4$ or a K3 surface. We have also defined in the second assumption $\mc{O}_{\nod{D}}(\vec{w}) := \mc{O}_{\mc{A}}(\vec{w})|_{\nod{D}}$.

As discussed in Section~\ref{sec:5brane_duality} above, the first assumption explicitly removes that small subset of cases with a non-zero $h^1(\nod{D},\mc{O}_{\nod{D}})$, which appears to present a genuine obstruction to moduli matching.

The second assumption explicitly restricts to the simplest case of conifold transitions, where the change in $h^{1,1}$ is minimal. It also restricts however to the case where a number of cohomologies vanish. In practice, one finds in fact that these vanishings always accompany the restriction $\Delta(h^{1,1})=1$, so that these are in fact no further restriction at all. Nonetheless, we have not proven this link, and so strictly this should be considered as an additional assumption.

\medskip

With the above two assumptions we can now straightforwardly prove the moduli matching result. First note that the curve $C$ is the intersection of the divisor $\nod{D}$ inside $\mc{A}$ with the hypersurface $[ \begin{array}{c | c} \mc{A} & ({\bf v}_0 + {\bf v}_1) \end{array} ]$, so that the Koszul resolution of $C$ inside $\nod{D}$ is a short exact sequence,
\be
0 \to \mc{O}_{\nod{D}}(-{\bf v}_0-{\bf v}_1) \to \mc{O}_{\nod{D}} \to \mc{O}_C \to 0 \,.
\ee
If we now tensor this with $\mc{O}_{\nod{D}}({\bf v}_0) \oplus \mc{O}_{\nod{D}}({\bf v}_1)$ we have
\be
0 \to \mc{O}_{\nod{D}}(-{\bf v}_0) \oplus \mc{O}_{\nod{D}}(-{\bf v}_1) \to \mc{O}_{\nod{D}}({\bf v}_0) \oplus \mc{O}_{\nod{D}}({\bf v}_1) \to \mc{O}_C({\bf v}_0) \oplus \mc{O}_C({\bf v}_1)\to 0 \,.
\ee
The third object is one whose zeroth cohomology we wish to compute. Noting that the canonical bundle of $\nod{D}$ is trivial, we see that the first two bundles are Serre dual. Hence, since by assumption the higher cohomologies of the second bundle vanish, we have in the long exact sequence in cohomology the following pattern of zeroes,
\be
\begin{array} { c c c c}
	& \mc{O}_{\nod{D}}(-{\bf v}_0) \oplus \mc{O}_{\nod{D}}(-{\bf v}_1)	& \mc{O}_{\nod{D}}({\bf v}_0) \oplus \mc{O}_{\nod{D}}({\bf v}_1)	& \mc{O}_C({\bf v}_0) \oplus \mc{O}_C({\bf v}_1)	\\
h^0	& 0	& ?	& ?	\\
h^1	& 0	& 0	& ?	\\
h^2	& ?	& 0	& 0	
\end{array}
\ee
and hence we can compute the required zeroth cohomology as an index on $\nod{D}$,
\be
h^0(C,\norm{\defm{C}}{\defm{X}}) = h^0\big( C,\mc{O}_C({\bf v}_0) \oplus \mc{O}_C({\bf v}_1) \big) = \ind\big(\nod{D}, \mc{O}_{\nod{D}}({\bf v}_0) \oplus \mc{O}_{\nod{D}}({\bf v}_1) \big) \,.
\ee
Next, we note that we can also rewrite $h^0(C,\norm{\res{C}}{\res{X}})$ in terms of an index on $\nod{D}$. Tensoring the Koszul resolution of $C$ inside $\nod{D}$ with $\mc{O}_{\nod{D}}({\bf v}_0 + {\bf v}_1)$, we have
\be
0 \to \mc{O}_{\nod{D}} \to \mc{O}_{\nod{D}}({\bf v}_0 + {\bf v}_1) \to \mc{O}_C({\bf v}_0 + {\bf v}_1) \to 0  \,.
\ee
By our first assumption, $\nod{D}$ is isomorphic to a K3 surface, so $h^0(\nod{D},\mc{O}_{\nod{D}}) = 1$ and ${h^1(\nod{D},\mc{O}_{\nod{D}}) = 0}$, and by our second assumption, the higher cohomologies of the second bundle vanish. Hence we have
\be
\begin{aligned}
h^0(C,\norm{\res{C}}{\res{X}}) 
&= h^0\big( C,\mc{O}_C \oplus \mc{O}_C({\bf v}_0 + {\bf v}_1) \big) 
= h^0\big( C,\mc{O}_C \big) + h^0 \big(\mc{O}_C({\bf v}_0 +{\bf v}_1) \big) \\
&= 1 + \big( \ind\big(\nod{D}, \mc{O}_{\nod{D}}({\bf v}_0 + {\bf v}_1) \big) - 1 \big) = \ind\big(\nod{D}, \mc{O}_{\nod{D}}({\bf v}_0 + {\bf v}_1) \big) \,.
\end{aligned}
\ee
Hence, we have expressed both cohomologies $h^0(C,\norm{\defm{C}}{\defm{X}})$ and $h^0(C,\norm{\res{C}}{\res{X}})$ as indices of bundles on $\nod{D}$. Why is this useful? Consider the Koszul resolution of $\nod{D}\cdot \nod{D}$  inside $\nod{D}$.
\be
0 \to \mc{O}_{\nod{D}}(-{\bf v}_0 - {\bf v}_1) \to \mc{O}_{\nod{D}}(-{\bf v}_0) \oplus \mc{O}_{\nod{D}}(-{\bf v}_1) \to \mc{O}_{\nod{D}} \to \mc{O}_{\nod{D} \cdot \nod{D}} \to 0 
\ee
If we tensor this with $\mc{O}_{\nod{D}}({\bf v}_0+{\bf v}_0)$, we have
\be
0 \to \mc{O}_{\nod{D}} \to \mc{O}_{\nod{D}}({\bf v}_1) \oplus \mc{O}_{\nod{D}}({\bf v}_0) \to \mc{O}_{\nod{D}}({\bf v}_0 + {\bf v}_1) \to \mc{O}_{\nod{D} \cdot \nod{D}} \to 0 \,.
\ee
Notably, from the definition of the surface $\nod{D}$, we see that its self-intersection $\nod{D} \cdot \nod{D}$ inside $Y$, where $Y$ is defined by
\be
Y = \left[ \begin{array}{c | c c c} \mc{A} & {\bf r}_0 & \ldots & {\bf r}_K \end{array} \right] \,,
\ee
equals the number of exceptional $\mbb{P}^1$s in the conifold transition,
\be
\nod{D} \cdot \nod{D} = \int_Y c_1 \big(\mc{O}_Y({\bf v}_0)\big)^2 \, c_1 \big(\mc{O}_Y({\bf v}_1)\big)^2 = \#(\mbb{P}^1\mathrm{s}) \,.
\ee
(See for example the discussion in Appendix~\ref{sec:exp_for_pn_split}.) Hence, if we take the index on the above four-term exact sequence, we get
\be
0 = \ind\big(\nod{D},\mc{O}_{\nod{D}}\big) - \ind\big(\nod{D},\mc{O}_{\nod{D}}({\bf v}_0) \oplus \mc{O}_{\nod{D}}({\bf v}_1)\big) + \ind\big(\nod{D},\mc{O}_{\nod{D}}({\bf v}_0 + {\bf v}_1)\big) - \#(\mbb{P}^1\mathrm{s}) \,,
\ee
so that if we note that, since $\nod{D}$ is a K3 surface, $\ind\big(\nod{D},\mc{O}_{\nod{D}}\big) = 2$, and if we recall that two indices in this expression are the required zeroth cohomologies, we find finally
\be
0 = 2 - h^0(C,\norm{\defm{C}}{\defm{X}}) + h^0(C,\norm{\res{C}}{\res{X}})  - \#(\mbb{P}^1\mathrm{s}) \,,
\ee
which is the relation we set out to prove, namely \eqref{eq:mod_match_req_gen} in the case that $\Delta(h^{1,1})=1$.

\section{Hecke moduli}
\label{app:hecke}

Consider a Hecke transform of the following form,
\begin{eqnarray} \label{genhecke}
0 \to V \longrightarrow V_0 \stackrel{f}{\longrightarrow} F \to 0 \;,
\end{eqnarray}
where $F$ is a sheaf supported on a curve and $V_0$ is a bundle. If $V$ is stable then the Zariski tangent space to the moduli space is given by $\textnormal{Ext}^1(V,V)$. We will compute this quantity, first in general and then in a special case relevant to this work.

\vspace{0.1cm}

We begin by reviewing some properties of $\textnormal{Ext}$ groups which will play a central role. Applying $\textnormal{Ext}^*(\;\_ \;,V)$ to a short exact sequences of sheaves $0 \to A\to B\to C\to0$ one obtains
\begin{eqnarray}
0 \to \textnormal{Ext}^0(C,V) \to \textnormal{Ext}^0(B,V) \to \textnormal{Ext}^0(A,V) \to \textnormal{Ext}^1(C,V) \to \ldots
\end{eqnarray}
and similarly for $\textnormal{Ext}^*(V,\;\_\;)$.
\begin{eqnarray}
0 \to \textnormal{Ext}^0(V,A) \to \textnormal{Ext}^0(V,B) \to \textnormal{Ext}^0(V,C) \to \textnormal{Ext}^1(V,A) \to \ldots
\end{eqnarray}
In addition to these properties we will use Serre duality, applied to our case where the dualizing sheaf is the trivial bundle on $X$,
\begin{eqnarray}
\textnormal{Ext}^i(E, {\cal O}_X) = H^{n-i}(E)^* \;,
\end{eqnarray}
where $n$ is the dimension of X. Finally, we will use the fact that $\textnormal{Ext}^i(A\otimes B, C) =\textnormal{Ext}^i( B, A^{\vee} \otimes C)$ if $A$ is locally free and $\textnormal{Ext}^i({\cal O}_X, A) = H^i(X, A)$.

\vspace{0.2cm}

Combining all of the properties of the previous paragraph, one can compute $\textnormal{Ext}^1(V,V)$ for (\ref{genhecke}) in terms of its component objects. One finds the following.
\begin{eqnarray} \label{genans}
\textnormal{Ext}^1(V,V) = \begin{array}{c} \textnormal{coker} \left( \textnormal{Ext}^0(F,F) \to \left(\textnormal{coker}\left(\textnormal{Ext}^0(V_0,V_0) \to \textnormal{Ext}^0(V_0,F)\right)\right)\right) \\ \oplus \\ \textnormal{ker}\left(\textnormal{Ext}^1(V_0,V_0) \to \textnormal{Ext}^1(V_0,F)\right) \\ \oplus \\ \textnormal{ker} \left( \textnormal{Ext}^1(F,F) \to \textnormal{coker}\left( \textnormal{Ext}^1(V_0,V_0) \to \textnormal{Ext}^1 (V_0,F) \right) \right) \\\oplus \\ \textnormal{ker} \left(\textnormal{ker} \left(\textnormal{Ext}^2(F,V_0) \to \textnormal{Ext}^2(F,F) \right) \to \textnormal{Ext}^2(V_0,V_0) \right)\end{array} 
\end{eqnarray}
In deriving this result we used the fact that $F$ is only supported on a curve. Obviously, the result in (\ref{genans}) is still somewhat involved, but some structure can be observed. In particular the first line details contributions associated to some elements of $\textnormal{Ext}^0(V_0,F)$, which is the space of possible maps $f$ in (\ref{genhecke}). The second line in (\ref{genans}) is associated to some elements of $\textnormal{Ext}^1(V_0,V_0)$, the bundle moduli of $V_0$. The third line is associated to elements of $\textnormal{Ext}^1(F,F)$, moduli of the sheaf $F$. Finally, the fourth line of (\ref{genans}) encodes moduli which do not fall in the previous three classes, and so we would expect them to correspond to those deformations which do not preserve the form (\ref{genhecke}). These are precisely the moduli which can be used to smooth $V$ from a sheaf into a bundle. Despite this coarse separation of moduli types, the structure of (\ref{genans}) is unpleasant to deal with, and so we will impose some addition properties of (\ref{genhecke}) which are relevant to our case. 

\vspace{0.3cm}

In the cases of interest in this work, $V_0$ takes the special form $V_0= \check{V}_0 \oplus {\cal O}$, where $\check{V}_0$ is a stable holomorphic bundle. In addition, we also have that $F={\cal O}_c$ for some curve $c$. Finally, $\check{V}_0$ is stable on restriction to $c$ and as such $\textnormal{Ext}^0(\check{V}_0, {\cal O}_c)=H^0(\check{V}_0^{\vee} \otimes {\cal O}_c)=0$. Note in such an instance we have from (\ref{genhecke}) that $V=\check{V_0} \oplus {\cal I}_c$ from the special structure imposed on the map $f$ by the above conditions and the defining sequence of an ideal sheaf.
\begin{eqnarray} \label{idealsheafdef}
0 \to {\cal I}_c \to {\cal O} \to {\cal O}_c \to 0
\end{eqnarray}

Using the properties of this special case, one can simplify (\ref{genans}) greatly to give the following\footnote{Note that although we have split up this expression into separate lines for convenience, in the following discussion these lines are not in one-one correspondence with those in (\ref{genans}).}.
\begin{eqnarray} \label{prefinal}
\textnormal{Ext}^1(V,V) = \begin{array}{c} H^1(\check{V}_0^{\vee} \otimes \check{V}_0) \\ \oplus \\ \textnormal{ker} \left( H^1(\check{V}_0^{\vee}) \to H^1(\check{V}_0^{\vee}|_c) \right)  \\ \oplus \\ H^1({\check V}_0) \oplus \textnormal{ker} \left(\textnormal{Ext}^2({\cal O}_c,\check{V}_0) \to \textnormal{Ext}^2({\cal O},\check{V}_0) \right) \\ \oplus \\ H^0({\cal N}|_c) \end{array} 
\end{eqnarray}
Here, ${\cal N}$ is the normal bundle associated to the curve $c$ and we have assumed that this curve is a complete intersection and thus admits a Koszul resolution of the following form.
\begin{eqnarray}
0 \to \wedge^2 {\cal N}^{\vee} \to {\cal N}^{\vee} \to {\cal O} \to {\cal O}_c \to 0
\end{eqnarray}
The middle two lines in (\ref{prefinal}) can be simplified in appearance greatly by using the properties of $\textnormal{Ext}$s and of our special case described above, as well as (\ref{idealsheafdef}). These allow us to show that $\textnormal{Ext}^1(\check{V}_0,{\cal I}_c) = \textnormal{ker} \left( H^1(\check{V}_0^{\vee}) \to H^1(\check{V}_0^{\vee}|_c) \right)$ and $\textnormal{Ext}^1({\cal I}_c,\check{V}_0) = H^1({\check V}_0) \oplus \textnormal{ker} \left(\textnormal{Ext}^2({\cal O}_c,\check{V}_0) \to \textnormal{Ext}^2({\cal O},\check{V}_0) \right)$ giving us our final result.
\begin{eqnarray} \label{final}
\textnormal{Ext}^1(V,V) = H^1(\check{V}_0^{\vee} \otimes \check{V}_0)  \oplus  \textnormal{Ext}^1(\check{V}_0,{\cal I}_c)   \oplus \textnormal{Ext}^1({\cal I}_c,\check{V}_0)  \oplus  H^0({\cal N}|_c)  
\end{eqnarray}

\vspace{0.3cm}

In terms of applying the above results to the examples in the main text it is worth noting that the first three terms in (\ref{final}) will be the same on both sides of the conifold transition. This is due to the nature of the spectator bundle $\check{V}_0$ on the two sides together with the fact that the curve $c$ is in the same class, viewed as a variety in the ambient space, on both the deformation and resolution geometries. That the middle two terms match in this manner is easiest to see from the form (\ref{prefinal}) and the simple behavior of cohomology under Leray.

\vspace{0.3cm}

As a check we can examine this result in the case of our canonical example based upon a $\mathbb{P}^1$-split of the quintic. In this example $h^1(\check{V}_0^{\vee} \otimes \check{V}_0) =124$, $\textnormal{ext}^1(\check{V}_0,{\cal I}_c) =32$ and $\textnormal{ext}^1({\cal I}_c,\check{V}_0)=132$ with these numbers indeed matching on both sides of the transition. The quantity $H^0({\cal N}|c)=38$ on the deformation side and $52$ on the resolution side of the conifold. These numbers lead to a totals for the Hecke moduli on the two sides of the transition of $326$ and $340$ respectively, with both numbers being one larger than the bundle moduli of the smooth gauge bundles that are obtained as smooth deformations of the Hecke sheaves $V$ as expected.

\section{Expressions for a general $\mbb{P}^n$-split}
\label{sec:exp_for_pn_split}

In the main text, as our prototypical example of a conifold transition between CY3s, we have considered a deformation geometry $\defm{X}$ and resolution geometry $\res{X}$ which are related by a `$\mbb{P}^1$-split' of a CICY, as introduced in Section~\ref{sec:con_trans}.

More generally, any `$\mbb{P}^n$-split' of a CICY, which involves the addition of an ambient $\mbb{P}^n$ space, will describe a conifold transition (as long as the splitting is `effective', meaning that the shrinking of this ambient $\mbb{P}^n$ corresponds to a wall of the K\"ahler cone). For a detailed discussion of the geometry of the associated conifold transitions, we refer the reader to \cite{Brodie:2021toe}, and we also refer the reader to the original works on splittings \cite{Candelas:1987kf,Green:1988wa,Green:1988bp}.

Even more generally, one can consider $\mbb{P}^n$-splits of toric complete intersections, which (if again the splitting is `effective') also correspond to conifold transitions. These $\mbb{P}^n$-splits of toric complete intersections are the broadest natural generalisations of the simple $\mbb{P}^1$-split setting in which we have constructed pairs of 5-brane theories across conifold transitions as described in the main text.

In particular, by following precisely the same logic as for the simple CICY $\mbb{P}^1$-split example in Section~\ref{sec:brane_thru_con}, in this general setting too one can straightforwardly construct the curves $\defm{C}$ and $\res{C}$ which will always be such that anomaly cancellation condition is ensured on both sides of the transition (up to the addition of `spectator' branes, as discussed in Section~\ref{sec:brane_thru_con}).
Additionally, the description of gauge-gravity pair creation, and the brane recombination allowing the 5-brane theory to traverse the transition, is entirely analogous in this very large class of examples to the discussion of that simple CICY $\mbb{P}^1$-split example. This hence provides a large class of examples in which one perform the same procedure as in the main text to describe the traversal of a 5-brane theory through a conifold transition between CY3s.

In this appendix, we simply collect the relevant formulae and results for this general case of $\mbb{P}^n$-splits of toric complete intersections.

\subsubsection*{The deformation and resolution geometries}

We consider the situation where the deformation geometry $\defm{X}$ is a complete intersection inside a smooth, compact toric variety $\mc{A}$. If one of the defining equations is tuned until it can be expressed as the determinant of some $(n+1) \times (n+1)$ matrix $M$, the resulting variety $\nod{X}$ will have a set of nodal points where the rank of the matrix drops to $n-1$, and a small resolution can be performed on these nodal points by fibering an additional $\mbb{P}^n[x]$ over the ambient space $\mc{A}$, and replacing the determinantal equation with the set of equations $M{\bf x} = {\bf 0}$, to give a resolved geometry $\res{X}$.

Let us make this explicit. Take any choice $Q$ of the defining polynomials of $\defm{X}$, which is some section of some line bundle $\mc{L}$. Now, for any $n\geq1$, choose any set of $2(n+1)$ effective line bundles $\mc{U}_{\hspace{.4pt}l}$ and $\mc{V}_l$, where $l = 0, 1, 2 , \ldots , n$, such that $\mc{L} = \mathrm{det}( \mc{U} )\otimes\mathrm{det}( \mc{V} )$, where we have defined
\be
\mc{U} = \bigoplus_{l=0}^n \mc{U}_{\hspace{.4pt}l} \,, \quad \mc{V} = \bigoplus_{l=0}^n \mc{V}_l \,.
\ee
Then, this defining polynomial $Q$ can be tuned to equal the determinant of an $(n+1) \times (n+1)$ matrix of sections $M$,
\be
Q \to \mathrm{det}(M) ~~ \mathrm{where} ~~  
M_{ij} \in \Gamma \big( \mc{V}_i \otimes \mc{U}_j \big) \,.
\label{eq:mat_m}
\ee
This produces a variety $\nod{X}$ which is singular at the set of points where $\mathrm{rank}\,(M) \leq n-1$. Then, a small resolution of this nodal variety can be described by a $\mbb{P}^n$-split, which consists of replacing this determinantal equation by the set of equations $M {\bf x} = {\bf 0}$, where ${\bf x} = (x_0 \,,\, \ldots \,,,\, x_n)^\mathrm{T}$, which are accommodated by introducing into the ambient space an additional $\mbb{P}^n$, whose coordinates $x_i$ have scalings under the weight system of $\mc{A}$ chosen to balance those of the entries of $M$. Explicitly, defining the quantities ${\bf u}_{\hspace{.2pt}i}$ and ${\bf v}_i$ by
\be
\mc{U}_i = \mc{O}_{\mc{A}}({\bf u}_{\hspace{.2pt}i}) \,, \quad \mc{V}_i = \mc{O}_{\mc{A}}({\bf v}_i) \,, 
\ee
as well as their sums ${\bf U}$ and ${\bf V}$,
\be
{\bf U} = \sum_{l=0}^n {\bf u}_{\hspace{.2pt}l} \,, \quad {\bf V} = \sum_{l=0}^n {\bf v}_l \,,
\ee
the deformation and resolution geometries $\defm{X}$ and $\res{X}^{\,\mc{U}}$ can be described as
\begin{equation}
\defm{X}: ~
\begin{array}{c|cccc}
y	      		&  		  		 &			&	       & \\ \hline \\[-1.1em]
\square  		&{\bf U} + {\bf V}   & {\bf r}_0  	& \ldots & {\bf r}_{K}
\end{array}
\end{equation}
\begin{equation}
\res{X}^{\,\mc{U}}: ~
\begin{array}{cccc|cccccc}
x_0^{\,\mc{U}}   	& \ldots 	& x_{n}^{\,\mc{U}}   & y	      &  		  &		&		    	    &			    	&	       & \\ \hline
1  			& \ldots 	& 1 			    & 0 	      & 1 	   	  & \ldots & 1 		 	    & 0 			& \ldots & 0 \\
-{\bf u}_{\hspace{.1pt}0}  	& \ldots 	& -{\bf u}_n 	    & \square  & {\bf v}_0 & \ldots 	& {\bf v}_n  	    & {\bf r}_0  		& \ldots & {\bf r}_{K}
\end{array}
\end{equation}
where we have schematically written $y$ for all the coordinates on $\mc{A}$, and $\square$ for the weight system of $\mc{A}$, and we have also written ${\bf r}_0$ through ${\bf r}_K$ for the weights of the remaining defining equations of $\defm{X}$. One can check that the weight system of the ambient space of $\res{X}^{\,\mc{U}}$ is indeed such that the defining equations $M {\bf x}^{\,\mc{U}} = 0$ are consistent. Additionally however, it is clear that we could have made a different consistent choice, namely
\begin{equation}
\res{X}^\mc{V}: ~
\begin{array}{cccc|cccccc}
x_0^\mc{V}   	& \ldots 	& x_{n}^\mc{V}   & y	      &  		  &		&		    	    &			    	&	       & \\ \hline
1  			& \ldots 	& 1 			    & 0 	      & 1 	   	  & \ldots & 1 		 	    & 0 			& \ldots & 0 \\
-{\bf v}_0  	& \ldots 	& -{\bf v}_n 	    & \square  & {\bf u}_{\hspace{.1pt}0} & \ldots 	& {\bf u}_n  	    & {\bf r}_0  		& \ldots & {\bf r}_{K}
\end{array}
\end{equation}
which corresponds to introducing equations $M^\mathrm{T} {\bf x}^{\mc{V}} = {\bf 0}$ instead of $M {\bf x}^{\,\mc{U}} = {\bf 0}$. These two possibilities reflect that there are two (generically) inequivalent ways to perform the small resolution. The two manifolds $\res{X}^{\,\mc{U}}$ and $\res{X}^\mc{V}$ are (generically) not isomorphic, and are in fact related by a flop, as discussed in detail in \cite{Brodie:2021toe}. Hence for a given deformation geometry, there are two possible resolution geometries associated with the same nodal tuning, and both paths are conifold transitions. Below we will only consider $\res{X}^{\,\mc{U}}$ since the discussion is entirely analogous for $\res{X}^\mc{V}$, involving just the replacement $\mc{U} \leftrightarrow \mc{V}$. For convenience, we also define
\be
\mc{R} = \bigoplus_{l=0}^K \mc{O}_\mc{A}({\bf r}_l) \,.
\ee

\medskip

The second Chern classes of the deformation and resolution geometries can be shown to be\footnote{Here and below we abuse notation slightly by continuing to write $\mc{U}$ etc.\ for the pullbacks of these line bundles from $\mc{A}$ to the ambient space of $\res{X}^{\,\mc{U}}$.}
\be
\begin{gathered}
c_2\big(\defm{X}\big) = 
\tfrac{1}{2}c_1^2(\mc{V}) + \tfrac{1}{2}c_1^2(\mc{U}) + c_1(\mc{U})\,c_1(\mc{V}) + \mathrm{ch}_2(\mc{R}) - \mathrm{ch}_2(\mc{A}) \,, \\
c_2\big(\res{X}^{\,\mc{U}}\big) = 
\big( 0 \big) +
\Big( J_0 \, \big(c_1(\mc{U}) + c_1(\mc{V}) \big)\Big) +
\big( \mathrm{ch}_2(\mc{V}) - \mathrm{ch}_2(\mc{U}) + \mathrm{ch}_2(\mc{R}) - \mathrm{ch}_2(\mc{A}) \big) \,,
\end{gathered}
\ee
where in the second line we have grouped terms according to whether they have two, one, or no powers of the hyperplane class $J_0$ of the ambient $\mbb{P}^n$. The curve class of the exceptional $\mbb{P}^1$s inside $\res{X}^{\,\mc{U}}$ is
\be
[\mbb{P}^1\mathrm{s}] = -J_0 \, \big(c_1(\mc{U}) + c_1(\mc{V}) \big) + \big( c_2(\mc{V}) - c_2(\mc{U}) + c_1^2(\mc{U}) + c_1(\mc{U}) \, c_1(\mc{V}) \big) \,,
\ee
which we note equals the difference between the two second Chern characters (where here and below we make a slight abuse of notation, made precise in \cref{fn:c2_rel}),
\be
\mathrm{ch}_2\big(\res{X}^{\,\mc{U}}\big) - \mathrm{ch}_2\big(\defm{X}\big) = [\mbb{P}^1\mathrm{s}] \,.
\ee
As an aside we also note that the number $\# (\mbb{P}^1\mathrm{s})$ of exceptional $\mbb{P}^1$s, or equivalently the number of nodal points on $\nod{X}$, is given in general by
\be
\begin{aligned}
\# (\mbb{P}^1\mathrm{s}) =& \; \int_{\mc{A}} \bigg[
\big(c_2(\mc{U}) - c_2(\mc{V})\big)^2 
- \big(c_1(\mc{U}) - c_1(\mc{V})\big) \big(c_3(\mc{U}) - c_3(\mc{V})\big) \\
&-  c_1(\mc{U}) \,c_1 \big( \mc{V}) (c_2(\mc{U}) + c_2(\mc{V}) \big)
+ c_1(\mc{V})^2 c_2(\mc{U}) + c_1(\mc{U})^2 c_2(\mc{V})
\bigg] c_{K+1}(\mc{R}) \,.
\label{eq:num_p1s_gen}
\end{aligned}
\ee
(This can be derived with the aid of the Thom-Porteous formula, as discussed below.)

\subsubsection*{The pair of 5-brane theories}

In the main text we have discussed how, as the deformation geometry is tuned to become the nodal variety, certain curves jump to become divisors, which fill out the new directions in the larger Picard group of the resolved geometry. (We note that these divisors on the nodal variety are special in that they are Weil but non-Cartier.) For a general $\mbb{P}^n$-split, these divisors are naturally described in terms of the matrix whose determinant describes the tuned nodal variety.

We can see this explicitly as follows. Define the matrices $M_{\hat{r}_i}$ and $M_{\hat{c}_j}$ as the $n \times (n+1)$ and $(n+1) \times n$ matrices resulting from removing the $i$th row or $j$th column from the matrix $M$ defined in \eqref{eq:mat_m}. In each case the locus where the rank of such a matrix drops below $n$ describes a divisor on the nodal variety $\nod{X}$,
\be
\begin{aligned}
\nod{D}_{\hat{r}_i} \colon \{ \mathrm{rank}(M_{\hat{r}_i}) \leq n-1 \} \subset \nod{X} \,, \\
\nod{D}_{\hat{c}_j} \colon \{ \mathrm{rank}(M_{\hat{c}_j}) \leq n-1 \} \subset \nod{X} \,.
\end{aligned}
\ee
Moving to the deformation geometry, these defining equations become independent of the equations describing the complete intersection, and so these divisors fall in dimension to become curves inside the deformation geometry $\defm{X}$.

Analogously to the main text, we now construct a natural pair of 5-brane theories on the deformation and resolution geometries. In particular, entirely analogously, we define the curves inside $\defm{X}$ and $\res{X}^{\,\mc{U}}$ on which the 5-branes are wrapped by beginning from one of the divisors $\nod{D}_{\hat{c}_j}$ associated with removing a column\footnote{If we instead constructed a pair of curves from one of the divisors $\nod{D}_{\hat{r}_j}$, we would find that the pair of 5-brane theories did not have the property that the remaining contributions to the anomaly cancellation conditions could be captured by spectator branes. Indeed, the divisors $\nod{D}_{\hat{r}_j}$ are instead the appropriate starting point to form a natural pair of 5-brane theories associated with the conifold transition $\defm{X} \to X^\mc{V}$.} from the matrix $M$.

The curve $\defm{C}_j$ on the deformation geometry naturally arises from the divisor $\nod{D}_{\hat{c}_j}$ as the geometry is deformed,
\be
\defm{C}_j \colon \{ \mathrm{rank}(M_{\hat{c}_j}) \leq n-1 \} \subset \defm{X} \,.
\ee
In contrast, the object $\nod{D}_{\hat{c}_j}$ remains a divisor as the small resolution is performed. In particular, we are interested in considering the proper transform $\pt{\nod{D}_{\hat{c}_j}}$. It is straightforward to check that this object is described simply and naturally as the zero locus of a particular corresponding coordinate in the ambient $\mbb{P}^n$ of $\res{X}^{\,\mc{U}}$,
\be
\pt{\nod{D}_{\hat{c}_j}} \colon \{ x^{\,\mc{U}}_j = 0 \} \subset \res{X}^{\,\mc{U}} \,.
\ee
To construct the curve on which we wish to wrap a 5-brane, we take the intersection of this object with the zero locus of a polynomial $\tilde{Q}$ of the same form that describes the deformation on the other side of the conifold transition, i.e.\ a section of the line bundle $\mathrm{det}(\mc{U}\oplus\mc{V})$. Hence
\be
\res{C}_j \colon \big( \{ x^{\,\mc{U}}_j = 0 \} \cap \{ \tilde{Q} = 0 \} \big)\subset \res{X}^{\,\mc{U}} \,.
\ee

Though the definition of the curve $\res{C}_j$ is the more involved, its class is easier to write down. It is simply
\be
[\res{C}_j] = \big(J_0 - c_1(\mc{U}_j) \big) \cdot \big( c_1(\mc{U}) + c_1(\mc{V}) \big) \,,
\ee
where we have noted that the class of the divisor with locus $\{ x^{\,\mc{U}}_j = 0 \} \subset \res{X}^{\,\mc{U}}$ is $J_0 - c_1(\mc{U}_j)$. On the other hand, the curve $\defm{C}_j$ is defined as a non-complete intersection, making its class more difficult to compute. However this computation is made possible by the Thom-Porteous formula. We explain this formula and perform the computation below. The result is that
\be
[\defm{C}_j] = c_2(\mc{V})-c_2(\mc{U})+c_1(\mc{U}) \,c_1(\mc{V})+c_1^2(\mc{U})
-c_1(\mc{U}_j) \, \big(c_1(\mc{U})+c_1(\mc{V})\big) \,.
\ee

Hence, analogously to what we saw in the simple example in the main text, recalling the expression for the class of the exceptional $\mbb{P}^1$s inside $\res{X}^{\,\mc{U}}$, we see that this pairing of curves precisely captures the difference in the second Chern classes of the deformation and resolution geometries,
\be
[\defm{C}_j] - [\res{C}_j] = [\mbb{P}^1\mathrm{s}] =
c_2\big(\defm{X}\big) - c_2\big(\res{X}^{\,\mc{U}}\big) \,.
\ee
Said differently, wrapping 5-branes on $\defm{C}_j$ and $\res{C}_j$ inside $\defm{X}$ and $\res{X}^{\,\mc{U}}$ leaves in the anomaly cancellation conditions only a `spectator' piece which can be trivially made up with spectator branes, namely
\be \label{eq:gen_spec_class}
\begin{gathered}
c_2(\defm{X}) - [\defm{C}_j] =
c_2(\res{X}) - [\res{C}_j] \\
= \mathrm{ch}_2(\mc{V}) - \mathrm{ch}_2(\mc{U}) + c_1(\mc{U}_j) \big(c_1(\mc{U})+c_1(\mc{V})\big) + \mathrm{ch}_2(\mc{R}) - \mathrm{ch}_2(\mc{A}) \,.
\end{gathered}
\ee
We see that the remaining difference depends on the index $j$, that is on the Weil non-Cartier divisor $\nod{D}_{\hat{c}_j}$ we used to construct the two 5-brane curves.

\subsubsection*{The Thom-Porteous formula}

It remains to compute the class $[\defm{C}_j]$ of the curve on the deformation side, which is described as a non-complete intersection, $\defm{C}_j \colon \{ \mathrm{rank}(M_{\hat{c}_j}) \leq n-1 \} \subset \defm{X}$.

For this purpose we can make use of the Thom-Porteous formula. Consider a morphism $U \to V$ between vector bundles on a smooth variety. The $k$th degeneracy locus $\big(k \leq \mathrm{min}(\mathrm{rk}\, U , \mathrm{rk}\,V) \big)$ of this morphism is the locus of points over which it has rank at most $k$. If all components of the degeneracy locus have the expected co-dimension $(\mathrm{rk}\,U - k)(\mathrm{rk}\,V-k)$, then the Thom-Porteous formula tells us that the fundamental class of the degeneracy locus is given by the determinant of the $(\mathrm{rk}\,U - k) \times (\mathrm{rk}\,U - k)$ matrix whose $(\alpha,\beta)$ entry is
\be
c_{\mathrm{rk}\,V - k + \alpha - \beta}(V/U) \,.
\ee

The $(n+1) \times (n+1)$ matrix $M$ defined in \eqref{eq:mat_m} can be viewed as a map between line bundle sums, namely
\be
M \colon \bigoplus_{l=0}^n \mc{U}_{\hspace{.4pt}l}^{\,-1} \;\to~\; \bigoplus_{l=0}^n \mc{V}_l \,.
\ee
The $(n+1) \times n$ matrix $M_{\hat{c}_j}$ resulting from deleting a column from $M$ can then be viewed as a map
\be
M_{\hat{c}_j} \colon \bigoplus_{l=0, \, l\neq j}^n \mc{U}_{\hspace{.4pt}l}^{\,-1} \;\to~\; \bigoplus_{l=0}^n \mc{V}_l \,.
\ee
Hence, defining for ease of notation $\mc{U}^- = \bigoplus_{l=0}^n \mc{U}_{\hspace{.4pt}l}^{-1}$, the class of the curve $\defm{C}_j \subset \defm{X}$ is given simply by
\be
[\defm{C}_j] = c_2 \bigg( \frac{\mc{V}}{\mc{U}^- / \mc{U}_j^{-1}} \bigg) \,.
\ee
Note that here that we implicitly take $c(\mc{V})$ etc.\ to be their restrictions to $\defm{X}$. After some algebra, noting in particular that $\mathrm{ch}(\mc{U}^-) = \mathrm{ch}_0(\mc{U})-\mathrm{ch}_1(\mc{U})+\mathrm{ch}_2(\mc{U}) - \ldots$, we find
\be
[\defm{C}_j] = c_2(\mc{V})-c_2(\mc{U})+c_1(\mc{U}) \,c_1(\mc{V})+c_1^2(\mc{U})
-c_1(\mc{U}_j) \, \big(c_1(\mc{U})+c_1(\mc{V})\big) \,.
\ee

As an aside we note that the computation of the number $\#(\mbb{P}^1\mathrm{s})$ of exceptional $\mbb{P}^1$s, or equivalently the number of nodal points on $\nod{X}$, can also be performed using the Thom-Porteous formula. In particular, these points correspond to the locus where $\mathrm{rank}(M) \leq n-1$, so that
\be
\#(\mbb{P}^1\mathrm{s}) = \int_{\mc{A}} \Big[ \big( c_2(\mc{V}/\mc{U})^2 - c_1(\mc{V}/\mc{U})\,c_3(\mc{V}/\mc{U}) \big) \Big] c_{K+1}(\mc{R}) \,,
\ee
(or equivalently with $\mc{U} \leftrightarrow \mc{V}$) which gives upon expansion the expression in \eqref{eq:num_p1s_gen}.

\end{document}